\documentclass[fleqn,usenatbib]{mnras}

\usepackage{newtxtext,newtxmath}
\usepackage{enumitem} % For customizing enumerate labels
\usepackage{esint}

\usepackage[T1]{fontenc}

\DeclareRobustCommand{\VAN}[3]{#2}
\let\VANthebibliography\thebibliography
\def\thebibliography{\DeclareRobustCommand{\VAN}[3]{##3}\VANthebibliography}

\usepackage{graphicx}	% Including figure files
\usepackage{amsmath}	% Advanced maths commands
\usepackage{subfig}     % Sub figure
\usepackage{placeins}   % Float barrier
\usepackage{float}      % Figure and table placement

\renewcommand{\d}{\ensuremath{\partial}}

\defcitealias{roberts_global_2025}{RLL25}
\defcitealias{iwasaki_dynamics_2024}{I24}
\defcitealias{jacquemin-ide_magnetic_2021}{J21}

\title[MHD simulations of protoplanetary discs II]{Global magnetohydrodynamic simulations of the inner regions of protoplanetary discs. II. Vertical-net-flux regime}

\author[M. J. O. Roberts, H. N. Latter and G. Lesur]{
Matthew J. O. Roberts$^{1}$,
Henrik N. Latter$^{1}$\thanks{E-mail: hl278@cam.ac.uk}
and Geoffroy Lesur$^{2}$
\\
$^{1}$Department of Applied Mathematics and Theoretical Physics, University of Cambridge, Centre for Mathematical Sciences, \\Wilberforce Road, Cambridge CB3 0WA, UK\\
$^{2}$Institut de Planétologie et d’Astrophysique de Grenoble, Université Grenoble Alpes, CNRS, IPAG, F-38000 Grenoble, France
}

\date{Accepted XXX. Received YYY; in original form ZZZ}

\pubyear{\the\year{}}

\begin{document}
\label{firstpage}
\pagerange{\pageref{firstpage}--\pageref{lastpage}}
\maketitle

\begin{abstract}
The inner regions of protoplanetary discs, which encompass the putative habitable zone, are dynamically complex, featuring a relatively well-ionised, turbulent active zone located interior to a poorly ionised `dead' zone. In this second paper, we investigate a model of the magnetohydrodynamic processes around the interface between these two regions, using five three-dimensional global magnetohydrodynamic simulations of discs threaded by a large-scale poloidal net-flux magnetic field. We employ physically motivated profiles for Ohmic resistivity and ambipolar diffusion, alongside a simplified thermodynamic model comprising a cool disc and hot corona. Our results show that, first, the interface acts as a one-way barrier to inward transport of large-scale magnetic flux from the dead zone. This leads to magnetic flux depletion throughout most of the active zone, whereby flux either advects inwards to the inner numerical boundary or accumulates just inside the interface. Second, two sources of strong variability emerge due to the difficulty of maintaining a constant, vertically integrated electrical current across the distinct and evolving magnetic-field states on either side of the interface. Third, despite the weak magnetothermal wind in the dead zone, a pressure maximum forms at the interface, leading to Rossby-wave-induced vortices. Fourth, unlike the model of Iwasaki et al.~(2024), there is no `transition zone' devoid of magnetic flux and magnetic winds. Instead, multiple outflow zones span all disc radii reflecting the radially varying launch conditions, with an inner turbulent wind impinging upon an outer, more laminar one. Fifth, a heated corona prevents the `puffing up' of poloidal-net-flux, active disc regions.
\end{abstract}


%%%%%%%%%%%%%%%%%%%%%%%%%%%%%%%%%%%%%%%%%%%%%%%%%%
%%%%%%%%%%%%%%%%% BODY OF PAPER %%%%%%%%%%%%%%%%%%

\section{Introduction}
\label{section:introduction}
% 0 - Link to first paper and hit the key acronyms.
Myriad and complex magnetohydrodynamic (MHD) processes operate in the inner regions of protoplanetary discs, thereby influencing the properties and evolution of the putative habitable zone. The most consequential feature is the dead--active zone interface, located at $\lesssim \!1\,\mathrm{au}$, beyond which turbulence driven by the magnetorotational instability \citep[MRI;][]{balbus_powerful_1991} is suppressed on account of the sharp drop-off in ionisation \citep{gammie_layered_1996}. Building on the zero-net-flux (ZNF) models presented in our first paper \citep[][hereafter \citetalias{roberts_global_2025}]{roberts_global_2025}, this study extends that analysis to the (weak) vertical-net-flux (VNF) regime, in which a large-scale poloidal magnetic field threads the disc, driving outflows and accretion in the MRI-dead region. \\
% 1 - Evidence for VNF regime 
\indent A growing body of evidence supports the conclusion that protoplanetary discs are threaded by a large-scale poloidal field. For example, (a) direct magnetic field measurements tentatively indicate a weak vertical component of $B_z\!\sim$ mG \citep[e.g.][]{vlemmings_stringent_2019, harrison_alma_2021, ohashi_observationally_2025, teague_radially_2025}; (b) theoretical arguments suggest a robust and efficient accretion mechanism is required in the MRI-dead zone; the magnetic-wind-driven paradigm \citep[][]{bai_wind-driven_2013, gressel_global_2015}, which presupposes a large-scale poloidal field threading the disc, satisfies this requirement and is consistent with observations of radially extended, chemically nested outflows originating from the inner disc \citep[e.g.][]{de_valon_alma_2020, pascucci_nested_2024}; and (c) the provenance of large-scale poloidal fields in protoplanetary discs is not problematic, whether they are inherited from the protostellar nebulae \citep[e.g.][]{crutcher_magnetic_2012}, as supported by observations of hourglass-shaped poloidal fields in Class 0 sources \citep[e.g.][]{stephens_magnetic_2013, tsukamoto_role_2023}, or generated \textit{in situ} via a dynamo process (\citealt{jacquemin-ide_magnetorotational_2024}, \citetalias{roberts_global_2025}). 
% 2 - The wealth of new scientific questions beyond those explored in the ZNF regime

\indent A VNF introduces a wealth of physics excluded from the ZNF regime. Now that the dead zone supports magnetic winds, the accretion mismatch across the dead--active zone interface may be less pronounced, which has implications for the formation of dust-trapping hydrodynamic (HD) structures -- rings and vortices -- that promote planetesimal growth \citep[e.g.][]{barge_did_1995, kretke_assembling_2009}. On the other hand, it imposes a new mismatch: in global magnetic flux transport between the `active' ideal MHD region (radially inward) and the `dead' non-ideal MHD region (radially outward) (e.g. \citealt{gressel_global_2020}; \citealt[hereafter \citetalias{jacquemin-ide_magnetic_2021}]{jacquemin-ide_magnetic_2021}; \citealt{lesur_hydro-_2023}). In attempting to accommodate this mismatch, the disc must regulate the strength of MRI turbulence \citep{hawley_local_1995} and the efficiency of its magnetic outflows \citep[e.g.][]{lesur_systematic_2021}. Generally, radial transitions between different large-scale magnetic-field structures -- such as in the VNF MRI-active and dead regions, as well as the disc's net-poloidal field and the protostar's dipole \citep[e.g.][]{romanova_accretion_2015} -- likely drive the ubiquitous accretion variability localised to the inner disc \citep[e.g.][]{cleaver_magnetically-activated_2023, fischer_accretion_2023}. Finally, the interplay between turbulent outflows from the MRI-active region and adjacent laminar outflows from the dead zone \citep[e.g.][]{tu_highly_2025} probably control the observed morphology of chemically nested conical outflows originating from the inner disc (see above). \\
% 3 - Challenges of global simulations and previous numerical work
\indent Whilst predictions based solely on either global MRI-active (\citealt{zhu_global_2018}; \citealt{mishra_strongly_2020}; \citetalias{jacquemin-ide_magnetic_2021}) or dead \citep[e.g.][]{bai_global_2017, bai_hall_2017, bethune_global_2017, gressel_global_2020, lesur_systematic_2021} models are informative, understanding the inner-disc dynamics requires global three-dimensional VNF non-ideal MHD models that include the dead--active zone interface. To date, only three studies -- conducted independently -- have done so: (i) \citet{flock_3d_2017}, whose limited meridional domain precludes a magnetic wind; (ii) \citet{iwasaki_dynamics_2024} (hereafter \citetalias{iwasaki_dynamics_2024}), which provides the only comparable point in global parameter space to this work and thus serves as the principal reference for our results; and (iii) \citet{tu_highly_2025,tu_yso_2025}, which examines inner-disc outflows, but with a poorly resolved disc. \\
% 4 - Paper aim and focus
\indent The aim of this second paper is to further investigate MHD-mediated processes near the dead--active zone interface. We use GPU-accelerated simulations performed with \textsc{idefix} \citep{lesur_idefix_2023} and a simplified numerical setup: an initially weak VNF magnetic field, physically motivated prescriptions for Ohmic and ambipolar diffusion, and a two-temperature structure representing a cool disc embedded in a hot corona. Within this framework we focus on magnetic flux transport, inner-disc variability, and the interplay between laminar and turbulent flows, particularly in relation to accretion, winds, and large-scale HD structure formation. \\
% 5 - Paper structure 
\indent The paper is structured as follows. In Section~\ref{section:methods_and_model} we outline the model and numerical methods, with additional details provided in \citetalias{roberts_global_2025}, before presenting an overview of the evolution in Section~\ref{section:results_overview}. We investigate the results in four parts: accretion flows and associated structures in Section~\ref{section:results_accretion}; current-sheet configurations at the interface in Section~\ref{section:nf_flux_current_sheet}; magnetic flux transport in Section~\ref{section:results_global_flux_transport}; and the outflow structure in Section~\ref{section:results_outflows}. Then, in Section~\ref{section:discussion} we compare our results with \citetalias{iwasaki_dynamics_2024}, before presenting the concluding remarks in Section~\ref{section:conclusion}. 

%%%%%%%%%%%%%%%%%%%%%%%%%%%%%%%%%%%%%%%%%%%%%%%%%%%%%%%%%%%%%%%%%%%%%%%%%%%%%%%%%%%%%%%%%%%%%%%%%%%%
%%%%%%%%%%%%%%%%%%%%%%%%%%%%%%%%%%%%%%%%%%%%%%%%%%%%%%%%%%%%%%%%%%%%%%%%%%%%%%%%%%%%%%%%%%%%%%%%%%%%

\section{Model and Numerical Methods}
\label{section:methods_and_model}

\subsection{Governing equations}
\label{section:equations}
The evolution of our inner-disc model is governed by the non-ideal MHD equations, with Ohmic and ambipolar diffusion included but the Hall effect neglected. These determine the evolution of the density $\rho$, velocity field $\mathbf{u}$, magnetic field $\mathbf{B}$, and energy density $E=e+\rho u^2/2+B^2/8\pi$, where $e$ is the internal energy density, through,
\begin{align}
    \partial_t{\rho} &= -\mathbf{\nabla} \cdot \left(\rho \mathbf{u}\right) 
     \label{equation:continuity_equation} 
     \\
    \partial_t{(\rho\mathbf{u})} &= - \mathbf{\nabla} \cdot \left( \rho\mathbf{u}\mathbf{u}\right)-\rho\mathbf{\nabla}\Phi - \mathbf{\nabla}{P} + \dfrac{\mathbf{J}\times\mathbf{B}}{c}
    \label{equation:momentum_equation} \\
    \partial_t\mathbf{B} &= -\nabla \times \left(\mathbf{\mathcal{E}_I} + \mathbf{\mathcal{E}_{NI}}\right) \label{equation:induction_equation} \\
    \label{equation:energy_equation}
    \partial_tE &= - \nabla \cdot \left[ \left(E + P +\dfrac{B^2}{8\pi}\right) \mathbf{u} - \dfrac{\mathbf{B} (\mathbf{B} \cdot \mathbf{u})}{4\pi} -\dfrac{\mathbf{\mathcal{E}_{NI}}\times\mathbf{B}}{4\pi}\right]\\&\;\,\,\;-\rho \mathbf{u} \cdot \nabla \Phi + \mathcal{L} \nonumber,
\end{align}
where the ideal and non-ideal components of the total electromotive field, $\mathcal{E} = \mathbf{\mathcal{E}_{I}} + \mathbf{\mathcal{E}_{NI}}$, are given by
\begin{equation}
	\mathbf{\mathcal{E}_{I}} = \mathbf{u}\times\mathbf{B} \;\;\; \textrm{and} \;\;\; \mathbf{\mathcal{E}_{NI}} = \dfrac{4\pi}{c} \left[  - \eta_\text{O} \mathbf{J}+ \eta_\text{A} (\mathbf{J} \times \mathbf{b}) \times \mathbf{b} \right]
\end{equation}
under the Gaussian unit convention. Here, $\Phi=-GM/r$ is the gravitational potential of a central spherical star of mass $M$ in spherical coordinates  $(r,\theta,\phi)$, $P$ the gas pressure, $c$ the speed of light, $\mathbf{J}= (c/4\pi)\mathbf{\nabla}\times\mathbf{B}$ the current density, $\mathcal{L}$ an imposed energy-density thermal-relaxation term, $\mathbf{b}$ the direction of the magnetic field, $\eta_\text{O}$ the Ohmic diffusivity, and $\eta_\text{A}$ the ambipolar diffusivity. \\
\indent The system of equations is closed using the equation of state for a perfect gas, $P=e(\gamma-1)$, where $\gamma$ is the adiabatic index. To ensure that the integration scheme is positive-definite with respect to $P$, particularly in low-density and highly magnetised regions, we adopt $\gamma=1.05$. The disc temperature is controlled through a $\beta$-cooling prescription: the system is relaxed towards a target temperature, $T_\textrm{eff}(R,z)$, using,
\begin{equation}
    \mathcal{L}= \dfrac{P - {T_{\textrm{eff}}\rho}}{\tau(\gamma-1)}.
    \label{equation:temperature_beta_model}
\end{equation}
 The characteristic thermal-relaxation timescale is $\tau=0.1\Omega_\text{K}^{-1}$, so that the vertical shear instability \citep[VSI;][]{nelson_linear_2013} is marginal, and $\Omega_\text{K}^2=GMR^{-3}$ is the squared Keplerian angular frequency. This rapid thermal readjustment leads to an effectively quasi-local isothermal equation of state, $P=c_\text{s}^2\rho$, where the local isothermal sound speed in the disc is $c_\text{s}=H\Omega_\text{K}$, for the disc scale height $H$. Finally, cylindrical coordinates $(R,\phi,z)$ are also used.

\subsection{Inner-disc model}
\label{section:inner_disc_model}

\begin{figure*}
    \centering  
    \subfloat{\includegraphics[height=0.4\textwidth]{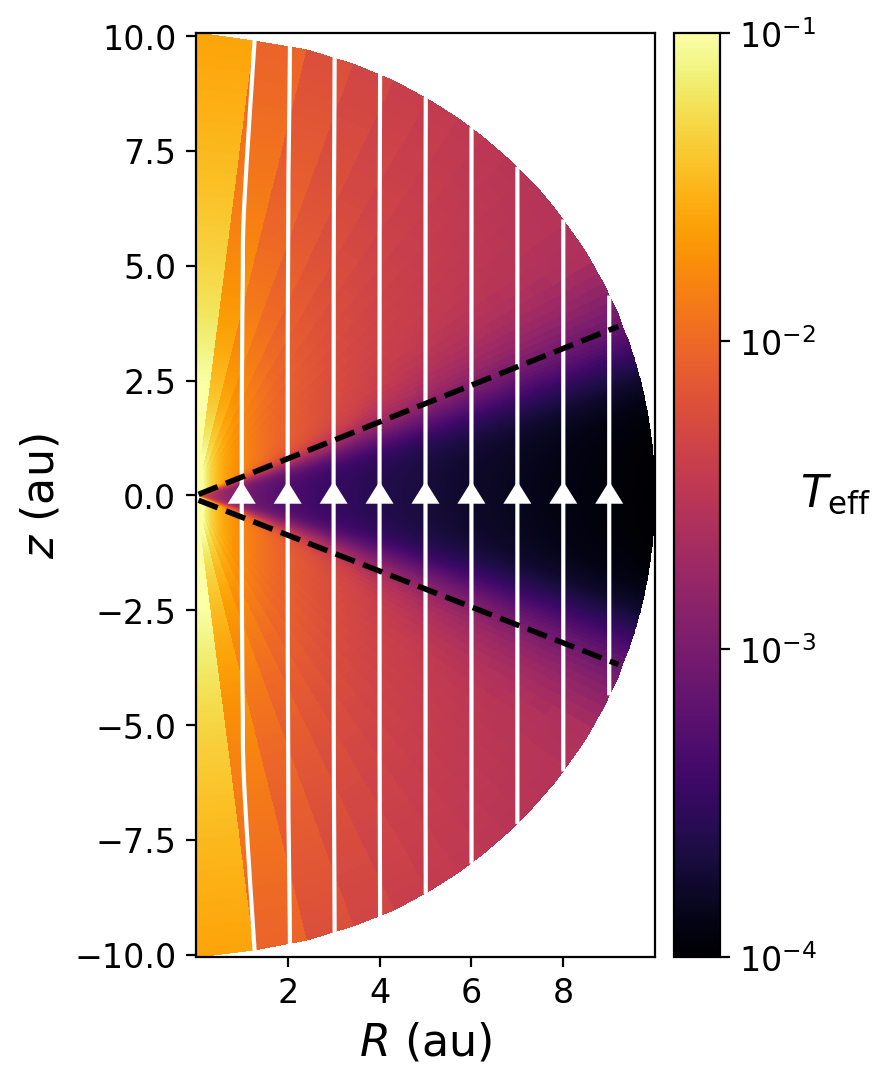}} 
    \subfloat{\includegraphics[height=0.4\textwidth]{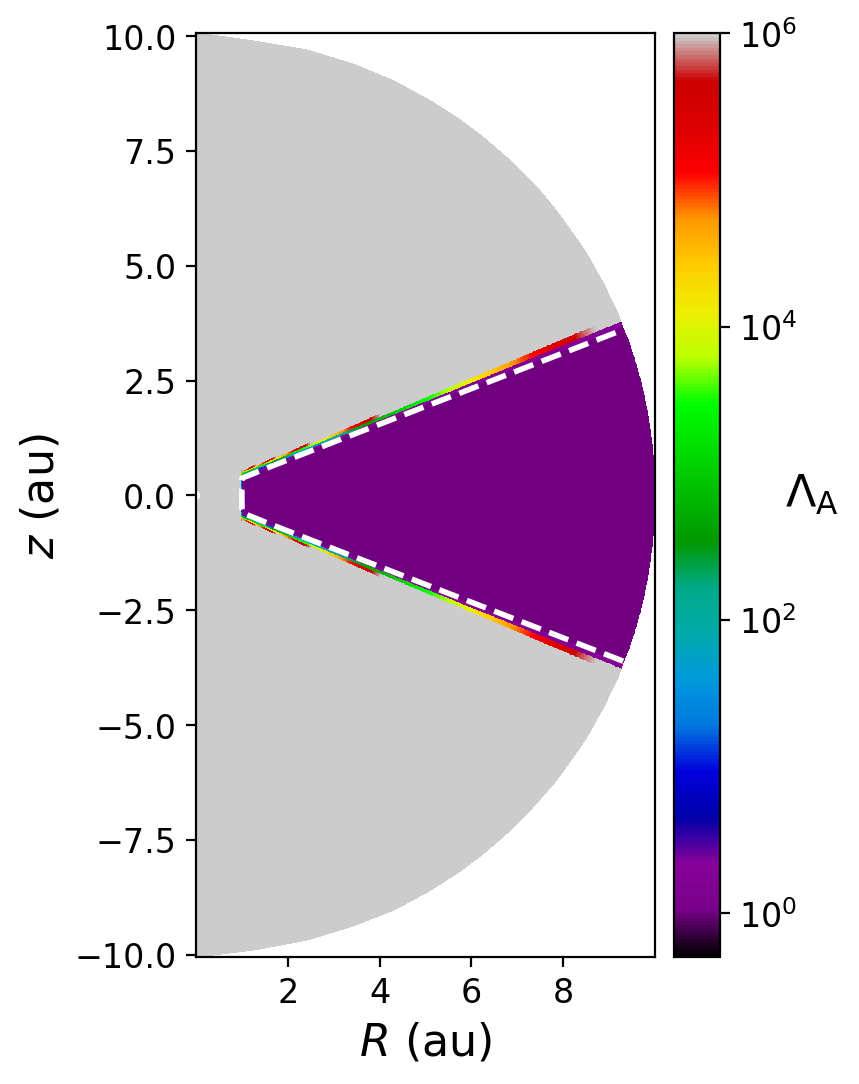}}
    \subfloat{\includegraphics[height=0.4\textwidth]{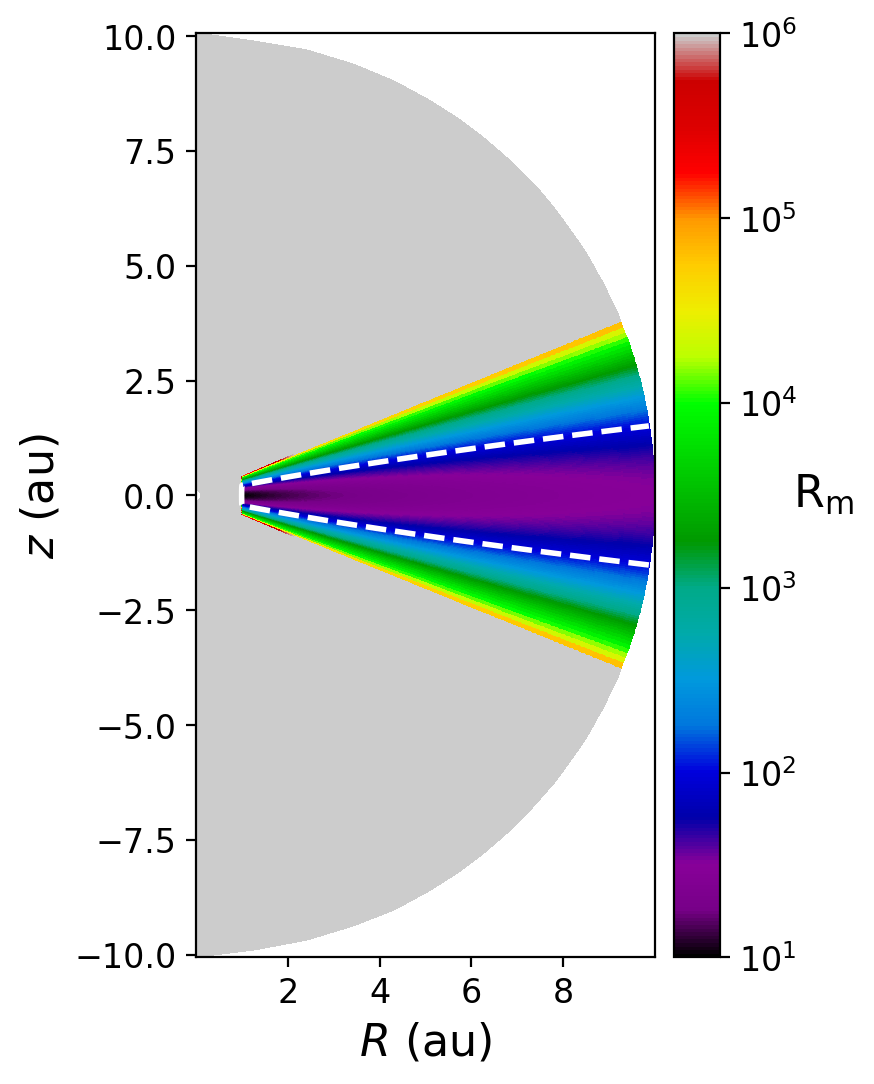}}
    \caption{Meridional $(R,z)$ plots of the prescribed target temperature $T_\text{eff}/T_0$ (left), ambipolar Elsässer number $\Lambda_\text{A}$ (middle) and magnetic Reynolds number $\text{R}_\text{m}$ (right) for \texttt{NF-BAZ}, as outlined in Section \ref{section:inner_disc_model}. The initial magnetic field configuration, which is defined in equation~\eqref{eqn:inital_magnetic_field}, is exhibited through certain magnetic field lines in the left panel (white arrowed lines), and the dashed black lines denote the disc--corona transition. Meanwhile, the dashed white lines delineate the following empirically derived stratified net-flux stability conditions for the MRI; R$_\text{m}\lesssim \sqrt{\beta} \sim 10^{2}$ \citep{latter_vertical_2022} and $\Lambda_\text{A}\lesssim 1$ \citep{bai_effect_2011, simon_turbulence_2013}, which is less certain. For clarity, the non-ideal MHD diffusivities are limited to $10^6$, and the Ohmic-resistive inner radial buffer region is not shown.}
    \label{fig:non_ideal_MHD_setup}
\end{figure*}

\subsubsection{Temperature prescription}
\label{section:temp_prescription}

The target temperature profile for the $\beta$-cooling prescription \eqref{equation:temperature_beta_model},
\begin{equation}
    T_{\textrm{eff}}(R,z) = \dfrac{1}{2}\left[ \left(T_{\textrm{d}}+T_{\textrm{c}}\right)+\left(T_{\textrm{d}}-T_\textrm{c}\right)\tanh\left(\dfrac{\left|z\right| - 4\varepsilon R_\text{min}}{R_0} \right)\right],
    \label{equation:target_temperature}
\end{equation}
is exhibited in the left panel of Fig. \ref{fig:non_ideal_MHD_setup}. The model splits the domain into a cold disc with temperature $T_\textrm{d}(R) = T_0R_0\varepsilon^2/R_\text{min}$, and a hot corona with temperature $T_\textrm{c}(R) = 16T_\textrm{d}(R)$, at a transition height of $|z|=4H$. Here $\varepsilon=H/R$ is the disc aspect ratio, the subscript $0$ refers to the midplane at the inner edge of the domain, in most cases, and $R_\text{min}=\max(R,R_0)$. Note that $T_0$ is a reference temperature, not the midplane temperature at the innermost radius.\\
\indent Thus the radial profile of the effective temperature scales as $T\propto R^{-1}$ when $R\geq R_0$. This enforces a constant aspect ratio. We adopt $\varepsilon=0.1$ as a compromise between numerical efficiency -- ensuring that the MRI is adequately resolved -- and the expected inner-disc thickness of roughly 0.03--0.05 at $R=1$ au \citep[e.g.][]{bitsch_structure_2015}. 
\subsubsection{Non-ideal MHD model and the dead--active zone interface}

We adopt a spatially static, physically motivated non-ideal MHD setup that includes both Ohmic and ambipolar diffusion to model the dead--active zone interface. The magnetic Reynolds and ambipolar Elsässer numbers are defined as,
\begin{equation}
\text{R}_\text{m} = \dfrac{\Omega_\text{K} H^2}{\eta_\text{O}}
\;\;\; \text{and} \;\;\;
\Lambda_\text{A} = \dfrac{v^2_\text{A}}{\Omega_\text{K} \eta_\text{A}},
\label{equation:non_ideal_MHD_diffusivities}
\end{equation}
where $v_\text{A}=B/\sqrt{4\pi\rho}$ is the Alfvén speed. Their profiles are prescribed as,
\small
\begin{align}
 \label{equation:non_ideal_MHD_prescription}
 \text{R}_\text{m}(R,z) =& \,\dfrac{4\text{R}_{\text{m}_0} \rho_0 R_0}{\rho R} 
\left[1 - \tanh\left(\dfrac{|z| - 4\varepsilon R}{0.2 R_0}\right)\right]^{-1} \\ &\times
\left[1 + \tanh\left(\dfrac{R - R_{\textrm{DZI}}}{0.02 R_0}\right)\right]^{-1},  \nonumber \\
\Lambda_\text{A}(R,z) =& \,4\Lambda_{\text{A}_0}\left[1 - \tanh\left(\dfrac{|z| - 4\varepsilon R}{0.2 R_0}\right)\right]^{-1} \left[1 + \tanh\left(\dfrac{R - R_{\textrm{DZI}}}{0.02 R_0}\right)\right]^{-1} \nonumber, 
\end{align}
\normalsize
\noindent
as shown in Fig.~\ref{fig:non_ideal_MHD_setup}, with $\Lambda_{\text{A}_0} = 1$ and $\text{R}_{\text{m}_0}$ chosen in each model such that $\text{R}_{\text{m}}=20$ at the dead--active zone interface ($R=R_{\textrm{DZI}}$), guided by dust- and metal-free estimates from fig. 8 in \citet{thi_radiation_2019} and fig. 10 in \citet{lesur_magnetohydrodynamics_2021}. The ionisation in the surface layers is assumed to become effective at the same height as the disc--corona temperature transition, $|z|\geq4H$. 

\subsection{Numerical Methods}
\label{section:numerical_methods}

The simulations presented in this work are performed using the \textsc{idefix} code, which integrates the compressible MHD equations with a finite-volume high-order Godunov method. The interpolation to cell interfaces is performed using a linear piecewise reconstruction scheme with a van Leer slope limiter, and inter cell-fluxes are computed using the HLLD approximate Riemann solver. The quantities are evolved in time using a total-variation diminishing second-order Runge--Kutta method with the Courant number set to $\text{C}=0.5$. The solenoidal condition ($\mathbf{\nabla}\cdot\mathbf{B}=0$) is enforced to machine precision via the constrained transport method \citep{evans_simulation_1988} and an electromotive-field reconstruction based on the $\mathcal{E}^\text{c}$ scheme \citep{gardiner_unsplit_2005}. Parabolic non-ideal MHD diffusivity terms are handled by a centred finite-difference method. All simulations were run on NVIDIA A100 GPUs using double-precision arithmetic.

\subsubsection{Simulations and numerical protocol}

\indent In total, we performed five simulations, outlined in Table~\ref{table:list_net_flux_global_simulations}: a fiducial \textit{big-active-zone} model, \texttt{NF-BAZ}, with $R_\text{DZI}=10R_0$, and four variants that modify the numerical setup, two of which also include a dead--active zone interface.\footnote{\texttt{NF-BAZ} used roughly $11{,}000$ GPU-hours to perform $1.3\times10^8$ integration cycles, achieving $2.8\times10^{9}$ cell updates in total per second across 32 GPUs, using a 32-1-1 decomposition in $(r,\theta,\phi)$.} \\
\indent The two VNF models without a dead--active zone interface are presented in Appendix~\ref{appendix:turbulent_versus_laminar_discs}: \texttt{NF-AZ} is a purely MRI-active disc, in which the non-ideal MHD prescription \eqref{equation:non_ideal_MHD_prescription} is neglected; and \texttt{NF-DZ} is a purely MRI-dead disc, in which the non-ideal MHD model is extended to the inner radial boundary. \\
\indent The \textit{small-active-zone} model, \texttt{NF-SAZ}, extends the evolution of the system at the interface and provides a second data point for large-scale HD structure formation. Rather than simply halve the resolution -- which would risk under-resolving the MRI -- we (i) reduce the radial domain size, resolution and interface location (see Table~\ref{table:list_net_flux_global_simulations}; right panel of Fig.~\ref{fig:BAZ_SAZ_grid_comparison}); and (ii) activate the non-ideal MHD prescription with $R_\text{DZI}=4R_0$ on top of the fully MRI-dead disc, \texttt{NF-DZ}, at $t_\text{in}=200$. This moment sets $t_\text{in}=0$ for \texttt{NF-SAZ}.\\
\indent Finally, \texttt{NF-SAZ-BC}, which is detailed and examined in Appendix~\ref{appendix:boundary_test}, tests the robustness of the results to a different inner boundary condition (BC). \\
\indent Throughout the rest of this section, we state the numerical setup for \texttt{NF-BAZ} and the key differences for \texttt{NF-SAZ}. Appendix~\ref{appendix:units} details the conversion from code to physical units, assuming a disc around a $1\,\text{M}_\odot$ star, with surface density \(\Sigma(R) = 300\,(R/1\,\text{au})^{-1/2}\,\mathrm{g}\,\mathrm{cm}^{-2}\) and an aspect ratio of $\varepsilon =0.1$. 

\begin{table*}
\centering
\caption{Simulation properties, each initialised with midplane $\beta=10^4$ and integrated to $t_{\textrm{in}}^{\textrm{end}}$ over a domain spanning $r\!\in\!\left[R_0,r_{\textrm{max}}\right]$, $\theta\!\in\!\left[0,\pi\right]$,  $\phi\!\in\!\left[0,\phi_{\textrm{max}}\right]$, with inner buffer width $r_\text{buf}$. The dead--active zone models are listed in the upper portion, and the total computational cost was $\sim\!20{,}000$ GPU-hours.}
% Stretch the table a bit
\renewcommand{\arraystretch}{1.3}
\setlength{\tabcolsep}{6pt}      % default ≈ 6pt; reduce gently
\begin{tabular}{c c c c c c c c c c c} 
\hline
 Model name & $\left(N_r,N_\theta,N_\phi\right)$ & $r_{\textrm{max}}$ & $r_{\text{buf}}$ & $\phi_{\text{max}}$ & $R_{\text{DZI}}$ & $\text{R}_{\text{m}_0}$ & $t_{\text{in}}^{\text{end}}$ & $t_{\text{DZI}}$ & $t_{\text{DZI}}^{\text{end}}$ & Used \\ [0.5ex] 
\hline 
\texttt{NF-BAZ} & $(1024,192,128)$ & $100R_0$ & $1.25R_0$ & $\pi/2$ & $10R_0$ & $\sqrt{10}$ & 2600 & $31.6t_\text{in}$ & $82$ & Sections~\ref{section:results_overview}--\ref{section:results_outflows} \\
\texttt{NF-SAZ} & $(512,192,128)$ & $40R_0$ & $1.5R_0$ & $\pi/2$ & $4R_0$ & 5 & 1950 & $8t_\text{in}$ & $244$ & Sections~\ref{section:emergence_MRI}, \ref{section:nf_flux_HD_str}, \ref{section:nf_flux_current_sheet} and \ref{section:mft_burps} \\
\texttt{NF-SAZ-BC} & $(512,192,128)$ & $40R_0$ & N/A & $\pi/2$ & $4R_0$ & 5 & 800 & $8t_\text{in}$ & $100$ & Sections~\ref{section:nf_flux_HD_str}, \ref{section:mft_burps} and Appendix~\ref{appendix:boundary_test}  \\
\hline 
\texttt{NF-AZ} & $(1024,192,64)$ & $100R_0$ & $1.25R_0$ & $\pi/4$ & N/A & N/A & 1200 & N/A & N/A & Appendix~\ref{appendix:turbulent_versus_laminar_discs} \\
\texttt{NF-DZ} & $(512,192,64)$ & $40R_0$ & $1.25R_0$ & $\pi/4$ & N/A & 5 & 700 & N/A & N/A & Appendix~\ref{appendix:turbulent_versus_laminar_discs} \\
\hline 
\end{tabular}
\label{table:list_net_flux_global_simulations}
\end{table*}

\subsubsection{Initial conditions} 

The initial HD fields are set to obey approximate radial and vertical hydrostatic equilibrium in the disc (after neglecting the Lorentz force) via \citep[e.g.][]{nelson_linear_2013},
\begin{align}
    \rho(R,z) &= \rho_0\left(\dfrac{R_\text{min}}{R_0}\right)^{-3/2}\exp\left[\dfrac{GM}{c_\text{s}^2}\left(\dfrac{1}{r}-\dfrac{1}{R}\right)\right], 
    \label{eq:density_initial_condition} \\
    \label{eq:uphi_initial_condition}
    u_\phi(R,z)  &=
    \begin{cases}
    \left(\dfrac{GM}{R}\right)^{1/2} \sqrt{\dfrac{R}{r} - \dfrac{5}{2}\left(\dfrac{H}{R}\right)^{\scriptscriptstyle 2} } , & \text{if $R\geq R_0$} \\
    \rule{0pt}{2.0em}
    R\left(\dfrac{GM}{R_{0}^3}\right)^{1/2}, & \text{otherwise},      \end{cases}
\end{align}
such that the initial gas surface density $\Sigma=\int^\infty_{-\infty}\rho\,\text{d}z\propto R^{-1/2}$. The above expressions are not at equilibrium in the corona because of the larger temperature in this region. The remaining velocity components are initialised with sub-sonic ($\sim\!10^{-3}c_\text{s}$) white noise in the inner disc. \\
\indent The initial VNF magnetic field configuration is set using the vector potential $\mathbf{A}$ as \citep[e.g.][]{zhu_global_2018},
\begin{equation}
	A_\phi =
	\begin{cases}
		\,\,B_\text{i}\left(\dfrac{R_{0}^2}{R} \left[\dfrac{1}{2} - \dfrac{1}{2 + m} \right] + \dfrac{R^{1 + m}}{R_0^m(2 + m)} \right),\; & \text{if } R \geq R_0 \\[2.0ex]
		\,\,\dfrac{B_\text{i} R}{2},\; & \text{otherwise},
	\end{cases}
	\label{eqn:inital_magnetic_field}
\end{equation}
where $m=-5/4$ for our choice of initial radial profiles for $T$ and $\rho$. Equation~\eqref{eqn:inital_magnetic_field} corresponds to an initial magnetic field of the form,
	\begin{equation}
	B_z = B_\text{i}\left(\dfrac{R}{R_0}\right)^{-5/4} ,\;  \text{if}\;R\geq{R}_0,
	\label{eqn:mean_field_paper_inital_magnetic_field_b_not_a}
	\end{equation}
which is portrayed by white lines in the left panel of Fig.~\ref{fig:non_ideal_MHD_setup}. We initialise the field strength $B_\text{i}$, to yield a uniform midplane plasma-$\beta$ of $10^4$, placing the system in the weak-field regime.\footnote{The initial vertical field strength in \texttt{NF-BAZ} is $B_z =2.1\,\text{mG}$ at $R = 10\,\text{au}$, roughly in line with tentative measurements \citep[e.g.][]{harrison_alma_2021}.}

\subsubsection{Grid, resolution and units}

\begin{figure}
    \centering
	\includegraphics[height=0.46\columnwidth]{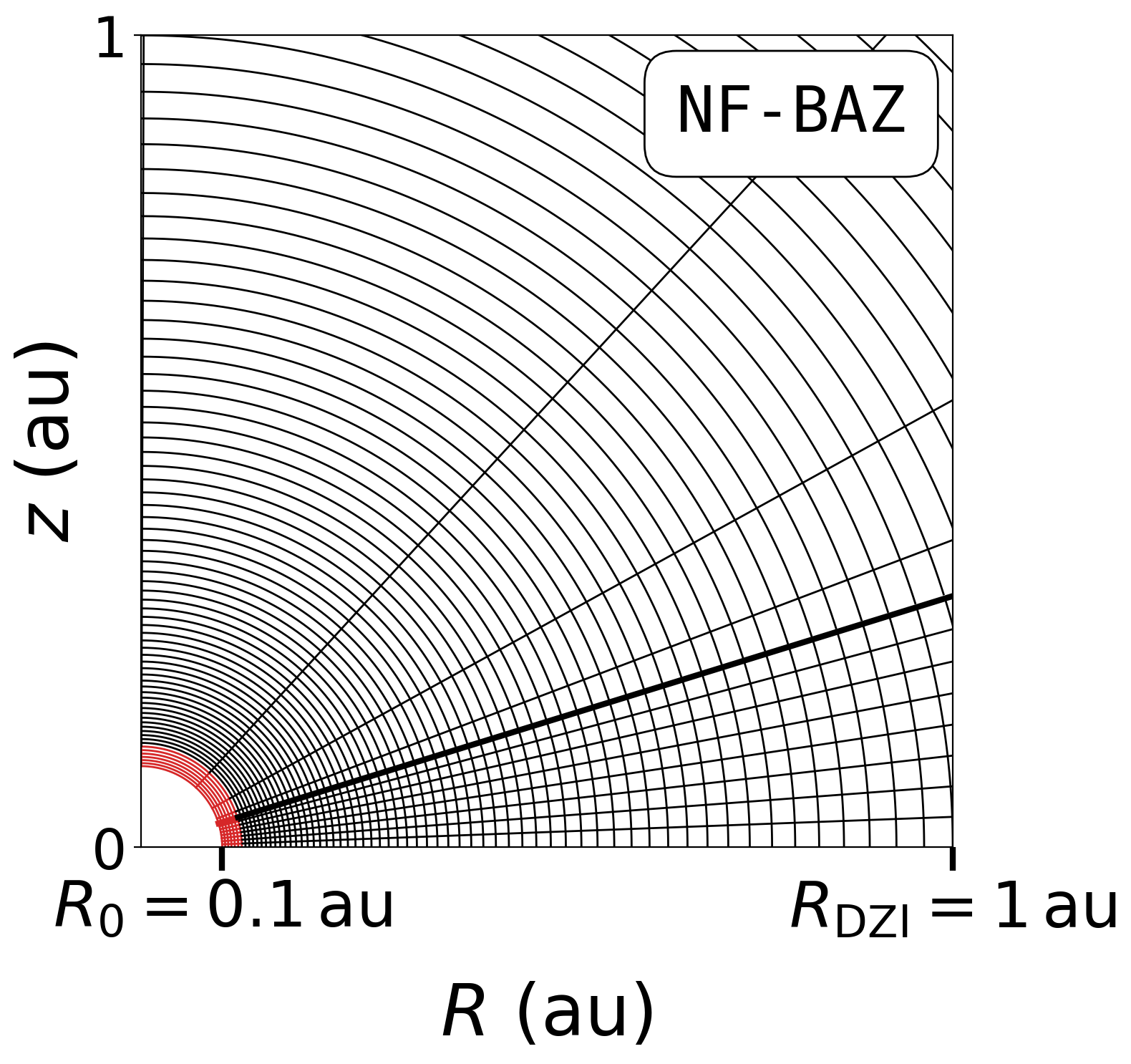}
	\includegraphics[height=0.46\columnwidth]{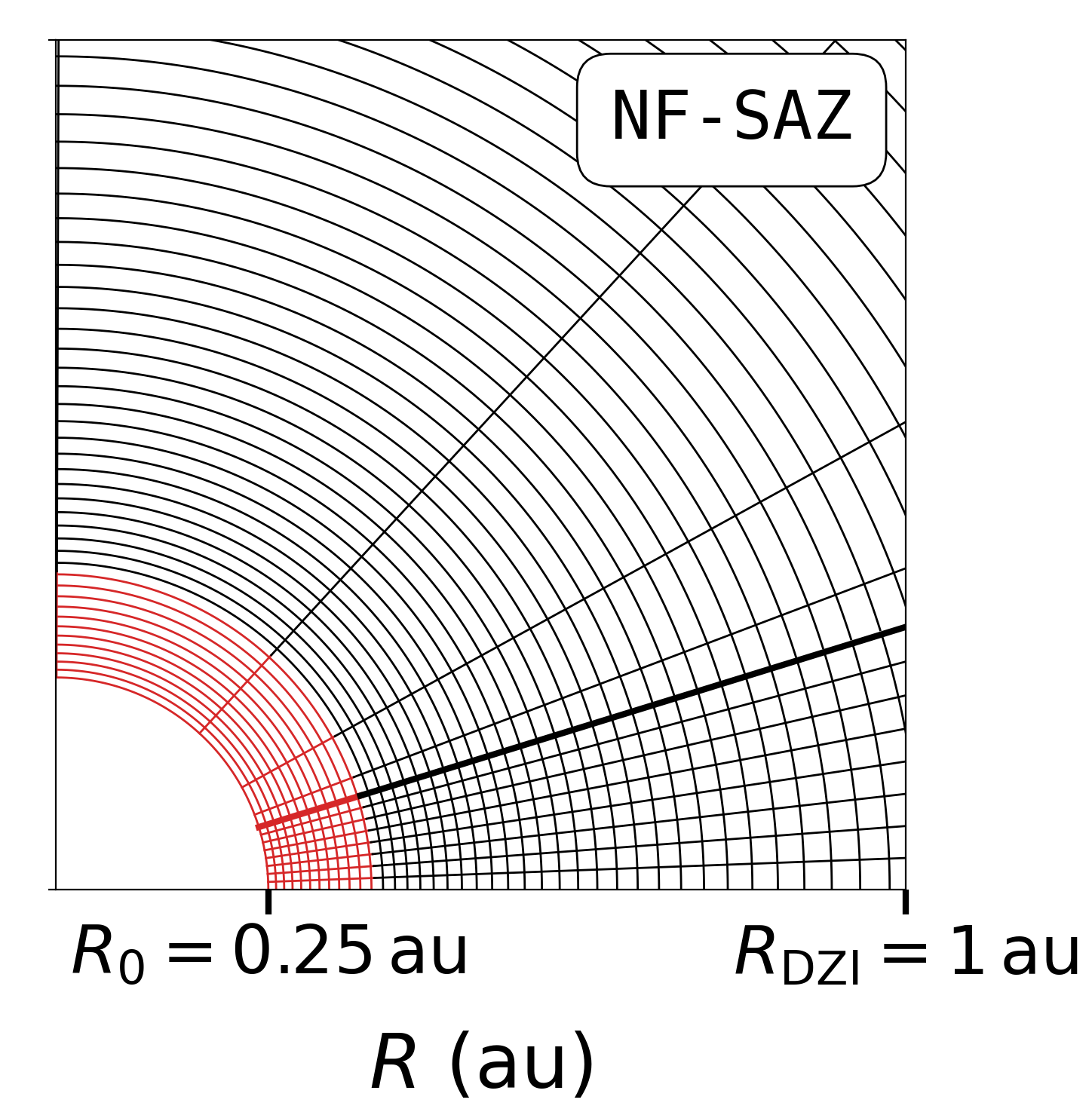}
    \caption{Meridional $(R,z)$ plots of the numerical grid in the upper half-plane, showing the inner regions for \texttt{NF-BAZ} (left) and \texttt{NF-SAZ} (right), with $R_{\text{DZI}}$ normalised to $1\,\text{au}$. The inner radial edge of the domain is $R_{0}=1$ in code units, whilst the buffer zone (red region) extends to $r_{\text{buf}}$ (see Table~\ref{table:list_net_flux_global_simulations}). The bold line denotes the disc--corona temperature transition at $|z|=4H$, beyond which the grid is geometrically stretched in the $\theta$ direction. The meshes are plotted by every eight cells in both directions for illustrative purposes.}
    \label{fig:BAZ_SAZ_grid_comparison}
\end{figure}

The numerical details of the fiducial simulation \texttt{NF-BAZ} are described here and illustrated in the left panel of Fig.~\ref{fig:BAZ_SAZ_grid_comparison}, with changes made in the other four simulations outlined in Table \ref{table:list_net_flux_global_simulations}. In spherical radius, a logarithmic grid with 1024 cells spans $r\!\in\![R_0,100R_0]$, such that the radial cell width scales as $\delta{r}\propto{r}$. The azimuthal domain covers $\phi \!\in\! [0, \pi/2]$ with 128 uniformly spaced cells. In the meridional direction, 128 uniformly spaced cells span $\theta \!\in\! [1.27, 1.87]$, corresponding to $z = \pm 4H$ about the midplane. This is flanked by two geometrically stretched buffer regions of 32 cells each in $\theta \!\in\! [0, 1.27]$ and $\theta \!\in\! [1.87, \pi]$, respectively, extending the grid to the polar axis. \\
\indent To prevent the time step being controlled by small cells at the poles, the grid coarsening level $\ell_c$, which is outlined in the \textsc{idefix} documentation, is set as
\begin{equation}
	\ell_c = \max\left( \left\lfloor \dfrac{1}{\sin(\theta)} \right\rfloor, \;6 \right)
	\label{eqn:mean_field_paper_grid_coarsening}
\end{equation}
in the azimuthal direction, where $\lfloor\,\rfloor$ is the floor function, such that there are only four effective cells in azimuth at the polar axis. \\
\indent We define two reference timescales:  $t_{\text{in}}=2\pi/\Omega_\text{K}(R_0)$ corresponding to the orbital period at the innermost radius, and $t_{\textrm{DZI}}=2\pi/\Omega_\text{K}(R_{\textrm{DZI}})$ corresponding to the orbital period at the dead--active zone interface. As outlined in Table~\ref{table:list_net_flux_global_simulations}, $t_{\textrm{DZI}}\!\approx\!31.6t_{\text{in}}$ in \texttt{NF-BAZ}, whilst $t_{\textrm{DZI}}=8t_{\text{in}}$ in the \texttt{NF-SAZ} models. To facilitate local radial comparisons between \texttt{NF-BAZ} and the \texttt{NF-SAZ} models, the number of orbits at radius $R_\text{j}$ is defined as $t_{R=R_\text{j}}$. Code units are chosen such that $GM=\rho_0=T_0=R_0=1$. \\
\indent Finally, the scale-free nature of the setup is broken by the dead--active zone interface, which imposes a fixed location for the sharp ionisation transition in the inner disc. Consequently, we normalise the length scales by ${R_{\textrm{DZI}}}=1$ au, thereby consistently emphasising that this model applies only to the inner disc.\footnote{Whilst numerically convenient, choosing $\varepsilon=0.1$ leads to an effective midplane temperature 2--3 times higher than the expected 800--1000 K at $R=R_\text{DZI}$ (see Appendix~\ref{appendix:units}). Given our very qualitative thermodynamic model, this will have little impact on the dynamics simulated.} However, we stress that $R_0=1$ for the purposes of interpreting the analysis of the fields, which are presented in code units.

\subsubsection{Boundary and internal conditions}
\label{section:boundary_conditions}

All the simulations, except for \texttt{NF-SAZ-BC} (see Appendix~\ref{appendix:boundary_test} for details), use the following boundary and internal conditions. At the inner radial boundary, the ghost cells take $\rho$, $P$, and $u_\theta$ from the innermost active cell; set $B_\theta$ and $B_\phi$ to zero; and fix $u_\phi=R\Omega_{\text{K}}(R_0)$ to impose solid-body rotation at the inner hole of the spherical domain. To prevent inflow from the boundary, $u_r$ is copied into the ghost cells if $u_r<0$ in the first active cell; otherwise, it is reflected symmetrically across the boundary. At the outer radial boundary, the ghost cells take $\rho$, $P$, $u_\theta$, $u_\phi$, and $B_\theta$ from the innermost active cell; set $B_\phi$ to zero to prevent toroidal-field loops embedding into the boundary \citep[e.g.][]{gressel_global_2020}; and copy $u_r$ if $u_r>0$ in in the first active cell, otherwise it is set to zero. In the meridional direction the axis-regularisation BC \citep[see appendix A in][]{zhu_global_2018} is used, due to the closure of the domain to the poles. Finally, the azimuthal boundaries are periodic. \\
\indent These BCs are completed by a buffer zone near the inner radius ($r<r_{\text{buf}}$), with $r_{\text{buf}}=1.25R_0$ for \texttt{NF-BAZ}, \texttt{NF-AZ} and \texttt{NF-DZ}, and $r_{\text{buf}}=1.5R_0$ for \texttt{NF-SAZ}. Within this region, $\mathbf{u}$ and $\rho$ are relaxed to their initial values on a timescale of 0.1 local orbital periods. In addition, an Ohmic-resistive damping term, $\eta_\text{O} = 0.05\varepsilon^2\max(r_{\text{buf}} - r, 0)$, is applied to suppress magnetic fluctuations. Finally, to avoid restrictive time steps in the low-density corona, a density floor of $\rho_\text{min} = 10^{-9}\rho_0$ is imposed, and the Alfvén speed is limited to $v_\text{A} \leq 10 \,R_0\Omega_{\text{K}}(R_0)$, by adjusting the local density. 
\subsection{Diagnostics and definitions}
\label{section:diagnostics_and_definitions}

\subsubsection{Averages and fields}
To characterise laminar and turbulent behaviour across a range of time and length scales, we define the following averages of a field $Q$:
\begin{align}
    \langle Q \rangle _\phi &= \dfrac{1}{\phi_{\textrm{max}}}
\int^{\phi_{\textrm{max}}}_{0} Q \;\text{d}\phi, \\
    \langle Q \rangle _{\phi,t} &=  \dfrac{1}{t_2-t_1} \int^{t_2}_{t_1} \langle Q \rangle_\phi\;\text{d}t, 
    \label{equation:spacetime_averages} \\
    \langle Q \rangle_{\phi,nH} &= \int^{\theta(z=-nH)}_{\theta(z=+nH)} r \sin(\theta)\langle Q\rangle_\phi\;\text{d}\theta.
\end{align}
\indent The poloidal component $\mathbf{Q_p}$, is defined via the decomposition:
\begin{equation}
    \mathbf{Q} = \mathbf{Q_p} + Q_\phi \mathbf{e_\phi},
    \label{eqn:poloidal_field_definition}
\end{equation}
where $\mathbf{e_\phi}$ is the azimuthal basis vector. Line integral convolution (LIC) is used to aid visualisation of two-dimensional vector fields.

\subsubsection{Accretion and stresses}
\indent Long-term behaviour of the accretion flow is characterised by the mass accretion rate,
\begin{equation}
\dot{M}_{\text{acc}}^{nH}(r) = -2\pi \int^{\theta(z=-nH)}_{\theta(z=+nH)} r^2\sin^2(\theta)  \, \langle \rho u_r \rangle_{\phi,t} \, \text{d}\theta ,
	\label{eq:mass_accretion_rate_mean_field}
\end{equation}
where $n$ is the number of scale heights integrated around the midplane, such that the radial mass accretion rate profile within the disc is $\dot{M}_{\text{acc}}^{4H}(r)$. Throughout the rest of this section, $\langle \cdot \rangle$ denotes a generic average, and its specific meaning is stated when used in the results.\\
\indent The $r\phi$ Reynolds and Maxwell stresses are defined as \citep[e.g.][]{mishra_strongly_2020},
\begin{equation}
    \mathcal{R}_{r\phi} = \langle\rho\delta{u_r}\delta{u_\phi}\rangle \;\;\; \text{and} \;\;\;
    \mathcal{M}_{r\phi} = \dfrac{-\langle B_r B_\phi\rangle}{4\pi},
    \label{eq:maxwell_rphi_stress}
\end{equation}
% \begin{align}
%     \mathcal{R}_{r\phi} &= \langle\rho\delta{u_r}\delta{u_\phi}\rangle, \\
%     \mathcal{M}_{r\phi} &= \dfrac{-\langle B_r B_\phi\rangle}{4\pi},
%     \label{eq:maxwell_rphi_stress}
% \end{align}
where $\delta Q$ denotes the fluctuation of $Q$ relative to its spatio-temporal average: $\delta Q = Q -\langle{Q}\rangle$. Following \citet{shakura_black_1973}, these can be written in dimensionless form by introducing an $\alpha$ parameter:
\begin{equation}
    \alpha = \dfrac{\mathcal{R}_{r\phi}+\mathcal{M}_{r\phi}}{\langle P \rangle} = \alpha_{\mathcal{R}} + \alpha_{\mathcal{M}},
    \label{eq:maxwell_alpha}
\end{equation}
where the subscripts refer to the Reynolds and Maxwell contributions. The coherent components of the Maxwell stresses are
\begin{equation}
    4\pi\, \mathcal{M}_{r\phi}^{\text{coh}} = -\langle{B_r}\rangle\langle{B_\phi}\rangle \;\;\; \text{and} \;\;\; 4\pi\,\mathcal{M}_{\theta\phi}^{\text{coh}} = -\langle{B_\theta}\rangle\langle{B_\phi}\rangle,
    \label{eq:maxwell_stress_components}
\end{equation}
in contrast to the turbulent components.

\subsubsection{Magnetic flux transport}
The secular evolution of the large-scale magnetic flux threading the disc can best be quantified via the flux function (e.g. \citetalias{jacquemin-ide_magnetic_2021}):
\small
\begin{equation}
	\Psi_{\textrm{mid}}(R,t) = \int\limits^{\pi/2}_0 \!R_{0}^2\sin(\theta)\langle{B_r}\rangle_\phi\;\text{d}\theta + \int\limits^{R}_{R_0}\!R\,\langle{-B_\theta} \rangle_\phi(\theta=\pi/2)\;\text{d}r.
	\label{eq:define_magnetic_flux_mid}
\end{equation}
\normalsize
This measures the flux threading the midplane, from the inner edge of the domain to radius $R$ including the inner shell. The magnetic flux transport velocity at the midplane is \citep[e.g.][]{lesur_systematic_2021},
\begin{equation}
	v_\Psi = \dfrac{\langle\mathcal{E}_{\phi}\rangle_{\phi,t}(z=0)}{\langle{B_{z}}\rangle_{\phi,t}(z=0)},
	\label{eqn:vb_definition_midplane}
\end{equation}
where $u_\text{K} = R \Omega_\text{K}$ is the Keplerian flow. Dropping braces for brevity, $v_\Psi$ can be decomposed into its ideal and non-ideal components:
\begin{equation}
	v_\Psi = -\dfrac{\left(u_zB_R-u_RB_z\right)}{B_z}\vphantom{\left(\dfrac{B^2}{B^2}\right)} + 
	\dfrac{1}{B_z}\left(\eta_\text{O}+\eta_\text{A}\dfrac{B_z^2}{B^2}\right)\left(\dfrac{\partial B_R}{\partial z} - \dfrac{\partial B_z}{\partial R }\right),
    \label{eqn:decompose_ephi}
\end{equation}
where the ambipolar term has been approximated for this regime \citep[e.g.][]{gressel_global_2020}.

\subsubsection{Magnetic outflows and current sheets}
\label{section:diagnostics_outflows_current_sheets}

The poloidal fluid speed, magnetic field strength, and current-density magnitude are $u_\text{p}$, $B_\text{p}$ and $J_\text{p}$ respectively. The total plasma-beta parameter, $\beta$, and its coherent poloidal component, $\beta_\text{p}^{\text{coh}}$, are
\begin{equation} 
    \beta = \dfrac{8\pi P}{B^2} \;\;\; \text{and} \;\;\; \beta_\text{p}^{\text{coh}} = \dfrac{8\pi\langle {P}\rangle}{\langle {\mathbf{B_p}\rangle^2}}.
    \label{eq:beta_poloidal_coherent} 
\end{equation}
\indent The angle between averaged poloidal magnetic field lines and fluid streamlines, $\chi_\text{w}$, is defined as
\begin{equation}
\cos\left({\chi_\text{w}}\right) = \dfrac{\langle\mathbf{u_p}\rangle\cdot\langle\mathbf{B_p}\rangle}{\langle u_\text{p} \rangle\langle{B_\text{p}}\rangle},
	\label{eqn:cos_chi_definition}
\end{equation}
such that $\cos\left({\chi_\text{w}}\right)=\pm1$ corresponds to perfect alignment. \\
\indent Following \citetalias{jacquemin-ide_magnetic_2021}, for an averaged field line with midplane foot point denoted by subscript f, the normalised mass loading, angular velocity, and angular momentum invariants, along with the Bernoulli function representing the energy content of the flow, are defined as:
\begin{align}
    k^* =& \,\dfrac{4\pi{R_\text{f}}{\Omega_{\text{K}_\text{f}}}\langle\rho\rangle\langle\mathbf{u_p}\rangle\cdot\langle\mathbf{B_p}\rangle}{B_{\text{p}_{\text{f}}}\langle\mathbf{B_p}\rangle\cdot\langle\mathbf{B_p}\rangle}
	\label{eq:normalised_mass_loading_parameter} \\
    w^* =& \, \dfrac{\langle u_\phi \rangle}{R \Omega_{\text{K}_\text{f}}} - \dfrac{k^*\langle B_\phi \rangle B_{\text{p}_\text{f}} }{4\pi R \langle \rho \rangle R_\text{f}\Omega_{\text{K}_\text{f}}^2} \\
    \ell^* =& \, \underbrace{\vphantom{\dfrac{\langle u^2 \rangle}{2 \Omega_{\text{K}_\text{f}}^2R_\text{f}^2}}\dfrac{R \langle u_\phi \rangle }{R_\text{f}^2\Omega_{\text{K}_\text{f}}}}_{\text{kinetic}} \,
    \underbrace{-\,\vphantom{\dfrac{\langle u^2 \rangle}{2 \Omega_{\text{K}_\text{f}}^2R_\text{f}^2}}\dfrac{R \langle B_\phi \rangle}
    {k^* R_{\text{f}} B_{\text{p}_\text{f}}}}_\text{magnetic}
	\label{eqn:ell_normalised} 
\end{align}
\vspace{-3mm}
\small
\begin{equation}
    \,e^* \;= \underbrace{\vphantom{\dfrac{\langle u^2 \rangle}{2 \Omega_{\text{K}_\text{f}}^2R_\text{f}^2}}\dfrac{\langle u^2 \rangle}{2 \Omega_{\text{K}_\text{f}}^2R_\text{f}^2}}_{\text{kinetic}} \, \underbrace{-\dfrac{R_\text{f}}{\sqrt{R^2+z^2}} {\vphantom{\dfrac{\langle u^2 \rangle}{2 \Omega_{\text{K}_\text{f}}^2R_\text{f}^2}}}}_{\text{gravitational}}  \,\underbrace{-{\vphantom{\dfrac{\langle u^2 \rangle}{2 \Omega_{\text{K}_\text{f}}^2R_\text{f}^2}}}\,\dfrac{w^* R \langle B_\phi \rangle}{k^* R_\text{f} B_{\text{p}_\text{f}}}}_{\text{magnetic}}\, \underbrace{\!\!{\vphantom{\dfrac{\langle u^2 \rangle}{2 \Omega_{\text{K}_\text{f}}^2R_\text{f}^2}}}+ \dfrac{1}{\Omega_{\text{K}_\text{f}}^2R_\text{f}^2}\int^{\infty}_{s}\! \dfrac{\nabla{P}}{\rho}\!\cdot\!\text{d}\mathbf{l}.}_{\text{thermal}} 
	\label{eqn:bernoulli_function_normalised}  
\end{equation}
\normalsize
\indent Finally, current sheets refer to regions of strong radial current density, $J_r$, in the disc, which correspond to localised zones where $|\partial_z B_\phi|$ is maximal and $B_\phi = 0$ in this system.

%%%%%%%%%%%%%%%%%%%%%%%%%%%%%%%%%%%%%%%%%%%%%%%%%%%%%%%%%%%%%%%%%%%%%%%%%%%%%%%%%%%%%%%%%%%%%%%%%%%%
%%%%%%%%%%%%%%%%%%%%%%%%%%%%%%%%%%%%%%%%%%%%%%%%%%%%%%%%%%%%%%%%%%%%%%%%%%%%%%%%%%%%%%%%%%%%%%%%%%%%

\section{Basic structure and evolution}
\label{section:results_overview}
\begin{figure}
    \centering
    \includegraphics[width=0.98\columnwidth]{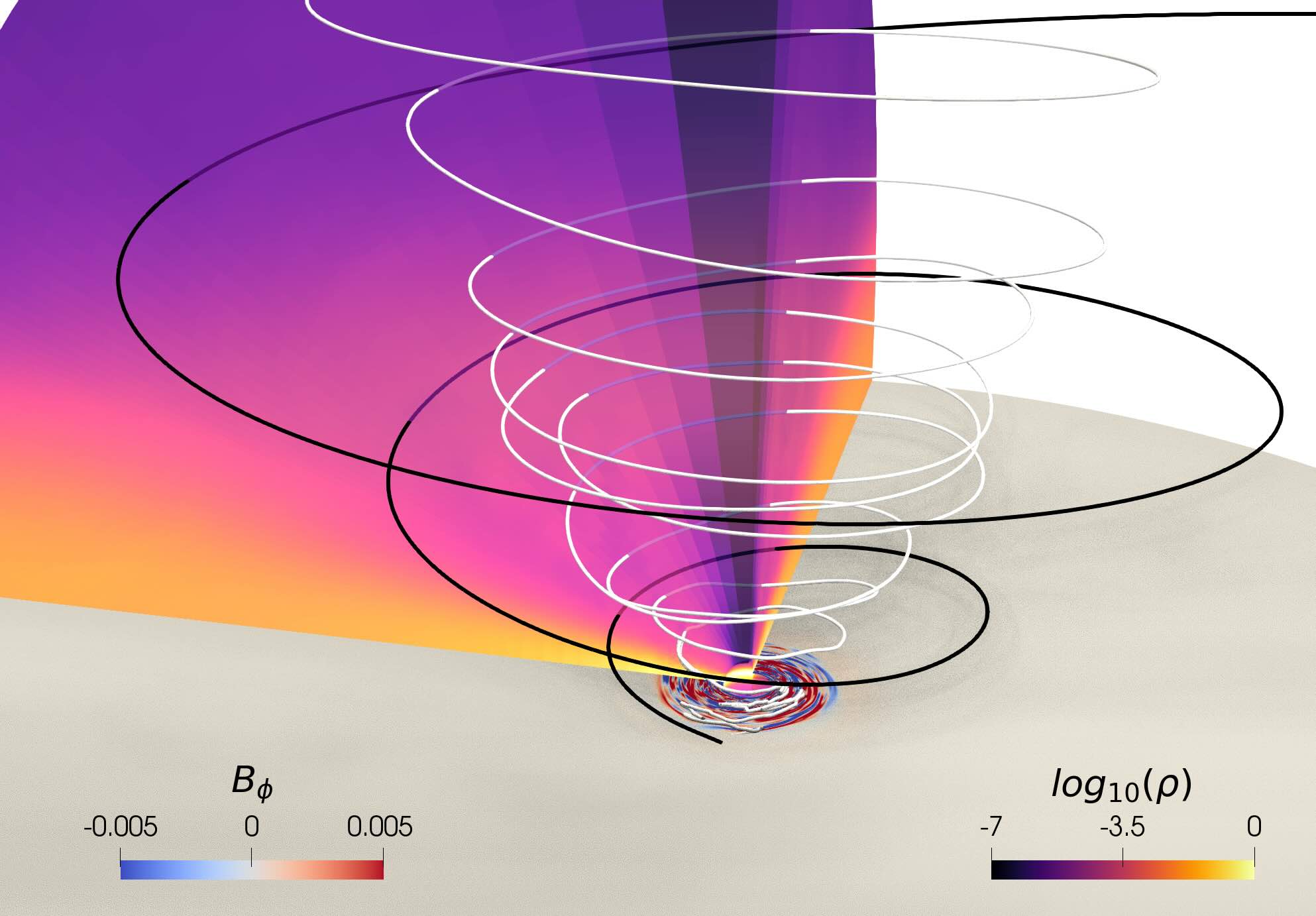}
    \caption{Zoomed-in rendering of an azimuthally extended model of \texttt{NF-BAZ} at $t_{\text{in}}=300$, comprising three components. First, a midplane cut reveals the toroidal field structure, $B_\phi$, and the small inner MRI-active zone, which extends to $R_{\text{DZI}} = 10R_0 = 1\,\text{au}$. Second, a spherical wedge, which defines the domain shape of \texttt{NF-BAZ}, shows the density $\rho$. Third, two magnetic field lines are traced from $R_\text{f} = 0.5\,\text{au}$ (white) and $R_\text{f} = 1.5\,\text{au}$ (black), exhibiting the highly helical field geometry essential for weak-magnetic-wind launching.}
  \label{fig:baz_render_2pi}
\end{figure}

An overview of the system's basic structure is provided in Fig.~\ref{fig:baz_render_2pi}, which presents a rendering of a $2\pi$-azimuthal extension of \texttt{NF-BAZ} at $t_{\text{in}}=300$. This comprises three components: a midplane cut that reveals the toroidal field, the density structure within the $\pi/2$-azimuthal wedge that defines the domain of \texttt{NF-BAZ}, and two magnetic field lines traced from midplane foot points at $R_\text{f} = 0.5$ and $1.5\,\text{au}$. In the ideal-MHD zone within $R_{\text{DZI}}=1\,\text{au}$, there is vigorous MRI-driven turbulence. Beyond the dead--active zone interface, the MRI is suppressed because the magnetic Reynolds number (see right panel of Fig.~\ref{fig:non_ideal_MHD_setup}) falls below the threshold for stratified VNF MRI linear stability, $\text{R}_\text{m} \lesssim \sqrt{\beta}\!\sim\!100$ \citep{latter_vertical_2022}, thus acting as a switch for MRI activity. Although the turbulence is confined to a tiny fraction of the domain, it guides the evolution of the entire system. During the first few orbits, the initially vertical and weak magnetic field is tightly wound in the direction opposite to the Keplerian flow, facilitating the launching of weak magnetothermal winds across all disc radii \citep[e.g.][]{bai_magneto-thermal_2016}, driven by a combination of magnetic and thermal pressure. \\

Throughout the remainder of this paper, the analysis focuses on \texttt{NF-BAZ}, but at times we also use \texttt{NF-SAZ}, interpreted as a plausible continuation of \texttt{NF-BAZ} from $82$ to $243t_{\mathrm{DZI}}$.

\subsection{Emergence and saturation of the MRI}
\label{section:emergence_MRI}

\begin{figure}
    \centering
    \includegraphics[width=\columnwidth]{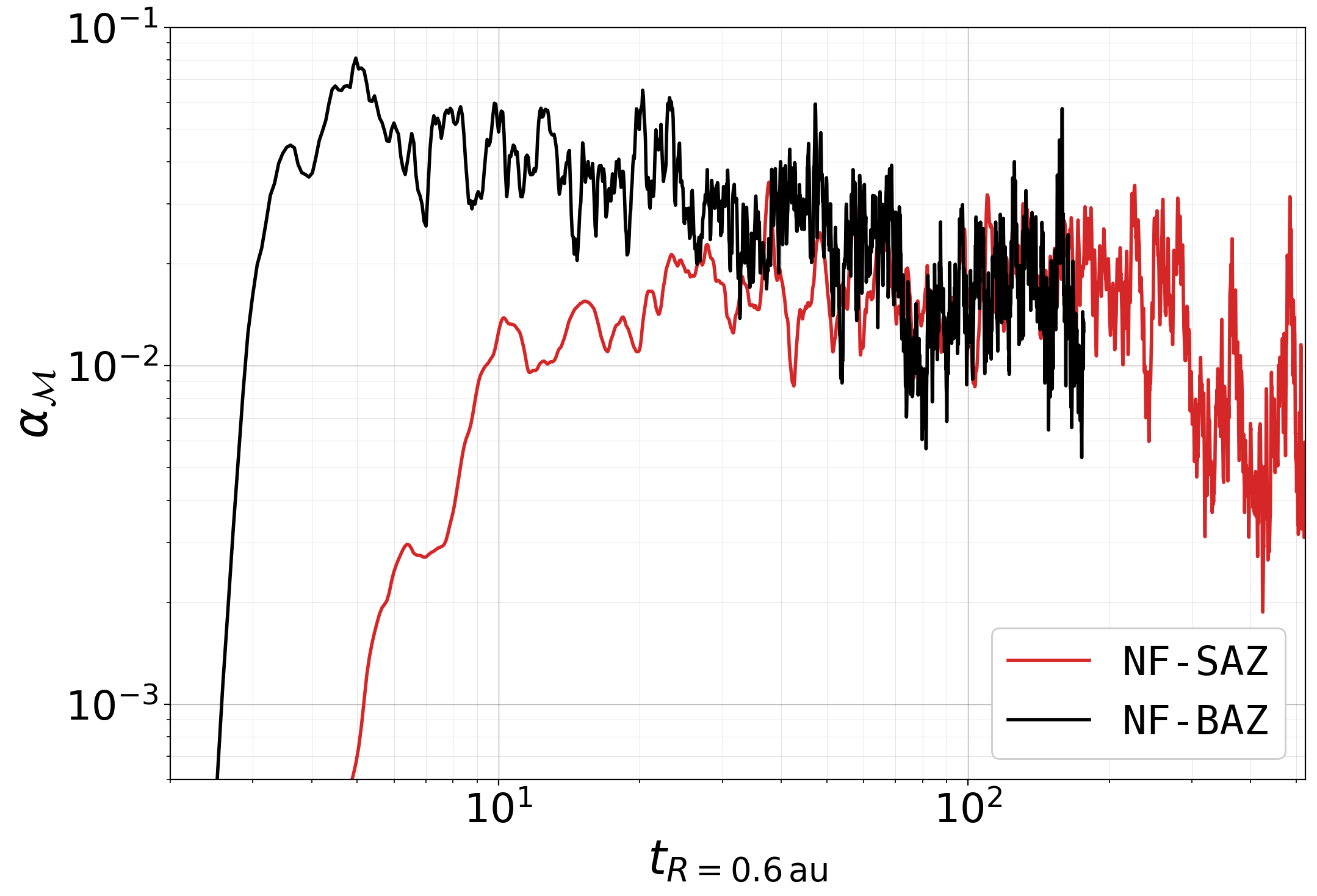}
    \caption{Evolution of the total Maxwell $\alpha$ in the active zone for \texttt{NF-BAZ} (black) and \texttt{NF-SAZ} (red), volume-averaged over: $r\!\in\![0.5,0.7]$ au, $\theta\!\in\![z=2H, z=-2H]$ and $\phi\!\in\![0,\pi/2]$. $\alpha_\mathcal{M}$ decays as magnetic flux is depleted from the region and $t_{R=0.6\,\text{au}}$ is the number of local orbits at $R=0.6\,\text{au}$.}
  \label{fig:alpha_volume_average_vnf}
\end{figure}

\begin{figure}
    \centering
    \includegraphics[width=\columnwidth]{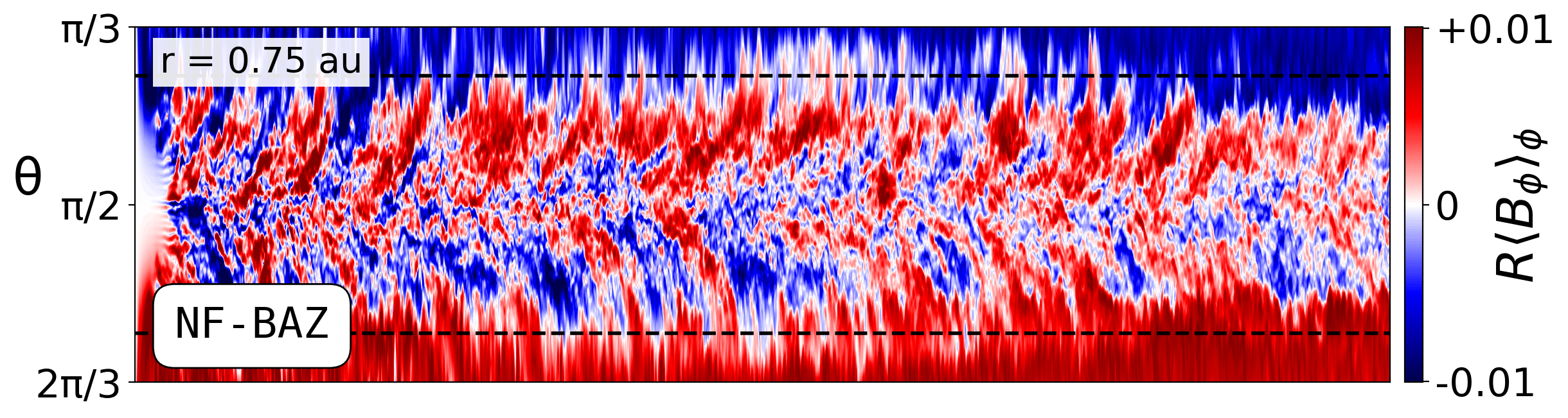} \\
    \includegraphics[width=\columnwidth]{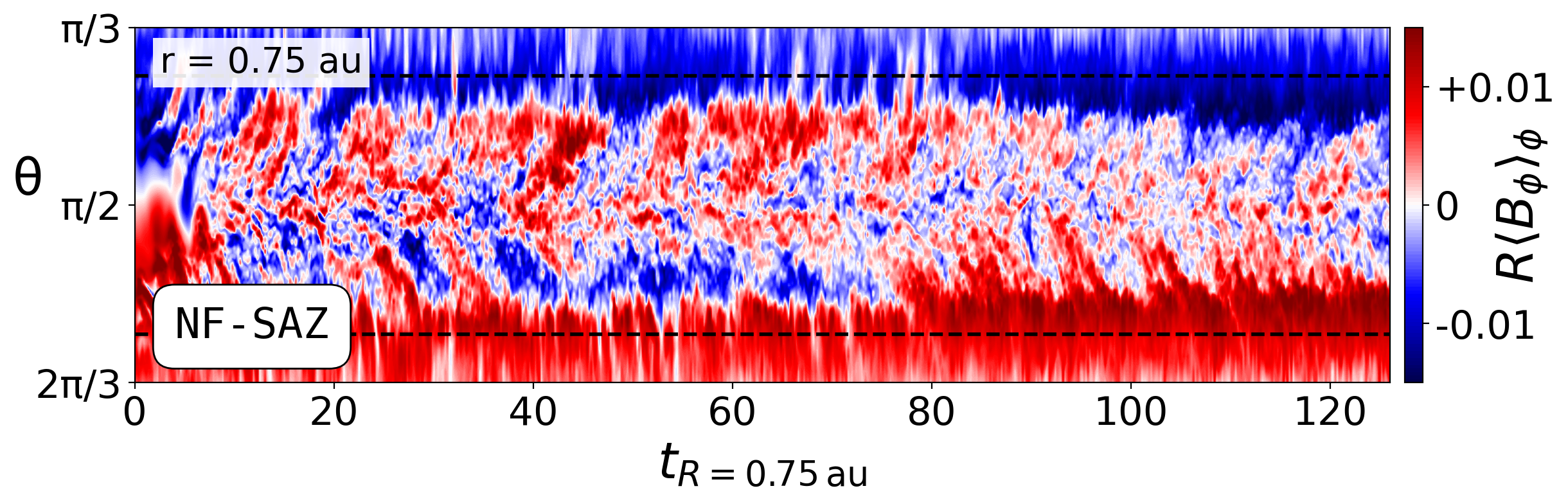}
    \caption{Space--time $(\theta,t)$ diagrams along the contour $r=0.75\,\text{au}$ for the normalised azimuthally averaged toroidal field, $R\langle B_\phi \rangle_\phi$, showing similar evolution in the active zone for \texttt{NF-BAZ} (top) and \texttt{NF-SAZ} (bottom). To aid the comparison, $t_{R=0.75\,\text{au}}$ is the number of local orbits at $R=0.75\,\text{au}$.}
    \label{fig:SAZ_BAZ_compare_bx3_0.75au}
\end{figure}

Fig.~\ref{fig:alpha_volume_average_vnf} presents the evolution of the Maxwell $\alpha$ components, $\alpha_\mathcal{M}$, for \texttt{NF-BAZ} and \texttt{NF-SAZ}, volume-averaged over the wedge: $r\!\in\![0.5,0.7]$ au, $\theta\!\in\![z=2H, z=-2H]$ and $\phi\!\in\![0,\pi/2]$.\footnote{The vertical integration is restricted to $z=\pm 2H$ to ensure that only the small-scale, MRI-driven stress contribution is included.} Each exhibits rapid exponential growth, saturating within a few local orbits at $\alpha_{\mathcal{M}}\!\sim\!4\times10^{-2}$ for \texttt{NF-BAZ} and $\alpha_{\mathcal{M}}\sim2\times10^{-2}$ for \texttt{NF-SAZ}, reflecting the sensitivity of the global VNF MRI to resolution at these levels (e.g. fig. 36 in \citetalias{iwasaki_dynamics_2024}). These initial saturated values, which depend on the initial midplane $\beta$, are consistent with previous local and global models \citep[e.g.][]{bai_local_2013, zhu_global_2018}, and are an order of magnitude greater than those in the ZNF models in \citetalias{roberts_global_2025}, highlighting the increased efficacy of the MRI in the VNF regime. Following the initial saturation, $\alpha_\mathcal{M}$ gradually decays in both simulations, driven by the depletion of large-scale magnetic flux throughout most of the active zone, a process examined in Section~\ref{section:mft_inner_az}. \\
\indent Fig.~\ref{fig:SAZ_BAZ_compare_bx3_0.75au} shows space--time $(\theta, t)$ diagrams along the contour $r = 0.75\,\text{au}$ for the azimuthally averaged toroidal field, $R\langle B_\phi \rangle_\phi$. The characteristic periodic cycles in azimuthal field (i.e., the butterfly pattern) \citep[e.g.][]{brandenburg_dynamo-generated_1995, salvesen_accretion_2016}, in conjunction with the time-series in Fig.~\ref{fig:alpha_volume_average_vnf}, indicate a sustained MRI-saturated state. In both simulations -- but particularly in \texttt{NF-SAZ} -- the butterfly pattern weakens noticeably over time, consistent with the decline of the local volume-averaged $\alpha_\mathcal{M}$ in Fig.~\ref{fig:alpha_volume_average_vnf}. \\
\indent The meridional structure of $R\langle B_\phi \rangle_\phi$ in the VNF active zone is close to the ZNF case (see fig. 6 in \citetalias{roberts_global_2025}). In both, strong toroidal fields are generated just below the disc surface layers, as the $\Omega$-effect shears large-scale radial fields near the disc--corona temperature transition. In the ZNF model, the radial fields are from buoyantly sustained loops of large-scale poloidal field generated by the MRI dynamo; whereas in the VNF models, they are due to the local inward pinching of large-scale vertical fields.
\begin{figure*}
    \centering
    \includegraphics[height=0.51\textwidth]{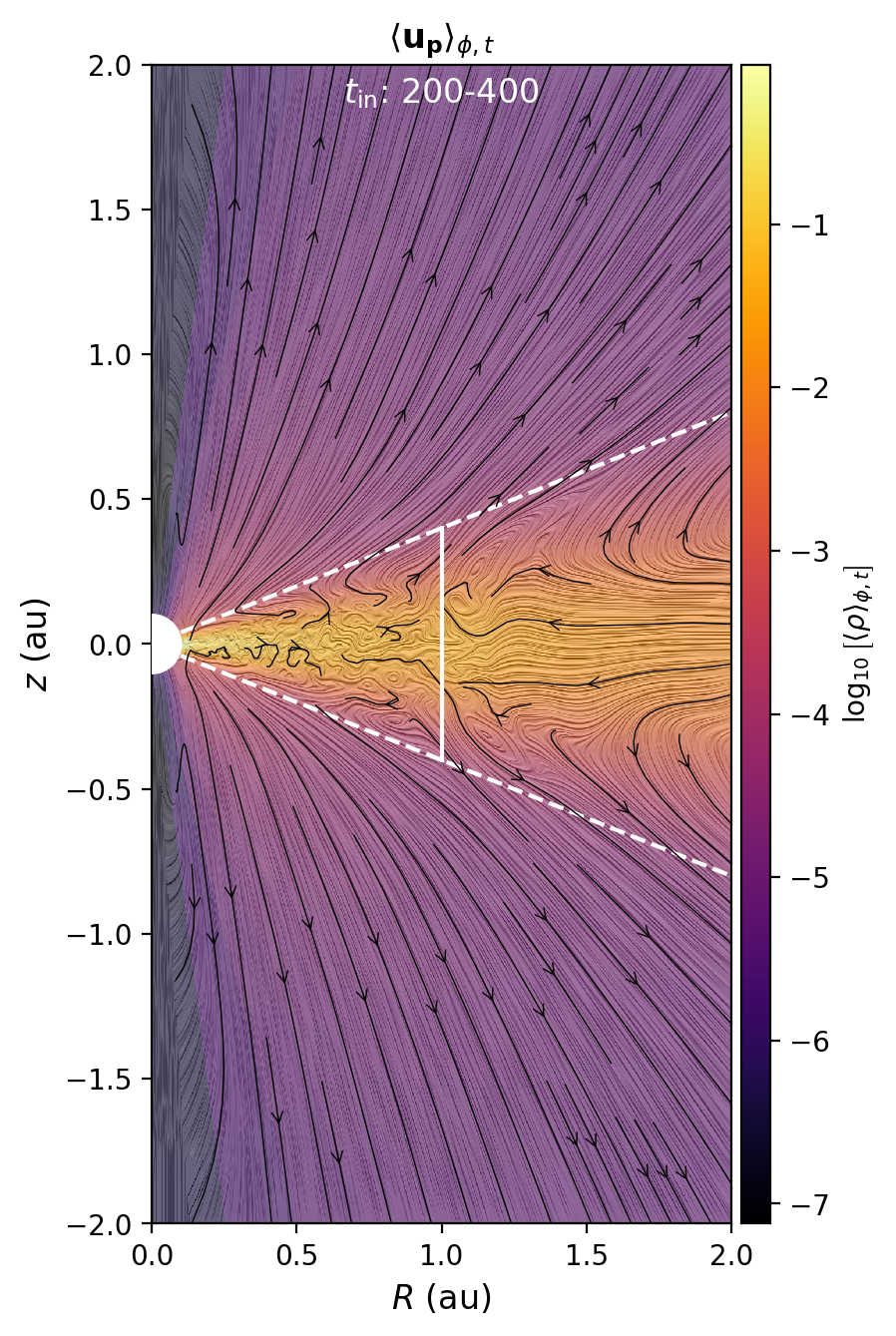}
    \includegraphics[height=0.51\textwidth]{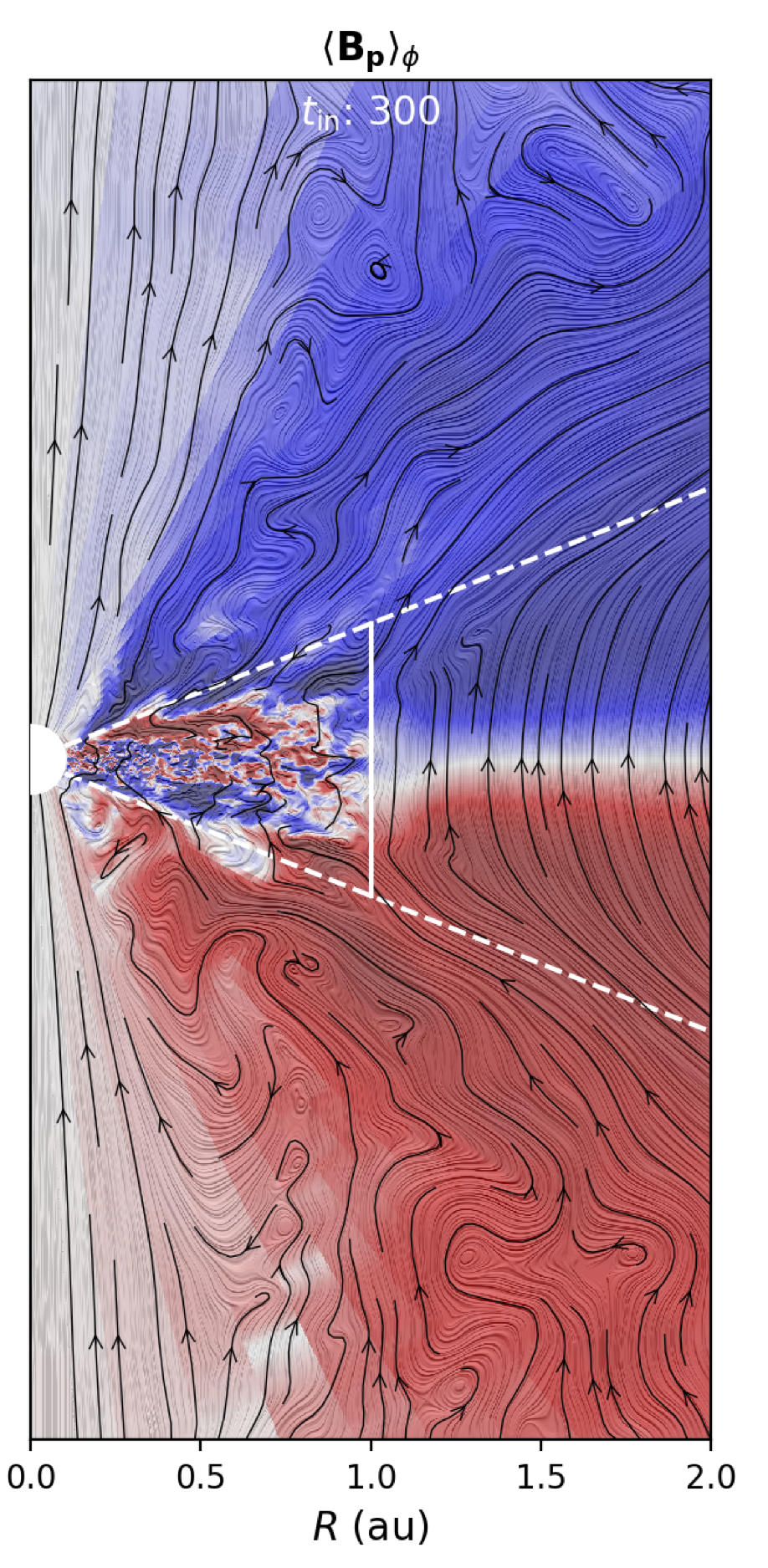}
    \includegraphics[height=0.51\textwidth]{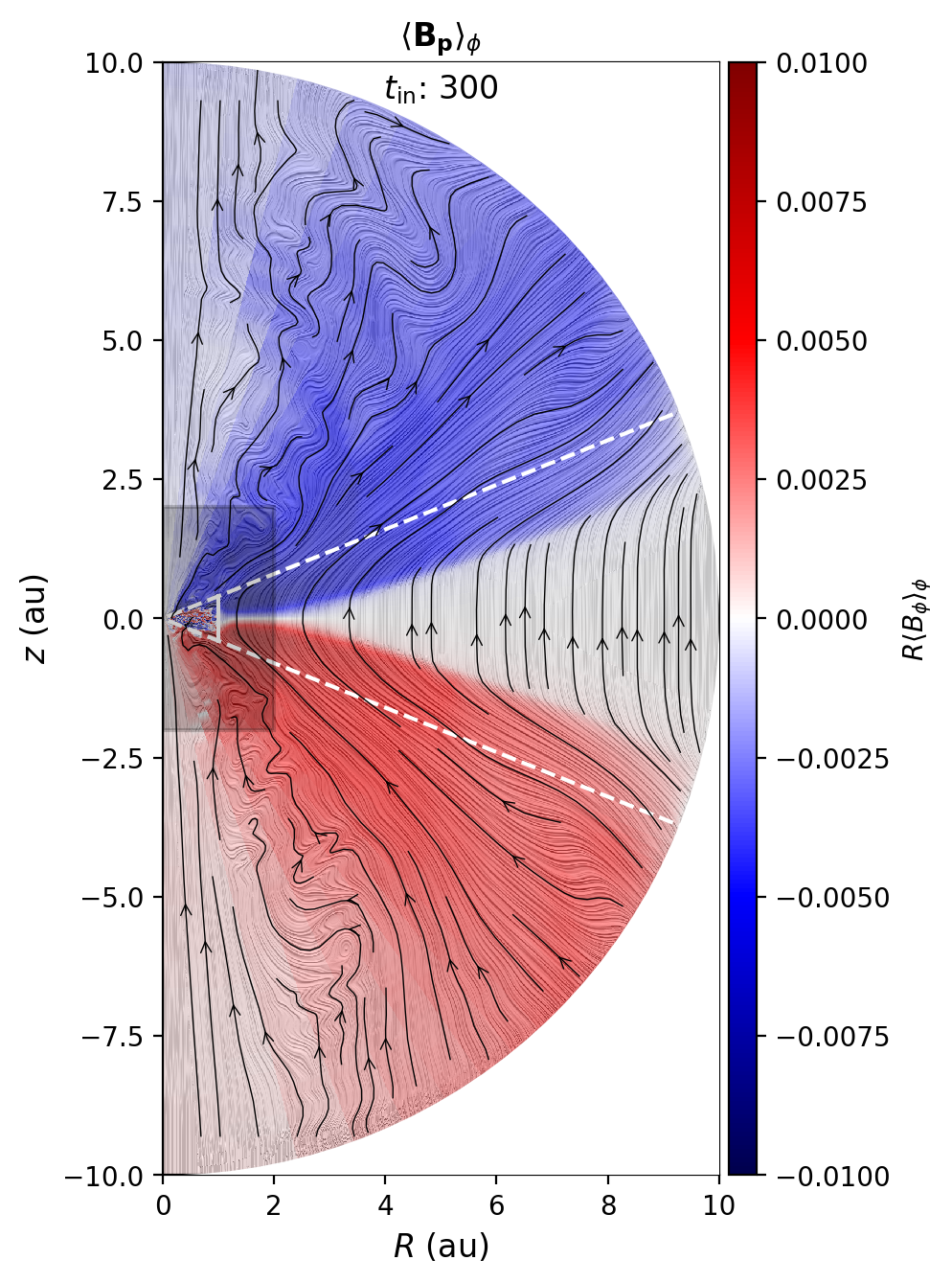}
    \caption{Meridional $(R,z)$ plots of the early-time configuration for \texttt{NF-BAZ}. Left: Poloidal velocity field, $\langle \mathbf{u_p} \rangle_{\phi,t}$, overlaid on density, $\langle \rho \rangle_{\phi,t}$, averaged in azimuth and over the interval $t_\text{in}\!\in\![200,400]$. Middle: Poloidal magnetic field, $\langle \mathbf{B_p} \rangle_{\phi}$, overlaid on normalised toroidal field, $R\langle B_\phi \rangle_\phi$, averaged in azimuth at $t_\text{in}=300$. Right: Equivalent to the middle panel -- corresponding to the grey square -- but showing the whole domain. As the system evolves, the current sheet extends throughout the entire dead zone. Black lines and the LIC overlay trace the respective poloidal vector fields in each panel, whereas the dead--active zone interface ($R=1\,\text{au}$) and disc--corona temperature transition ($|z|=4H$) are denoted by the solid and dashed white lines, respectively.} 
  \label{fig:baz_wide_scale_early_time}
\end{figure*}

\subsection{Large-scale disc and outflow structure at early times}

Fig.~\ref{fig:baz_wide_scale_early_time} presents meridional ($R,z$) plots of the early-time structure for \texttt{NF-BAZ}. The left panel displays the poloidal velocity field, $\langle \mathbf{u_p} \rangle_{\phi,t}$, overlaid on the density, $\langle \rho \rangle_{\phi,t}$, both averaged in azimuth and over the interval $t_{\text{in}}\!\in\![200,400]$. The panel shows magnetothermal outflows across all disc radii, surface-layer accretion in the active zone, and midplane accretion within the dead zone. \\
\indent The other panels display the poloidal magnetic field, $\langle \mathbf{B_p} \rangle_{\phi}$, overlaid on the normalised toroidal field, $R\langle B_\phi \rangle_{\phi}$, at $t_\mathrm{in}=300$. This is averaged only in azimuth, thereby preserving some of the large-scale turbulent variations in time. The poloidal magnetic field is characterised by an hourglass-shaped, laminar morphology in the dead zone, that becomes disordered when it threads the active region. This transition produces a multi-zone outflow system, examined in Section~\ref{section:results_outflows}, with an inner turbulent wind bordering, and impinging upon, an outer more laminar wind. Meanwhile, the toroidal field fluctuates significantly in the active region, thus revealing the MRI-driven turbulence. In the dead zone, on the other hand, it exhibits a sign reversal across a single, diffuse midplane current sheet -- still developing in its outer regions and seeking a path of least resistance -- where accretion is localised. In contrast, within the inner part of the active zone there are two, surface-layer current sheets, which are roughly associated with nulls in $B_\phi$, and thus the boundaries between blue and red in the middle panel of Fig.~\ref{fig:baz_wide_scale_early_time}. \\
\indent In summary, when sufficiently far away from the dead--active zone interface, the early-time disc structure resembles that of unconnected pure active and dead-zone states (see Appendix~\ref{appendix:turbulent_versus_laminar_discs}).

\subsection{Overview of variability at the dead--active zone interface}
\label{section:overview_reconfiguration}

\begin{figure*}
    \centering
    \includegraphics[width=0.98\textwidth]{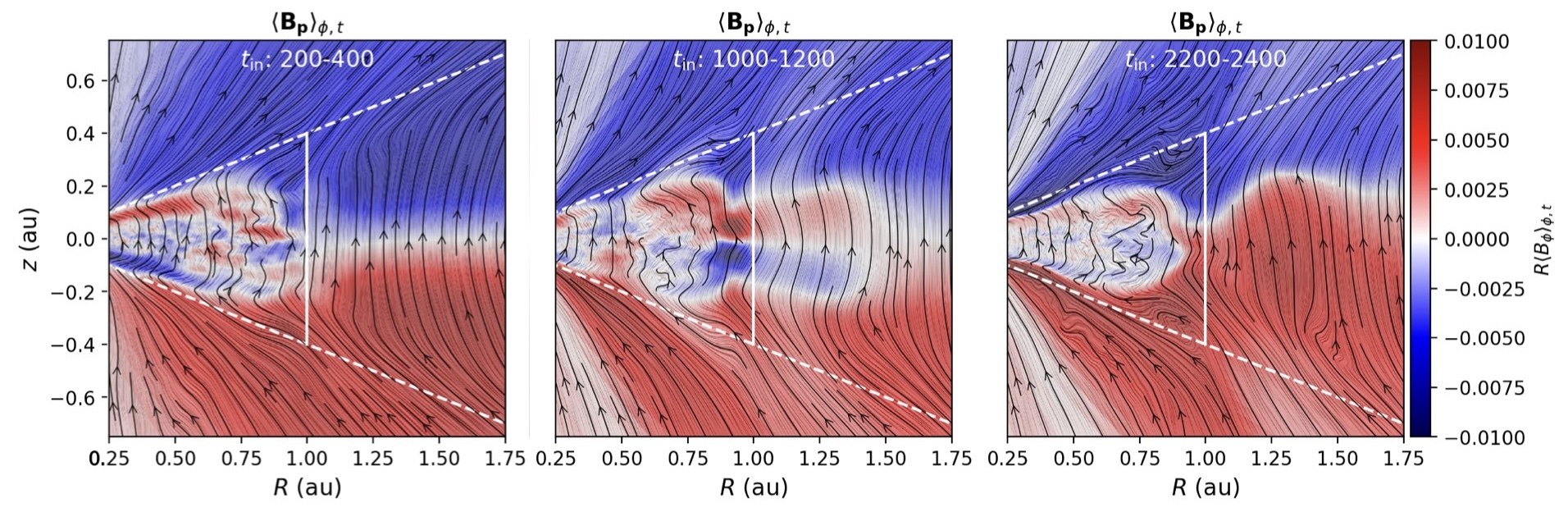}
    \caption{Meridional plots of the averaged magnetic field configuration in the inner disc at the epochs: $t_\text{in}\!\in\![200,400], [1000,1200]$ and $[2200,2400]$ for \texttt{NF-BAZ}. The poloidal field lines, $\langle\mathbf{B_p}\rangle_{\phi,t}$, are shown by black lines and a LIC overlay. The background colour is the normalised toroidal field, $R{\langle B_\phi}\rangle _{\phi,t}$, which is used in Section~\ref{section:overview_reconfiguration} to describe the current-sheet evolution at the interface (solid white line): from an early-time disordered state (left), through a transitional, trident-like state (middle), to a vertically asymmetric state (right). Animations, without time averaging, are available as ancillary files for \texttt{NF-BAZ} and \texttt{NF-SAZ} (see \texttt{BAZ\_bphi.mp4} and \texttt{SAZ\_bphi.mp4}).}
    \label{fig:baz_connection_magnetic}
\end{figure*}

The previous section established an early-time mismatch in the number of current sheets: two in the active zone and one in the dead zone. Therefore, whilst the system is expected to evolve towards a preferred configuration, the inability of a single current sheet to cleanly split in two (since $\nabla\cdot\mathbf{J}=0$), makes the manner of this rearrangement far from obvious. The remainder of this section summarises the evolution at the interface, whilst Section~\ref{section:results_accretion} examines the implications for accretion variability and Section~\ref{section:nf_flux_current_sheet} investigates the underlying mechanism. \\
\indent Fig.~\ref{fig:baz_connection_magnetic} shows the azimuthally and time-averaged magnetic field configuration in the inner disc for \texttt{NF-BAZ} across the following epochs: $t_\text{in}\!\in\![200,400]$, [1000,1200], and [2200,2400]. These averages accentuate the current-sheet structure around the interface but suppress the turbulent fluctuations evident in the middle panel of Fig.~\ref{fig:baz_wide_scale_early_time}. The normalised toroidal field structure, $ R\langle B_\phi \rangle_{\phi,t}$ in Fig.~\ref{fig:baz_connection_magnetic} reveals a progressive evolution of this structure -- present in all simulations with an interface -- that can be categorised into three distinct phases: (i) an initial \textit{disordered state}, with two surface-layer current sheets in the active zone, and a single, diffuse midplane current sheet in the dead zone (left panel); (ii) a transitional \textit{trident-like state}, consisting of three vertically stacked current sheets that extend across the interface into the dead zone (middle panel); and (iii) a vertically \textit{asymmetric state}, in which one surface-layer current sheet in the active zone is favoured (right panel). This state ultimately settles to a midplane current sheet within the dead zone, showing no further evolution on the (short) timescale of \texttt{NF-BAZ}. 

\subsection{Overview of magnetic flux transport}
\label{section:overview_magnetic_flux_transport}

\begin{figure}
    \centering
    \includegraphics[width=1\columnwidth]{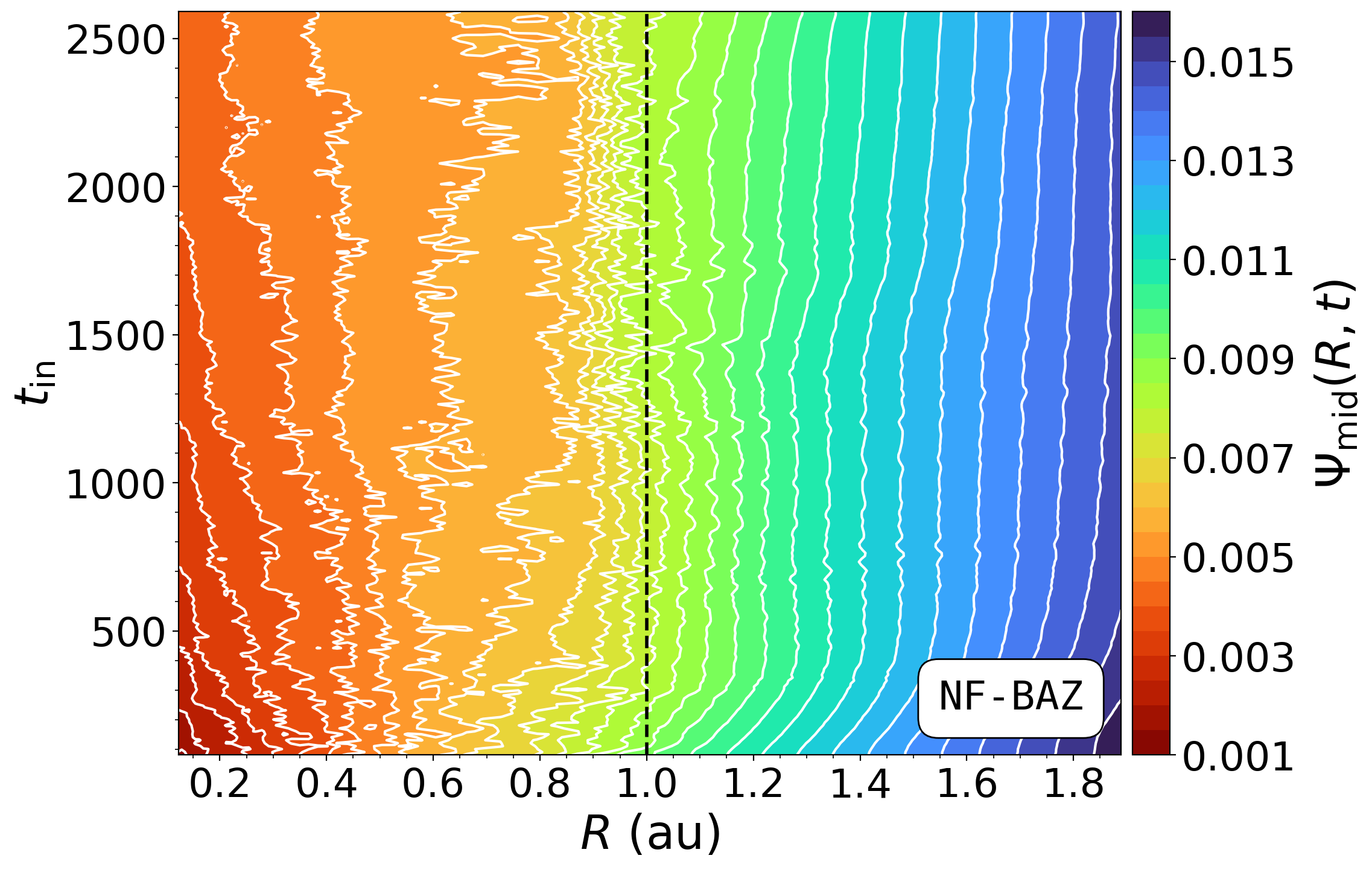}
    \caption{Evolution of the magnetic flux threading the midplane $\Psi_{\text{mid}}(R,t)$, defined in equation \eqref{eq:define_magnetic_flux_mid}, for \texttt{NF-BAZ}. The transport of poloidal magnetic field (white contour lines) is inward in the inner part of the active region ($R\lesssim 0.5$ au) and outward in the dead region. In the outer part of the active region, flux accumulates close to the dead--active zone interface (dashed black line).}
    \label{fig:baz_magnetic_flux_transport_plot}
\end{figure}

The large-scale magnetic flux transport exhibits a complex radial evolution, best characterised by the flux threading the disc, $\Psi_{\text{mid}}(R,t)$, defined in equation~\eqref{eq:define_magnetic_flux_mid}. Fig.~\ref{fig:baz_magnetic_flux_transport_plot} shows its evolution for \texttt{NF-BAZ}, revealing three behaviours occurring in distinct radial locations: inward radial transport in the inner part of the active zone ($R\lesssim 0.5\,\text{au}$); flux concentration near the interface in the outer part of the active zone ($0.5\,\text{au}\lesssim R\lesssim 1\,\text{au}$); and outward transport in the dead zone ($R\gtrsim 1\,\text{au}$). The accumulation of flux in the outermost part of the active region ultimately places it in the \textit{strong-field} VNF ideal-MHD regime, where $\beta_\text{p}^{\text{coh}}(z=0)\!\lesssim\!10^2$ (e.g. \citetalias{jacquemin-ide_magnetic_2021}), whilst the remainder of the active zone remains in the \textit{weak-field} regime, where $\beta_\text{p}^{\text{coh}}(z=0)\!>\!10^2$. \\
\indent Crucially, the dead--active zone interface is a \textit{one-way barrier} to inward transport of magnetic flux from the dead zone in all three models, consistent with \citetalias{iwasaki_dynamics_2024}. This leads to gradual flux depletion throughout most of the active zone and a concomitant, gradual reduction in the MRI-driven torques, inner-disc accretion rate, and inward-directed flux transport velocity. Appendix~\ref{appendix:boundary_test} confirms that these results are robust to a different, physically motivated, choice of inner radial BC and \texttt{NF-SAZ} supports these conclusions, except in the innermost part of the active zone, whilst also exhibiting additional interface-localised flux variability. The global magnetic flux transport is examined further in Section~\ref{section:results_global_flux_transport}.

%%%%%%%%%%%%%%%%%%%%%%%%%%%%%%%%%%%%%%%%%%%%%%%%%%%%%%%%%%%%%%%%%%%%%%%%%%%%%%%%%%%%%%%%%%%%%%%%%%%%
%%%%%%%%%%%%%%%%%%%%%%%%%%%%%%%%%%%%%%%%%%%%%%%%%%%%%%%%%%%%%%%%%%%%%%%%%%%%%%%%%%%%%%%%%%%%%%%%%%%%

\section{Accretion flows and associated structures}
\label{section:results_accretion}

\subsection{Evolution of accretion structure}
\label{section:vnf_evolution_acc_struct}
The interface-localised current-sheet states introduced in Section~\ref{section:overview_reconfiguration} partly control the accretion structure around the interface. The conundrum is that despite the strong, relatively short-timescale variability at the interface, accretion is inherently a slow process. Accordingly, our strategy is to take slightly longer averages over the disordered, trident-like and asymmetric states than before, $t_{\text{in}}\in [200,600]$, [800,1200] and [2000,2400]. The focus is on \texttt{NF-BAZ}, since \texttt{NF-SAZ} exhibits a similar evolution during this epoch. 

\subsubsection{Overview of accretion structure}
\begin{figure*}
    \centering
    \includegraphics[height=0.36 \textwidth]{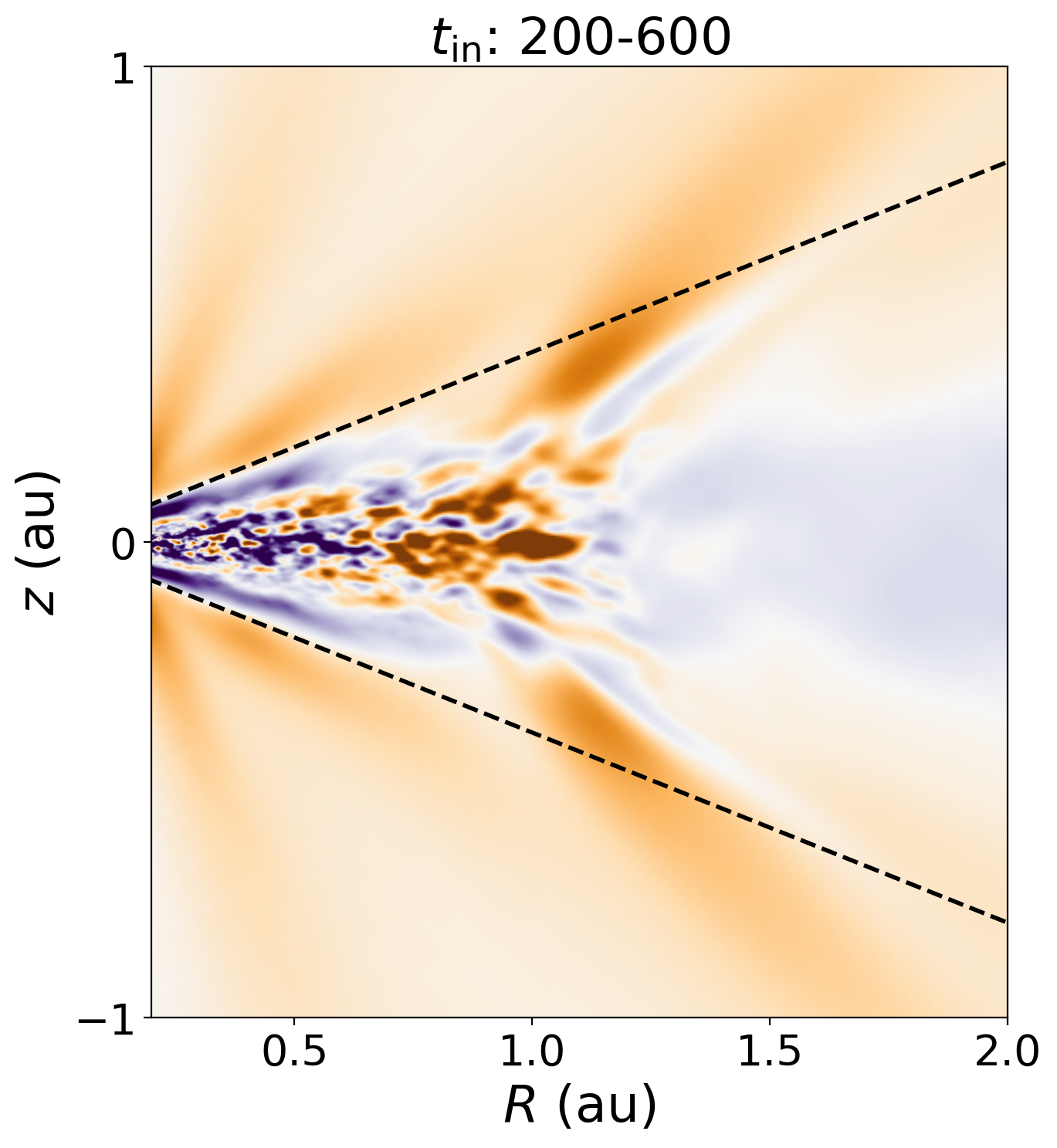}
    \includegraphics[height=0.36 \textwidth]{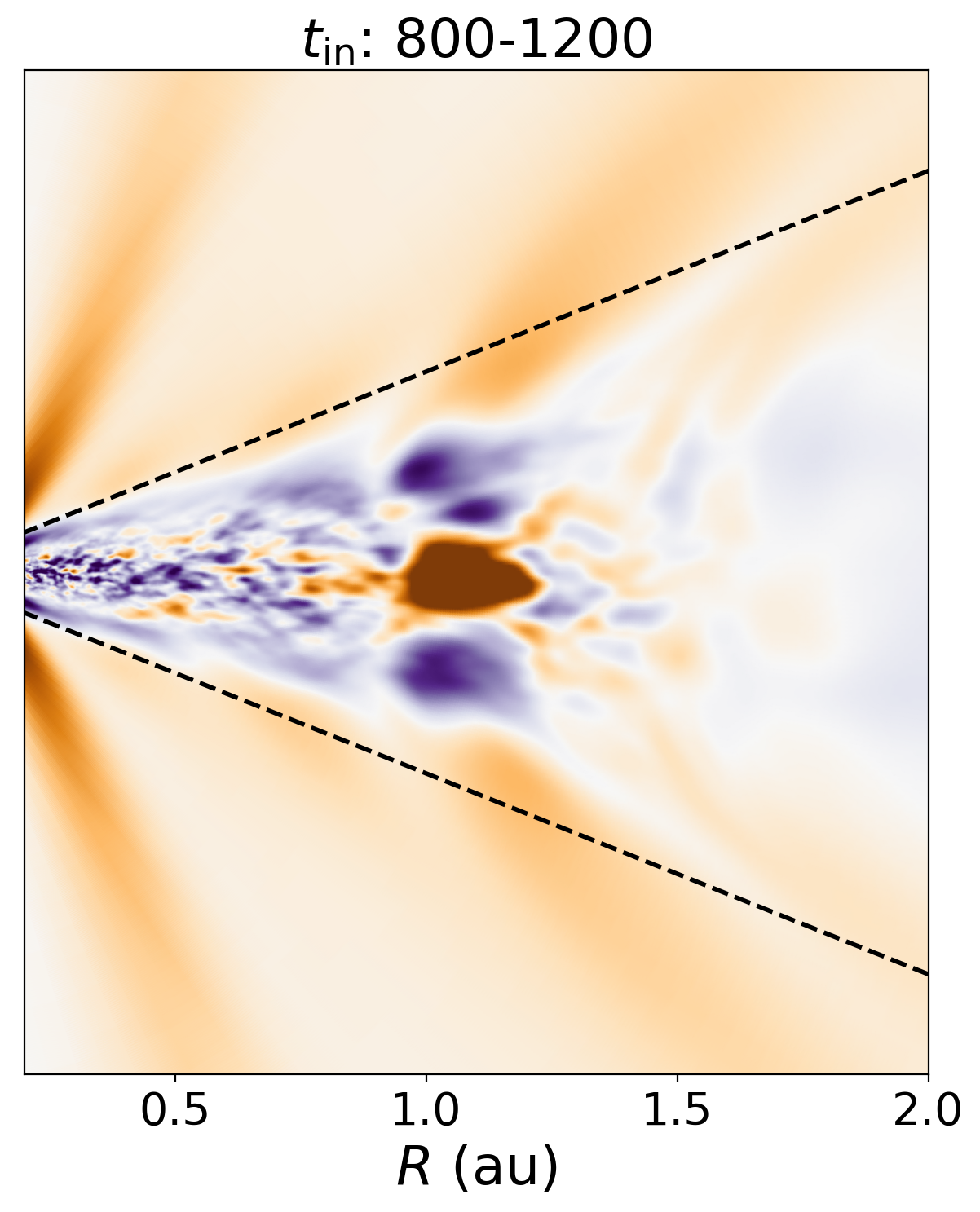}
    \includegraphics[height=0.36 \textwidth]{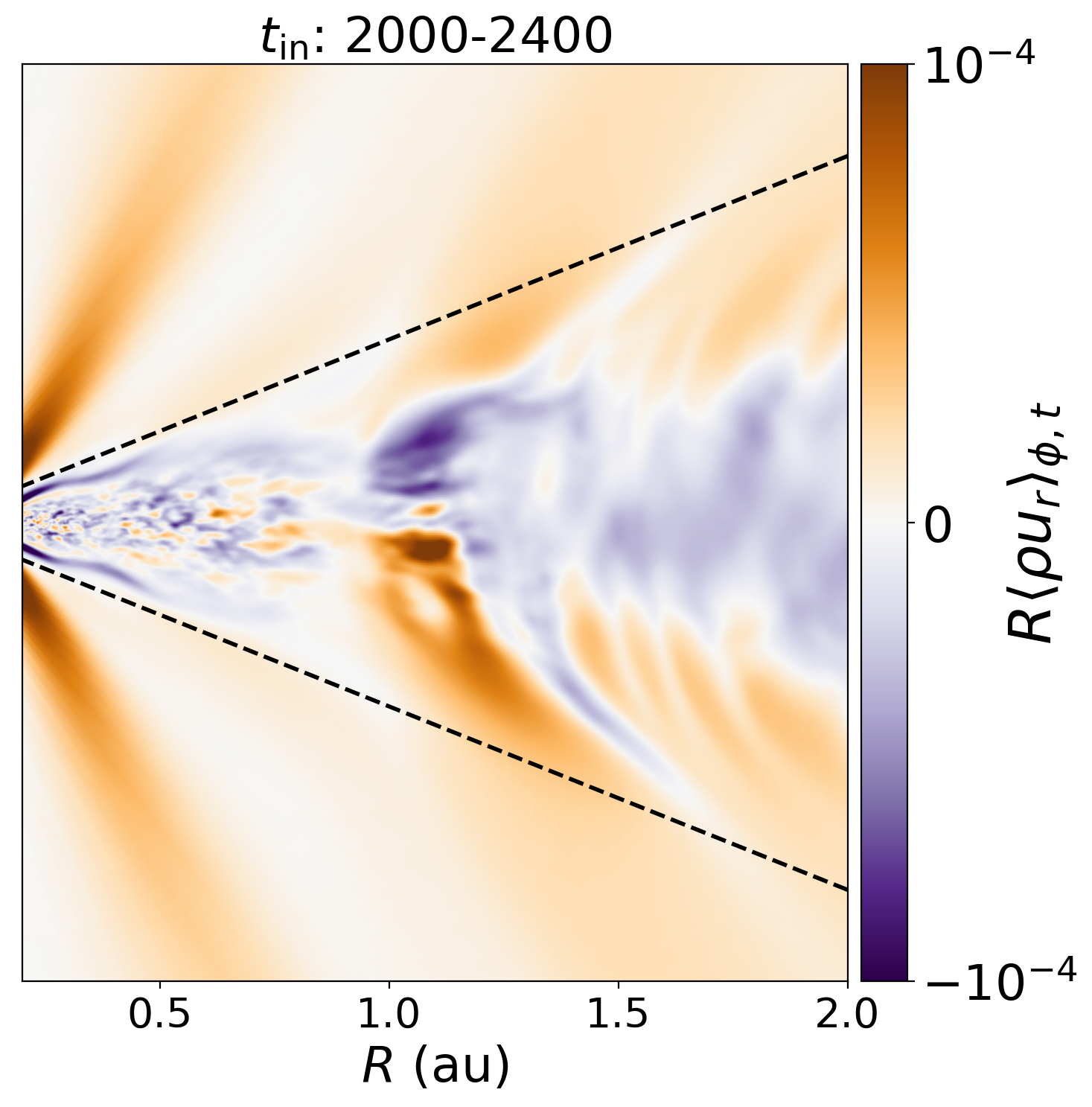}
    \caption{Meridional $(R,z)$ plots of the accretion flow, quantified by $R\langle \rho u _r \rangle _{\phi,t}$, for the three epochs: $t_\text{in} \in \{[200,600], [800,1200], [2000,2400]\}$ for \texttt{NF-BAZ}. The flow is characterised by a varying, meridionally dependent structure around the dead--active zone interface, and sustained surface-layer, and midplane, accretion flows in the inner part of the active zone ($R\lesssim 0.6\,\text{au}$) and bulk dead zone ($R\gtrsim 1.5\,\text{au}$), respectively.}
    \label{fig:baz_rho_ur_evolution}
\end{figure*}
\begin{figure*}
    \centering
    \includegraphics[height=0.2 \textwidth]{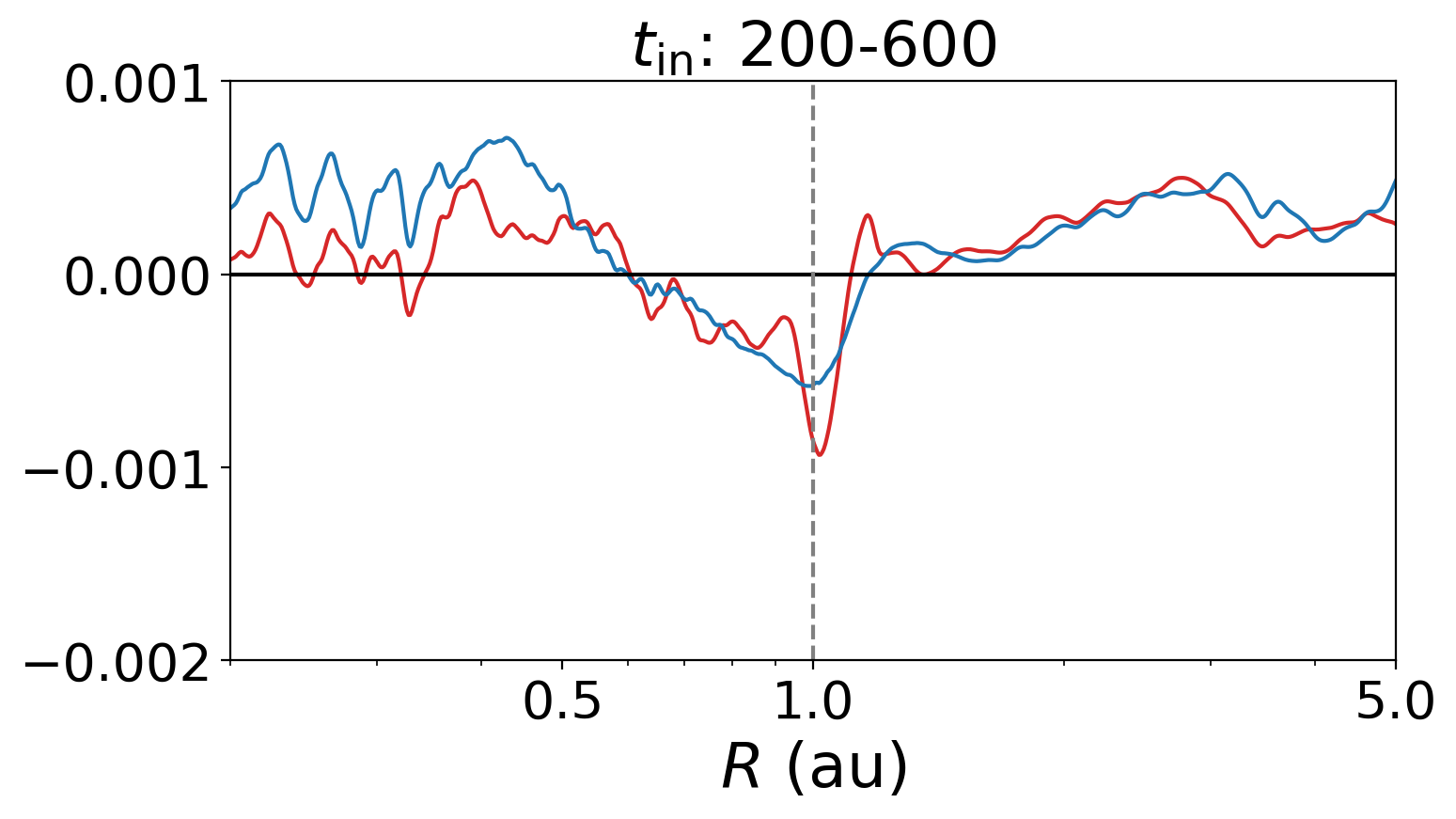}
    \includegraphics[height=0.2 \textwidth]{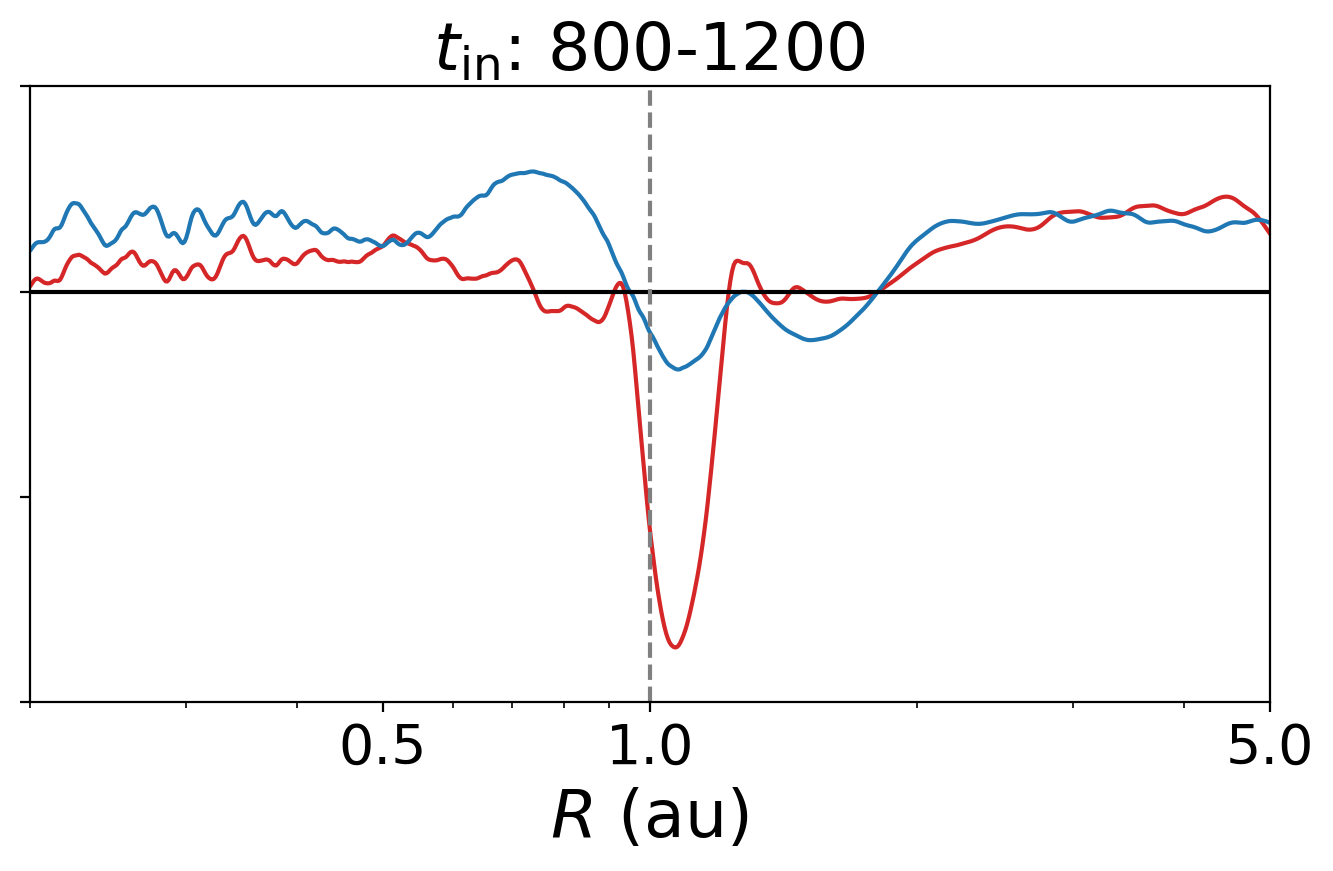}
    \includegraphics[height=0.2 \textwidth]{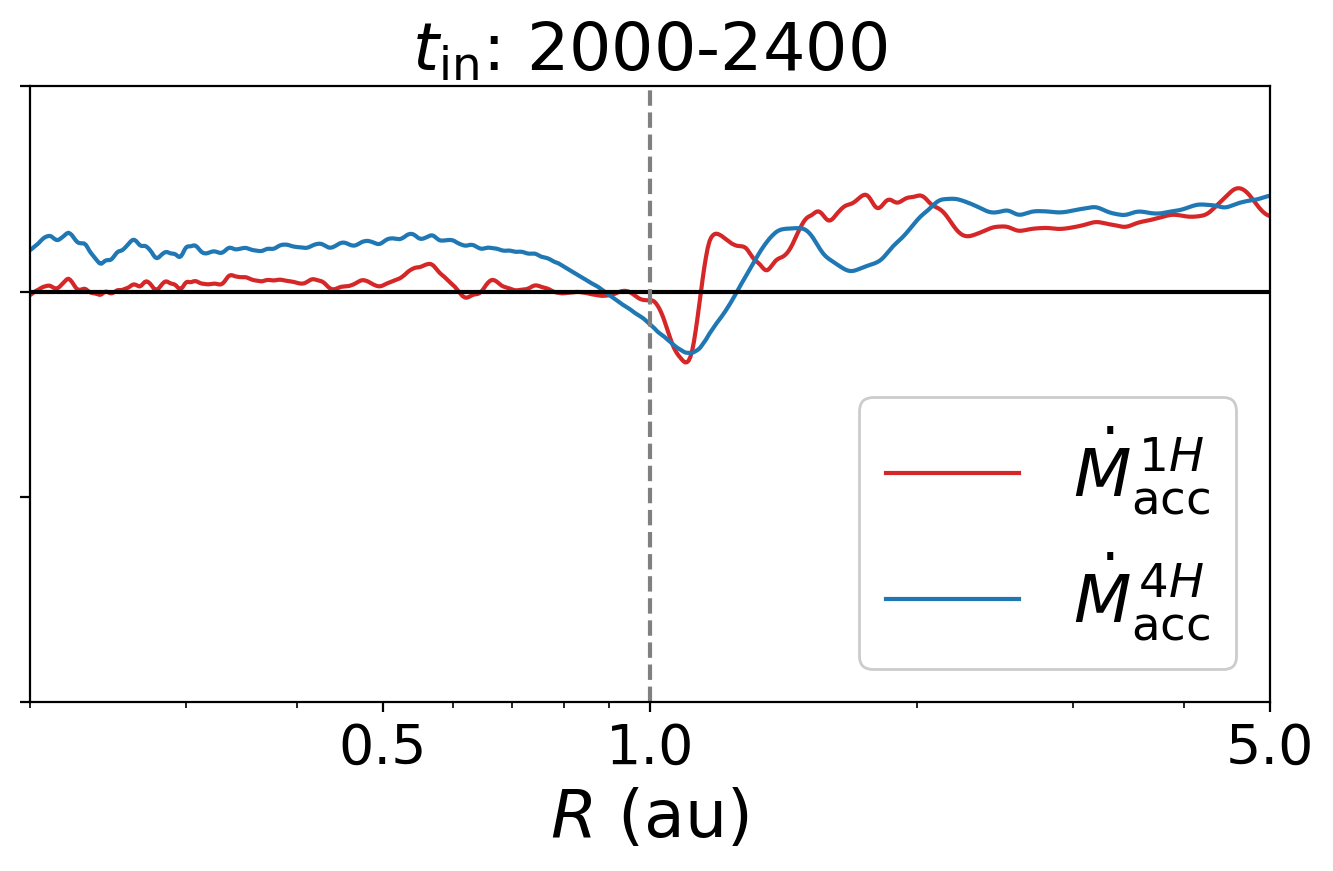}
    \caption{Radial profiles of the vertically integrated accretion rate, $\dot{M}^{nH}_\text{acc}$, defined in equation~\eqref{eq:mass_accretion_rate_mean_field}, for n = 1 (red) and n = 4 (blue), over the three epochs of interest in \texttt{NF-BAZ}. Around the midplane, there is decretion across the interface (dashed grey line) for all epochs. The full meridional structure is shown in Fig.~\ref{fig:baz_rho_ur_evolution}.}
    \label{fig:BAZ_Mdot_evolution}
\end{figure*}

The evolving accretion structure over these epochs is shown by meridional plots of $R\langle \rho u_r \rangle _{\phi,t}$ in Fig.~\ref{fig:baz_rho_ur_evolution}. Whilst this is not a perfect proxy for accretion, as it accounts for all radial mass movement including outflows, the mass flux is broadly categorised by: (i) strong surface-layer accretion in the inner part of the active zone, which is responsible for the surface-layer dragging, and bending, of poloidal magnetic field lines (see Fig.~\ref{fig:baz_connection_magnetic}); (ii) persistent decretion through the dead/active interface, despite strong variability in the local accretion structure around the interface; (iii) midplane accretion in the dead zone; and (iv) sustained outflows above the disc in all three regions. \\
\indent The current-sheet structure impacts the local flow, because in the absence of other mechanisms, the magnetic-tension force $B_z\partial_zB_\phi$ provides the dominant torque \citep[see][]{lesur_magnetohydrodynamics_2021}. Local regions where $\partial_z B_\phi\!<\!0$, shown in Fig.~\ref{fig:baz_connection_magnetic}, broadly correspond to regions of strong accretion, indicated in purple in Fig.~\ref{fig:baz_rho_ur_evolution}. For example, the trident-like state (middle panels) roughly corresponds to strong radial mass flows along each `tine'. Meanwhile, in the asymmetric state (right panel), the radial accretion flow in the dead zone is deflected upwards as it approaches the interface, reflecting the vertical asymmetry of the current sheet in the inner dead zone within this epoch. As such, the broad categorisation above reflects: a two-current-sheet structure in the active zone, a single diffuse current sheet in the dead zone, and a complex, evolving morphology around the interface. 

\subsubsection{Mass accretion rates}

\begin{figure}
    \centering
    \includegraphics[width=\columnwidth]{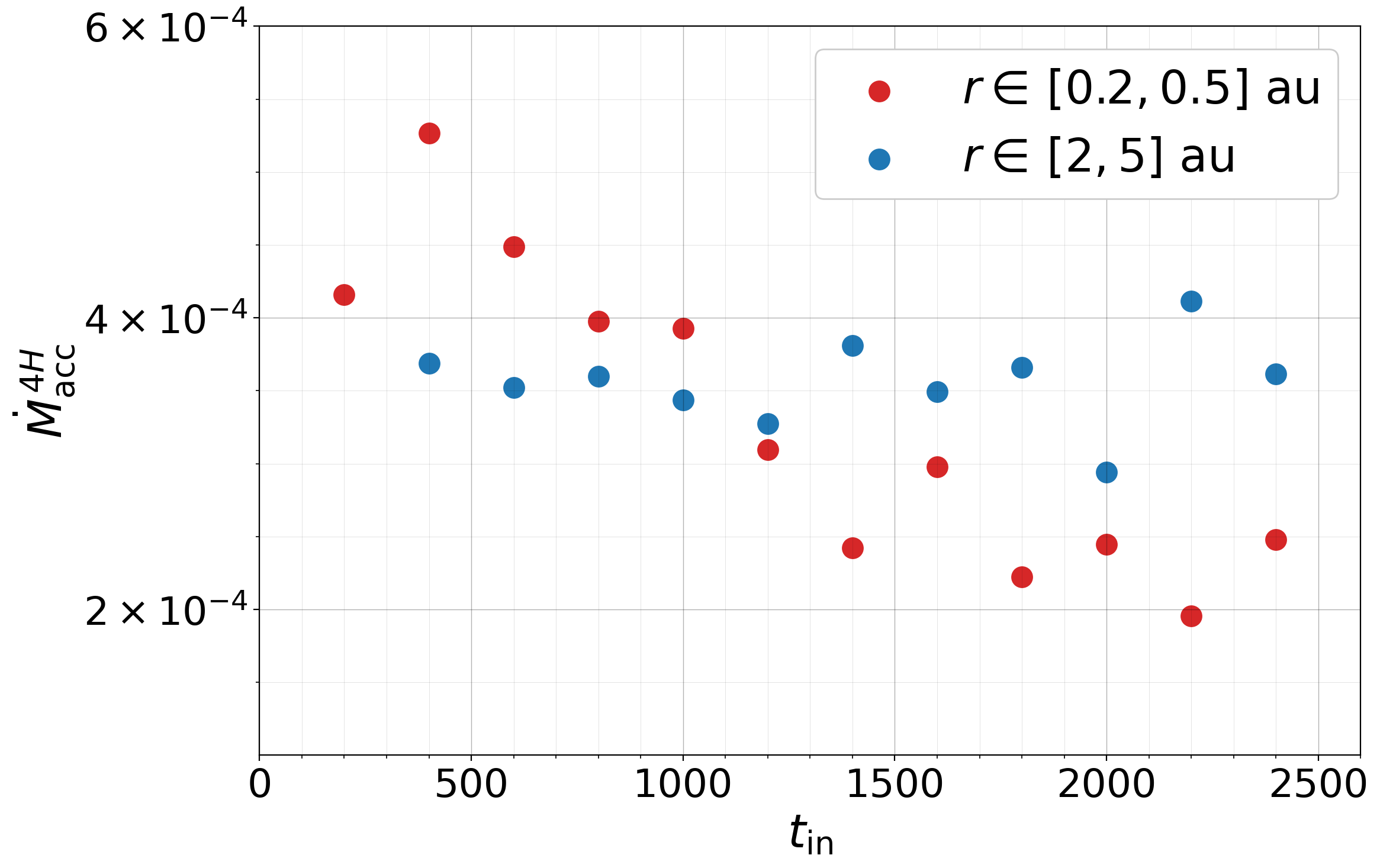}
    \caption{Evolution of the mass accretion rate, $\dot{M}^{4H}_\text{acc}$ defined in equation~\eqref{eq:mass_accretion_rate_mean_field}, shown in code units, and smoothed with radial averaging over intervals of $t_\text{in}=200$.
    The accretion rate decays in the inner part of the active zone (red; $r\!\in\![0.2,0.5]$ au) as large-scale magnetic flux is evacuated, and is constant in the dead zone (blue; $r\!\in\![2,5]$ au). The radial structure is shown in Fig.~\ref{fig:BAZ_Mdot_evolution} and the absent blue dot reflects the wind-up time of accretion driven by magnetic winds.}
  \label{fig:BAZ_accretion_rate_time}
\end{figure}

To characterise the accretion further, Fig.~\ref{fig:BAZ_Mdot_evolution} shows the radial profile of the mass accretion rate, defined in equation~\eqref{eq:mass_accretion_rate_mean_field}, vertically integrated over the entire disc ($\dot{M}_{\text{acc}}^{4H}$, blue line) and over a scale height on either side of the midplane ($\dot{M}_{\text{acc}}^{1H}$, red line) for the three epochs of interest. Note that positive values indicate accretion. \\
\indent The accretion structure around the interface is extremely complex and time dependent, with strong variations issuing from the evolution of the current-sheet structure at the interface (shown in Fig.~\ref{fig:baz_connection_magnetic}). Despite that, across all epochs, there is sustained decretion around the midplane at the interface (red lines in Fig.~\ref{fig:BAZ_Mdot_evolution}), in agreement with the results of \citetalias{iwasaki_dynamics_2024}. This is intriguing because it shows that, notwithstanding the evolution of the current-sheet structures, which are controlling the accretion elsewhere, it is still the viscous-driven outward movement of material that dominates overall in this region, akin to the ZNF models examined in \citetalias{roberts_global_2025}. For example, the asymmetric state in the right panel of Fig.~\ref{fig:baz_connection_magnetic} corresponds to an outward-directed current sheet at the midplane. If this was the sole driver of accretion it would force inward accretion within this region. Instead it only moderates the radial outflow of material. Meanwhile, during the trident-like state shown in the middle panel of Fig.~\ref{fig:BAZ_Mdot_evolution}, the current-sheet structure exacerbates the decretion system around the midplane (see the red curve), whilst supporting accretion nearer the surface layers, since $|\dot{M}_{\text{acc}}^{1H}| \gg |\dot{M}_{\text{acc}}^{4H}|$, in agreement with the middle panel of Fig.~\ref{fig:baz_rho_ur_evolution}. \\ 
\indent In contrast, away from the interface, the accretion rate exhibits less dramatic variation. Fig.~\ref{fig:BAZ_accretion_rate_time} complements Fig.~\ref{fig:BAZ_Mdot_evolution} and shows the evolution of $\dot{M}^{4H}_\text{acc}$ for the inner part of the active zone and the dead zone, smoothed with radial averaging over intervals of $t_\text{in}=200$. In the inner part of the active zone, the accretion rate gradually decreases from $5\times 10 ^{-4}$ to $2\times 10 ^{-4}$, driven by the depletion of flux. Crucially, this reduces, and eventually reverses, the accretion mismatch sufficiently far from the interface (see Fig.~\ref{fig:BAZ_accretion_rate_time}). Expressed in physical units, the decline corresponds to $8.0$--$2.6\times10^{-7}\,\mathrm{M}_\odot\,\mathrm{yr}^{-1}$ over \texttt{NF-BAZ}, which is consistent with \citetalias{iwasaki_dynamics_2024} who report $2.8\times10^{-7}\,\mathrm{M}_\odot\,\mathrm{yr}^{-1}$, and with observed accretion rates \citep[e.g.][]{manara_demographics_2023}. This active-zone accretion is dominated by the mass flux in the disc surface layers since $|\dot{M}_{\text{acc}}^{4H}| \gg |\dot{M}_{\text{acc}}^{1H}|$ (see Fig.~\ref{fig:BAZ_Mdot_evolution}), in agreement with the two surface-layer current sheets and the large-scale $r$-$\phi$ stresses in the disc's surface layers \citep[see][]{zhu_global_2018}.\\
\indent Finally, in the dead zone, accretion is confined close to the midplane since $\dot{M}_{\text{acc}}^{4H}\!\sim\! \dot{M}_{\text{acc}}^{1H}$, as expected for laminar magnetic-wind-driven accretion when the current sheet is located at the midplane. The coherent Maxwell stress components, $\mathcal{M}_{r\phi}^{\mathrm{coh}}\!\sim\!\mathcal{M}_{\theta\phi}^{\mathrm{coh}}\!\sim\!2\times10^{-8}$ (not shown), predict an accretion rate of $4\pi R^{-1}\Omega_\text{K}^{-1}R^2\mathcal{M}_{\theta\phi}^{\mathrm{coh}}\!\sim\!4\times10^{-4}$ at $R\!\simeq\!20R_0$ according to thin-disc, self-similar magnetic-wind models \citep[e.g.][]{lesur_magnetohydrodynamics_2021}. This agrees closely with the measured, roughly steady value (blue dots in Fig.~\ref{fig:BAZ_accretion_rate_time}), indicating that other sources of angular momentum transport in the dead zone -- such as density waves or marginal VSI -- are negligible.

\subsubsection{Implications for long-term inner-disc evolution}
\label{section:implications_long_term_inner_disc_evolution}

\indent Across all three epochs the accretion rate in the inner part of the active zone is comparable to, or slightly less than, the rate in the dead zone beyond $R\!\sim\!2\,\mathrm{au}$ (see Fig.~\ref{fig:BAZ_accretion_rate_time}); the contrast is even stronger when the integration is restricted to a single scale height about the midplane (red lines in Fig.~\ref{fig:BAZ_Mdot_evolution}). This constitutes a clear departure from ZNF simulations. \\
%\citep[e.g.][]{gammie_layered_1996}. 
\indent In the `classical' route to a (possible) steady state -- an $\alpha$-disc with $\alpha=\alpha(R)$ -- the system decreases the active-zone torque to the required level by simply depleting the active zone of mass \citep[e.g.][]{terquem_new_2008}. Our results indicate an alternative, possibly complementary, pathway: the active-zone torque is decreased `magnetically', via the depletion of large-scale flux, and by the associated increase in the poloidal plasma-$\beta$ (see Section~\ref{section:mft_inner_az}). The latter route need not necessarily lead the system to a `transition-disc state' \citep{pinilla_can_2016} -- i.e., a low-$\beta$, gas-depleted inner disc \citep[e.g.][]{martel_magnetised_2022-1} -- at least on the short timescales accessible to global numerical models. Instead, the accretion rate in the inner part of the active zone gradually declines, whilst remaining relatively steady in the dead zone. Should this persist, the local disc configuration could approach a regime resembling, in some respects, an outer dead--active zone interface \citep[e.g.][]{delage_steady-state_2022}, where the total torque is higher outside the interface. \\
\indent Having said all that, near the interface it appears that a persistent bottleneck remains. In all epochs there is a local net mass-flux directed radially into the dead-zone (blue lines in Fig.~\ref{fig:BAZ_Mdot_evolution}), sustained by meridional circulation, which promotes vertical mixing in the outer part of the active zone. Yet the density enhancement at the midplane does not grow monotonically with time, as one might conclude. This is shown in Fig.~\ref{fig:SAZ_BAZ_compare_HD_density}, which presents snapshots of the azimuthally averaged midplane density, $\langle \rho \rangle_\phi$, for \texttt{NF-BAZ} and \texttt{NF-SAZ}. If this is robust, it has important implications for variability, as limiting the \mbox{(gas-)mass} build-up at the interface suppresses one pathway that leads to episodic outbursts (e.g. \citealt{wunsch_two-dimensional_2006}; \citealt{cecil_variability_2024}). 

\subsection{Large-scale hydrodynamic structures}
\label{section:nf_flux_HD_str}

A key motivation for these simulations is to test the robustness of pressure-maximum formation at the dead--active zone interface when accretion in the dead zone is driven by weak magnetic pressure winds. Unlike the `layered-accretion' paradigm of \citet{gammie_layered_1996}, or the early configuration of our ZNF simulations in \citetalias{roberts_global_2025}, the VNF setup supports a non-negligible source of accretion within the dead zone that complicates the accretion mismatch concept.

\subsubsection{Morphology and evolution}
\label{section:hd_morphology_evolution}

In all VNF dead--active zone models -- \texttt{NF-BAZ}, \texttt{NF-SAZ} and \texttt{NF-SAZ-BC} -- an axisymmetric pressure maximum forms just outside the interface, developing a local vortensity minimum that triggers the Rossby wave instability within a few local orbits \citep{lovelace_rossby_1999}. Ultimately, the instability forms a single vortex by $t_\text{DZI}\!\sim\!8$, which eventually decays due to numerical dissipation \citepalias{roberts_global_2025} in a comparable manner to other global simulations \citep[e.g.][]{hsu_rossby_2024}. \\
\begin{figure}
    \centering
    \includegraphics[height=0.5 \columnwidth]{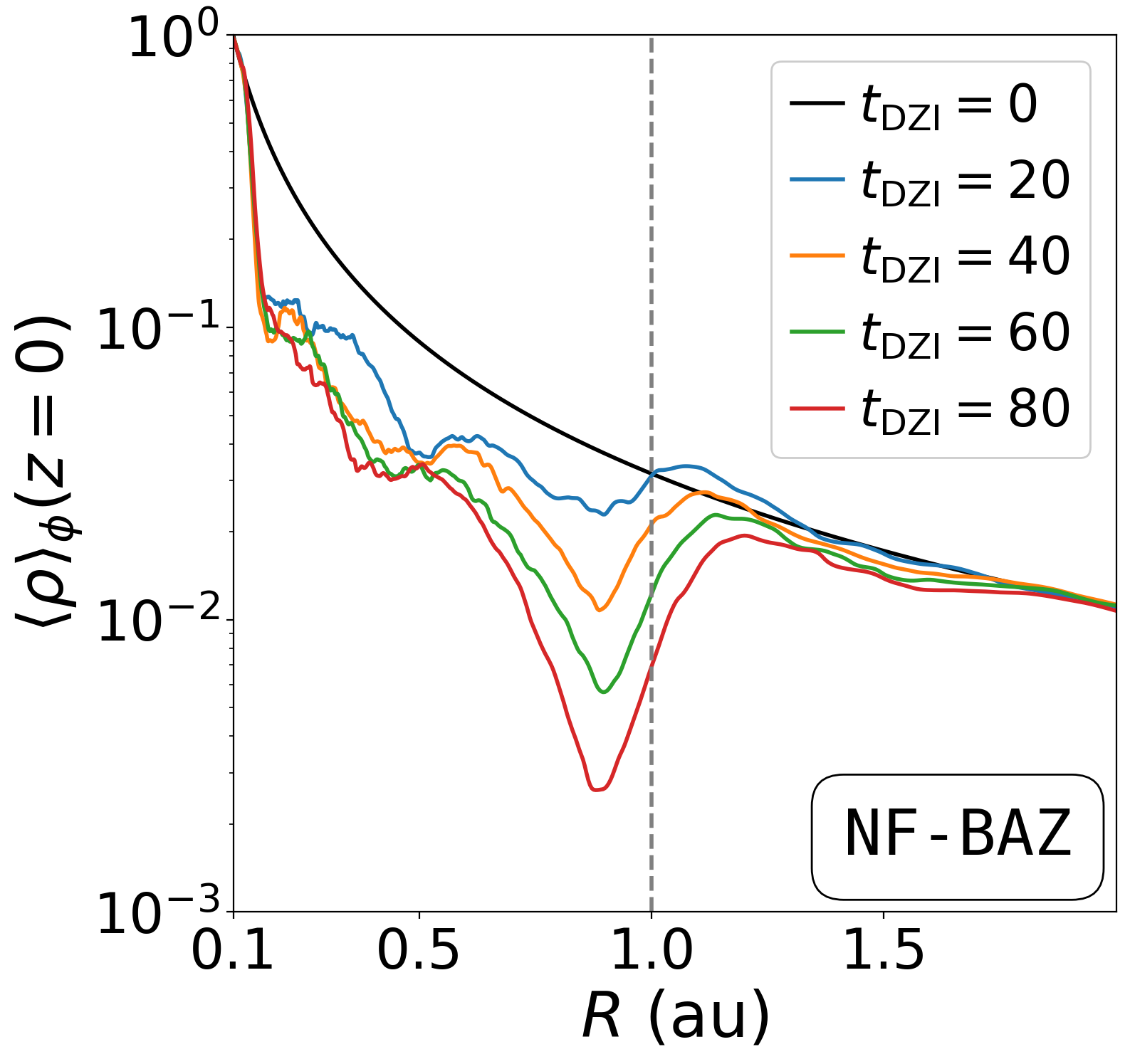}
    \includegraphics[height=0.491875 \columnwidth]{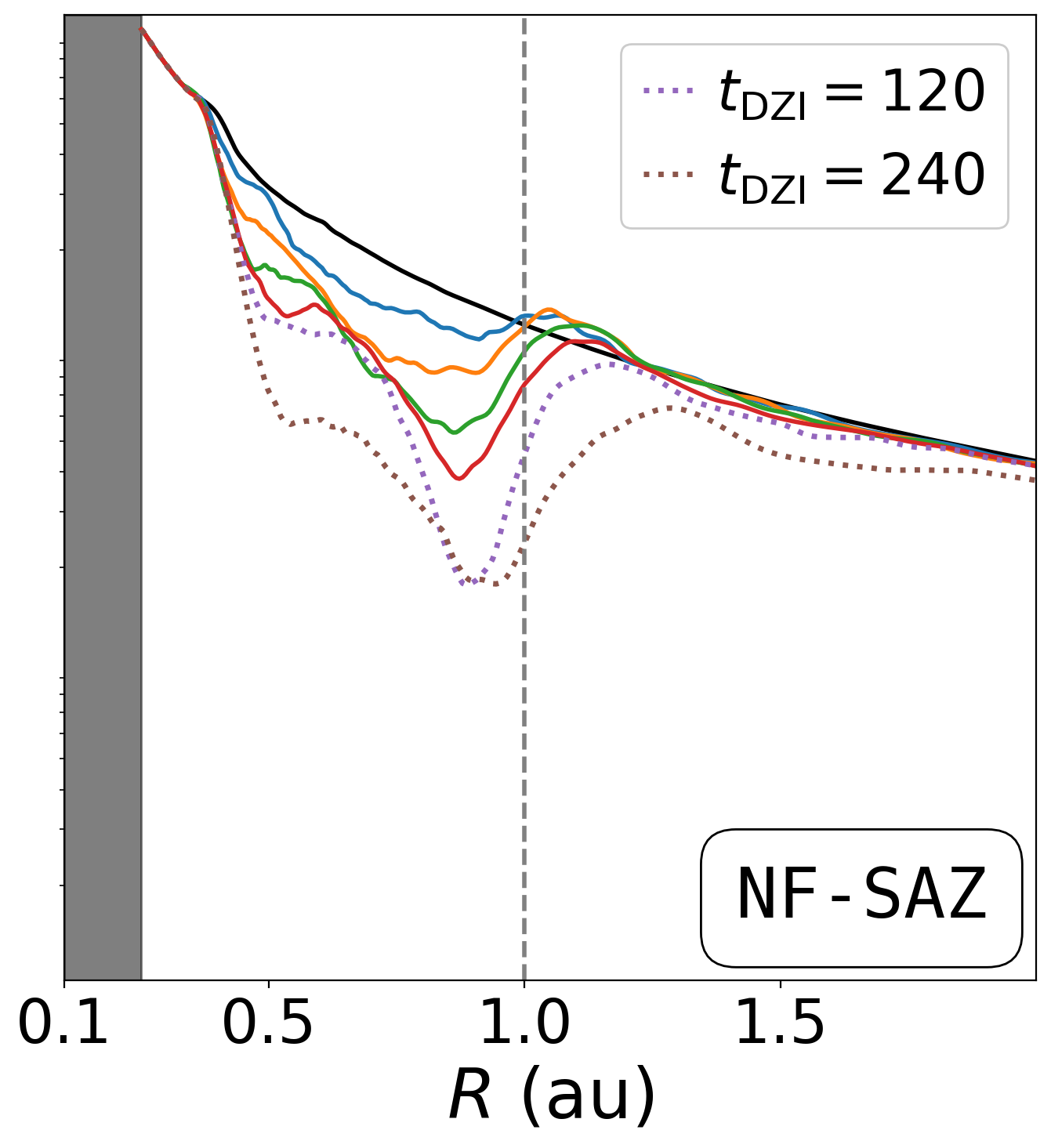}
    \caption{Radial profiles of the azimuthally averaged midplane density, $\langle \rho \rangle _\phi$, for \texttt{NF-BAZ} (left) and \texttt{NF-SAZ} (right). The dotted curves in the right panel indicate times not reached by \texttt{NF-BAZ} and the dark-grey shaded region marks the domain edge in \texttt{NF-SAZ}.}
    \label{fig:SAZ_BAZ_compare_HD_density}
\end{figure}
\indent Although the initial disc structures differ in each model (see Section~\ref{section:numerical_methods}), the evolution of the density structure around the interface is remarkably similar. It evolves in two phases: an early-time density enhancement just outside the interface, followed by the gradual formation of a deep density minimum in the outer part of the active zone \citep[in contrast to the ZNF models of][]{roberts_global_2025}. Specifically, at early times (blue and orange lines in Fig.~\ref{fig:SAZ_BAZ_compare_HD_density}) a local axisymmetric density enhancement forms just outside the interface, due to the viscous outward movement of material. After $t_{\text{DZI}}\!\sim\!40$ (e.g. green and red lines in Fig.~\ref{fig:SAZ_BAZ_compare_HD_density}) the radial density structure gradually develops a deep minimum at the outer edge of the active zone, whilst the initial density enhancement erodes away. Even at $t_{\text{DZI}}=100$ in the ZNF models examined in \citetalias{roberts_global_2025}, there is no evidence of such a pronounced, local minimum, and thus this `carving out' is a consequence of the VNF in the active zone. In fact, the evacuation is driven by the large-scale poloidal magnetic flux that accumulates at the outer edge of the active zone (see Section~\ref{section:mft_outer_az}), in a fashion reminiscent of zonal flows \citep[e.g.][]{johansen_zonal_2009}, or planet-carved gaps \citep[e.g.][]{wafflard-fernandez_planet-disk-wind_2023}. \\
\indent We emphasise that the erosion of the early-time density enhancement just outside the interface (also discussed at the end of Section~\ref{section:implications_long_term_inner_disc_evolution}), should not be conflated with the existence of a local density/pressure maximum, which is present throughout the evolution of each model. Relatedly, the strong density evacuation may encourage the existence of a coherent pressure maximum and vortensity minimum, even if the dead zone has a very high accretion rate. \\
\indent Finally, despite the complex vertical accretion structure at the interface and the local variability, each simulation maintains a coherent pressure maximum up to $|z| = 4H$ due to the suppression of the MRI in the dead-zone surface layers by strong ambipolar diffusion. Therefore, dust accumulation from either side of the interface throughout the disc's vertical extent remains a likely outcome in the VNF regime, provided a pressure maximum forms.

\subsubsection{Ubiquity of pressure-maximum formation}
\label{section:ubiquity_pressure_maximum_formation}
\indent It is challenging to determine the generality of pressure-maximum formation at the interface in the VNF regime. The disc torques are sensitive to parameters such as $\beta_\text{p}$, and the accretion rates in the active and dead zones likely scale differently with each parameter, making predictions hard. Unfortunately, the computational cost of these global simulations limits our ability to explore this parameter space, and our parameter setup is similar to \citetalias{iwasaki_dynamics_2024}. Therefore, we hypothesise that if parameters were fine-tuned to maximise the dead-zone wind torque -- such as by reducing $\beta_\text{p}$, $\text{R}_{\text{m}_0}$, $\Lambda_{\text{A}_0}$ and $T_{\text{c}}$ (e.g. \citealt{lesur_systematic_2021}) -- the pressure maximum at the interface might not emerge at all. \\
\indent Moreover, the depletion of large-scale magnetic flux throughout most of the active region, examined in Section~\ref{section:results_global_flux_transport}, progressively weakens the torques in the inner region, thereby reducing -- and eventually reversing -- the accretion mismatch, when sufficiently removed from the interface (see right-half of Fig.~\ref{fig:BAZ_accretion_rate_time}). This secular evolution introduces further doubt, because ubiquitous strong outbursts \textit{reset} the inner-disc structure \citep[e.g.][]{fischer_accretion_2023}, necessitating the \mbox{(re-)formation} of a pressure maximum at the interface under a (possibly) less favourable torque configuration.

%%%%%%%%%%%%%%%%%%%%%%%%%%%%%%%%%%%%%%%%%%%%%%%%%%%%%%%%%%%%%%%%%%%%%%%%%%%%%%%%%%%%%%%%%%%%%%%%%%%%
%%%%%%%%%%%%%%%%%%%%%%%%%%%%%%%%%%%%%%%%%%%%%%%%%%%%%%%%%%%%%%%%%%%%%%%%%%%%%%%%%%%%%%%%%%%%%%%%%%%%

\section{Current-sheet configurations at the dead--active zone interface}
\label{section:nf_flux_current_sheet}

\subsection{Reprisal of the current-sheet evolution}
\begin{figure}
    \centering
    \includegraphics[width=\columnwidth]{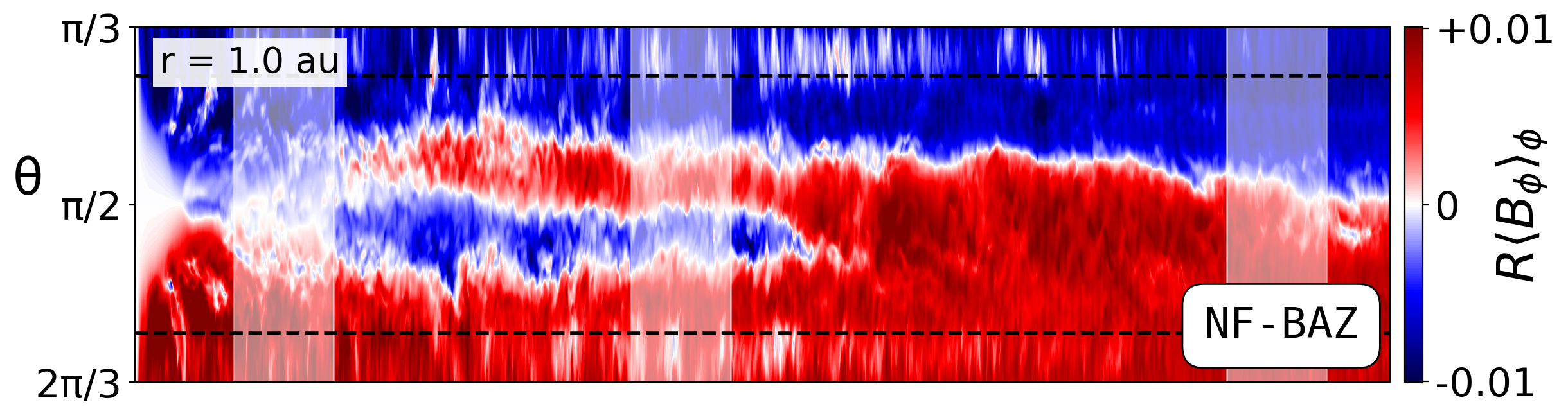} \\
    \includegraphics[width=\columnwidth]{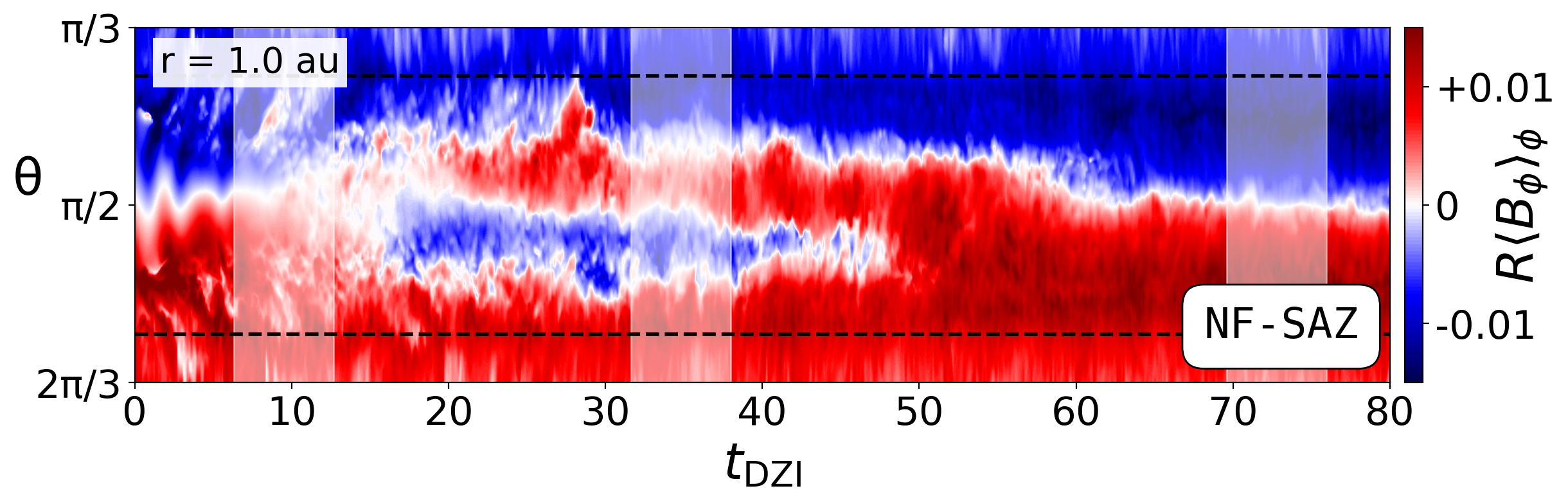}
    \caption{Space--time $(\theta,t)$ diagrams along the contour $r=1\,\text{au}$ for the normalised azimuthally averaged toroidal field, $R\langle B_\phi \rangle_\phi$, where the white-shaded regions denote the three chosen current-sheet evolution epochs, as shown in Fig.~\ref{fig:baz_connection_magnetic}. The evolution at the dead--active zone interface in \texttt{NF-BAZ} (top) and \texttt{NF-SAZ} (bottom) is nearly indistinguishable, despite the differing numerical protocol (see Table~\ref{table:list_net_flux_global_simulations}), and \texttt{NF-SAZ} evolves for a further $164t_{\text{DZI}}$.}
	\label{fig:SAZ_BAZ_compare_bx3}
\end{figure}

The current-sheet configuration at the dead--active zone interface exhibits a complex topological rearrangement over time, shown previously in Fig.~\ref{fig:baz_connection_magnetic}. This is corroborated by Fig.~\ref{fig:SAZ_BAZ_compare_bx3}, which shows space--time $(\theta, t)$ diagrams along the contour $r = 1\,\text{au}$ for the azimuthally averaged toroidal field, $R\langle B_\phi \rangle_\phi$, for \texttt{NF-BAZ} and \texttt{NF-SAZ}. The three white-shaded regions correspond to the epochs introduced in Section~\ref{section:overview_reconfiguration} and, as before, the nulls of $B_\phi$ (white) are proxies for strong radial currents: the current sheets. Animations of $R\langle B\rangle_\phi$ around the dead--active zone interface are available as ancillary files for \texttt{NF-BAZ} and \texttt{NF-SAZ} (see \texttt{BAZ\_bphi.mp4} and \texttt{SAZ\_bphi.mp4}). \\
\indent This evolution is remarkable as the toroidal field spontaneously develops finer $z$-structure in the presence of strong Ohmic diffusion, without relying on the formation of large-scale poloidal field loops. For example, the transitional trident-like state comprises three stacked current sheets that penetrate into the dead zone (see middle panel of Fig.~\ref{fig:baz_connection_magnetic}). These subsequently diffuse to the later-time, asymmetric state -- sustained for the remainder of \texttt{NF-SAZ} -- on the Ohmic-diffusion timescale, $t_\text{O}\!\sim\!L^2 / \eta_\text{O}\!\sim\!25t_\text{DZI}$, where $L\!\sim\!4R_0$ is the (vertical) length scale of the structure. Furthermore, the current-sheet evolution in the two panels is indistinguishable; and when viewed in conjunction with the radial density structure in Fig.~\ref{fig:SAZ_BAZ_compare_HD_density}, suggest that \texttt{NF-SAZ} plausibly extends \texttt{NF-BAZ} to longer times. \\
\indent In the remainder of this section, we hypothesise that the current-sheet rearrangement is a direct consequence of the challenges in \textit{matching} poloidal electric currents across the interface.

\subsection{Mechanism of current-sheet evolution}
\label{section:currents_connection_mechanism}
\begin{figure*}
    \centering
    \includegraphics[width=\textwidth]{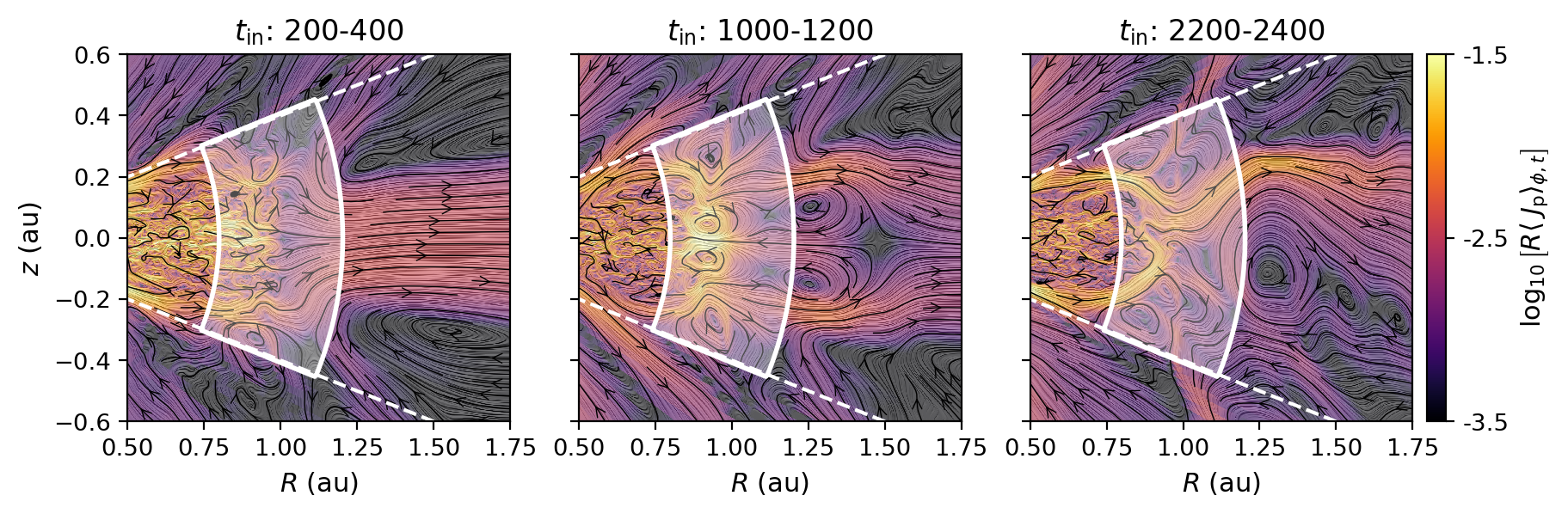}
    \caption{Meridional plots of the poloidal current density, $\langle \mathbf{J_p}\rangle_{\phi,t}$ (black lines and LIC overlay) and its normalised magnitude, $R\langle J_\mathrm{p}\rangle_{\phi,t}$  (background), at three epochs of \texttt{NF-BAZ}: $t_\text{in} \in$ [200,400], [1000,1200] and [2200,2400]. The sequence traces the evolution from an early-time disordered state (left), through a transitional, trident-like state (middle), to a vertically asymmetric state (right), in line with the stages shown in Fig.\ref{fig:baz_connection_magnetic}. The chosen closed contour $\partial S$ (solid white) is the boundary of $r\!\in\![0.8,1.2]$ au and $\theta\!\in\![z=-4H, z=4H]$.}
    \label{fig:baz_connection_electric}
\end{figure*}

\begin{figure}
    \centering
    \includegraphics[width=\columnwidth]{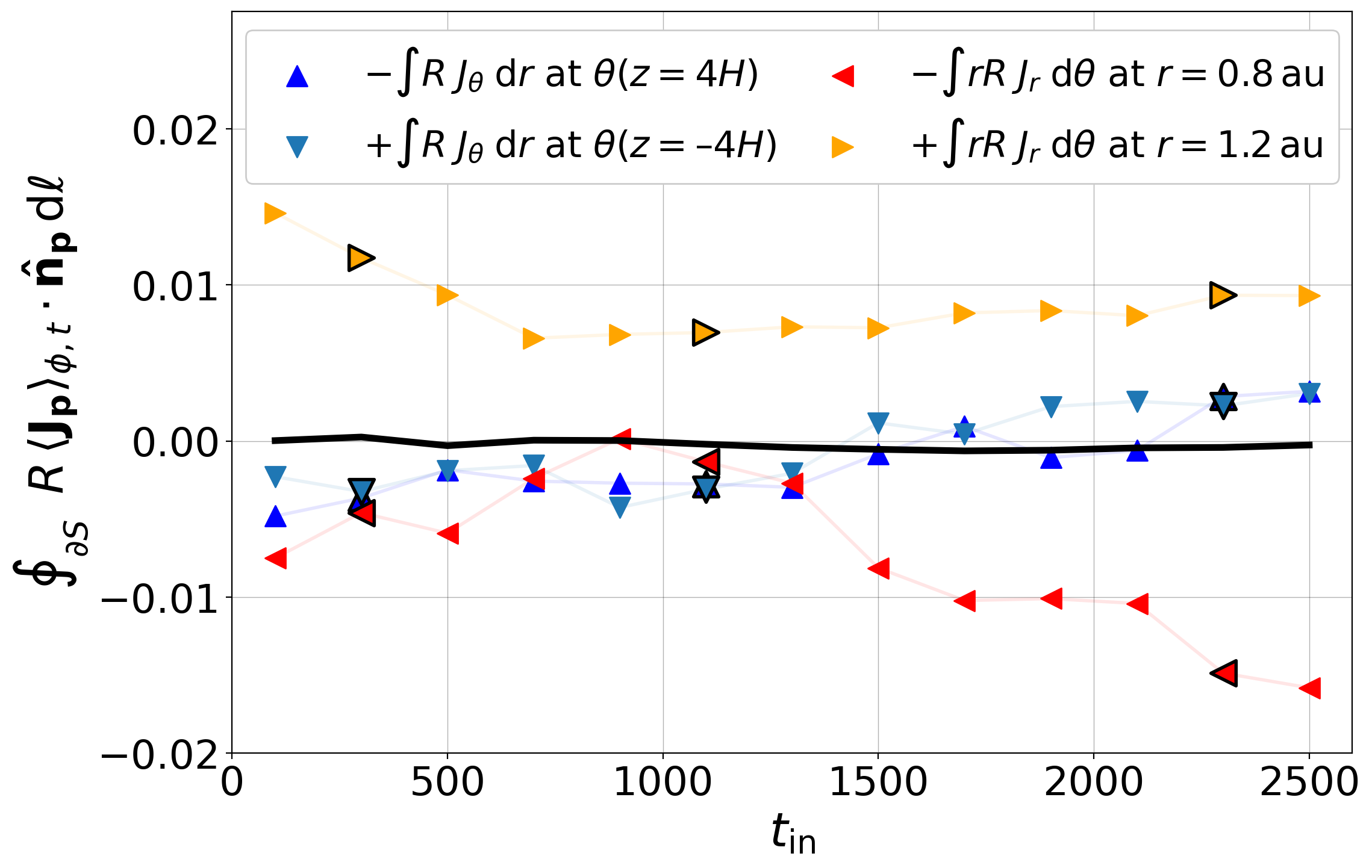}
    \caption{Evolution of $\oint_{\partial S} R \,\langle \mathbf{J_p} \rangle_\phi \cdot \mathbf{\hat{n}_p}\;\text{d}\ell$, derived in equation~\eqref{eqn:integrated_current}, for \texttt{NF-BAZ} integrated over the closed contour $\partial S$ shown in Fig.~\ref{fig:baz_connection_electric}. The total (black line) is conserved and the data are averaged over intervals of $t_\text{in}=200$. The four components of the integral are shown in colour and those with black outlines correspond to the plots shown in Fig.~\ref{fig:baz_connection_electric}.} 
    \label{fig:BAZ_eq32_component_evolution}
\end{figure}

The current-sheet evolution at the interface shows several ways of accommodating the matching of vertically integrated currents. Unlike mass, electric charge cannot accumulate in the system because the net current through any closed volume $V$ vanishes. Applying the divergence theorem to $\nabla\cdot\mathbf{J}=0$ over a control volume $V$ that is $\phi$-independent, and accounting for periodicity in $\phi$ yields
\begin{equation}
    \iiint_V \nabla\cdot\mathbf{J}\;\text{d}V =
    \oiint_{\partial V} \mathbf{J} \cdot \mathbf{\hat{n}} \;\text{d}S = \phi_\text{max}\oint_{\partial S} R \,\langle \mathbf{J_p} \rangle_\phi \cdot \mathbf{\hat{n}_p}\;\text{d}\ell =0,
    \label{eqn:integrated_current}
\end{equation} 
where $S=\partial V$, $\mathbf{\hat{n}}$ and $\mathbf{\hat{n}_p}$ are the respective outward-directed unit normals, $\text{d}\ell$ is the infinitesimal arc length along the closed contour $\partial S$ in the meridional plane, and the factor of $R$ is from the cylindrical Jacobian. Therefore, in the meridional $(R,z)$ plane, $R\langle\mathbf{J_p}\rangle_\phi$ must be conserved within any closed contour.  \\
\indent Fig.~\ref{fig:baz_connection_electric} shows the normalised poloidal current-density magnitude, $R\langle J_\text{p} \rangle _{\phi,t}$, for the three epochs of interest during the rearrangement process: $t_\text{in}\!\in$ [200,400], [1000,1200], and [2200,2400]. We choose a contour $\partial S$ that encloses the interface, defined by the boundary of $r\!\in\![0.8,1.2]$ au and $\theta\!\in\![z=-4H, z=4H]$. It is overplotted as a solid white line in  Fig.~\ref{fig:baz_connection_electric}. Fig.~\ref{fig:BAZ_eq32_component_evolution} (black curve) shows the evolution of $\oint_{\partial S} R \,\langle \mathbf{J_p} \rangle_\phi \cdot \mathbf{\hat{n}_p}\;\text{d}\ell$ , averaged over intervals of $t_\text{in}=200$, confirming that equation~\eqref{eqn:integrated_current} is satisfied.\\
\indent The left panel of Fig.~\ref{fig:baz_connection_electric} describes the early-epoch, disordered current-sheet structure: in the active zone, two weak surface-layer current sheets flank a chaotic interior and loosely connect with the single thick dead-zone `current block' at high altitudes ($|z|\approx0.2$ au). Unfortunately, this configuration cannot be sustained because the active-zone current structure evolves as a result of magnetic flux accumulation just inside the interface, and the current must remain matched across the interface. Additionally, there are (four) distinct large-scale poloidal current-loop systems that close in the corona (e.g. top-right of left panel), which are essential for the operation of the magnetic wind. The ensuing variability at the interface is driven by the evolution of the current sheets. This is a complex process, as a spontaneous bifurcation of a single current sheet is improbable in a laminar, ordered regime given $\nabla \cdot \mathbf{J} = 0$; therefore, the evolution is likely to be turbulent and time-dependent, or to rely on the single dead-zone current block splitting into three.  \\
\indent Meanwhile, in the middle panel of Fig.~\ref{fig:baz_connection_electric}, which corresponds to the trident-like state shown in the middle panel of Fig.~\ref{fig:baz_connection_magnetic}, a hyperbolic stagnation point is visible in the inner dead zone at $(R,z)\approx(1.5,0)$ au. This feature is critical, as it facilitates the splitting of the single dead-zone current block into three distinct sheets: a radially inward-directed midplane current sheet, flanked by two radially outward-directed narrower sheets at intermediate altitudes, in order to accommodate the current matching. \\
\indent The right panel of Fig.~\ref{fig:baz_connection_electric} show the subsequent evolution of the disfavoured, radially outward-directed lower current sheet as it bends downwards and disappears into the corona, resulting in a single, diffuse current sheet in the outer part of the active zone that aligns with the structure in the dead zone, thereby establishing a preferred, off-midplane configuration within the disc interior. Eventually, the dead-zone current sheet settles to the midplane, and the current morphology is sustained for several hundred local orbits in \texttt{NF-SAZ}, underscoring its dynamical favourability. During this evolution ($t_\text{in}\!\gtrsim\!1000$), the radial current in the outer part of the active zone (red line in Fig.~\ref{fig:BAZ_eq32_component_evolution}) strengthens, consistent with the continual accumulation of magnetic field there and the loss of the local inward-directed midplane current. The required current matching is primarily facilitated by the meridional current components (blue lines in Fig.~\ref{fig:BAZ_eq32_component_evolution}), in contrast to the early-epoch evolution, highlighting the crucial role of the vertical current structure in global magnetic wind models. \\
\indent Whilst this is not a simple dynamical system, this process is reminiscent of the unfolding of a subcritical pitchfork bifurcation into a generic saddle-node bifurcation plus a smooth continuous curve, as the control parameter is varied \citep[see fig. 11 in][]{crawford_introduction_1991}. The latter is exactly what is observed in the top part of the right panel of Fig.~\ref{fig:baz_connection_electric} with the smooth single diffuse current sheet crossing through the interface, whilst the saddle-node is visible in the bottom part of the panel at roughly $(R,z) = (1.2,-0.2)$ au. \\
\indent Of course, this procedure relies on the morphology and number of current sheets in the dead zone, which are primarily governed by the (highly uncertain) non-ideal diffusivities. The vast majority of global non-ideal magnetic-driven wind simulations sustain a single current sheet -- located either at the midplane (e.g. fig. 2 in \citealt{bai_hall_2017}; fig. 5 in \citealt{lesur_systematic_2021}), at a disc surface (e.g. fig. 1 in \citealt{riols_ring_2020}), or in a more complex configuration (e.g. fig. 7 in \citealt{bai_global_2017}; fig. 9 in \citealt{gressel_global_2020}; fig. 8 in \citetalias{iwasaki_dynamics_2024}), whereby a single sheet changes altitude as it goes through the disc. The few systems with multiple dead-zone current sheets (e.g. fig. 20 in \citealt{bethune_global_2017}), would only introduce additional variability to the current-sheet evolution at the interface.

\subsection{Generality of current-sheet evolution}
\label{section:currents_generality_connection}
\indent Whilst it is tempting to focus solely on the later-time state shown in the right panel of Fig.~\ref{fig:baz_connection_magnetic} -- which is also sustained during the more mature evolution of \texttt{NF-SAZ} -- and to discount the current-sheet evolution at the interface examined in Section~\ref{section:currents_connection_mechanism} as a rearrangement of the initial conditions, we instead suggest that this phase is more prevalent and general. Specifically, major accretion events can disrupt the evolved disc structure, returning the system to the early-time disordered state shown in the middle panel of Fig.~\ref{fig:baz_wide_scale_early_time}. Such strong events -- ranging from FU Ori outbursts to violent interactions with the stellar dipole \citep[e.g.][]{fischer_accretion_2023} -- are ubiquitous in these systems, implying that \textit{resets} are frequent and that the associated current-sheet evolution is more representative than these short simulations indicate.  \\
\indent It is also noteworthy that a remarkably similar evolution -- including a transitional phase characterised by a trident-like morphology -- is present in global non-ideal MHD transition-disc simulations (e.g. fig. 5 in \citealt{martel_magnetised_2022-1}; fig. 1 in \citealt{sarafidou_global_2025}). This suggests the current-sheet evolution is a more general mechanism associated with the radial matching of different magnetic-field states; in the transition-disc cases, this is between a gas-depleted, low-$\beta$ inner disc, and a denser higher-$\beta$ outer disc. Consequently, the conceptual framework of current-matching is an important new tool for assessing a broad range of MHD disc simulations.

%%%%%%%%%%%%%%%%%%%%%%%%%%%%%%%%%%%%%%%%%%%%%%%%%%%%%%%%%%%%%%%%%%%%%%%%%%%%%%%%%%%%%%%%%%%%%%%%%%%%
%%%%%%%%%%%%%%%%%%%%%%%%%%%%%%%%%%%%%%%%%%%%%%%%%%%%%%%%%%%%%%%%%%%%%%%%%%%%%%%%%%%%%%%%%%%%%%%%%%%%

\section{Magnetic flux transport}
\label{section:results_global_flux_transport}

\subsection{Magnetic flux conservation}
\label{section:mft_flux_cons}

\indent Before analysing the large-scale magnetic flux transport in \texttt{NF-BAZ}, which was introduced in Section~\ref{section:overview_magnetic_flux_transport}, we verify that magnetic flux is globally conserved. Fig.~\ref{fig:BAZ_psi_mid_split_shell} shows the evolution of $\Psi_{\text{mid}}(R,t)$ split into three components: the flux on the inner shell (orange), which increases as flux is transported inward and accumulates at the boundary; the flux threading the active zone (red), which gradually decreases over time since no flux is resupplied from the dead zone; and the flux threading the dead zone (blue), which remains approximately constant. The sum of these contributions (black), which represents the total flux threading the disc, is roughly conserved throughout both simulations, increasing by only 0.4 per cent in \texttt{NF-BAZ}. \\
\indent We now examine the radially varying flux-transport behaviour in \texttt{NF-BAZ} (see Fig.~\ref{fig:baz_magnetic_flux_transport_plot}) in Sections~\ref{section:mft_inner_az}--\ref{section:mft_dead_zone}, before discussing \texttt{NF-SAZ} in Section~\ref{section:mft_burps}.

\subsection{Inner part of the active zone}
\label{section:mft_inner_az}

\begin{figure}
    \centering
    \includegraphics[width=\columnwidth]{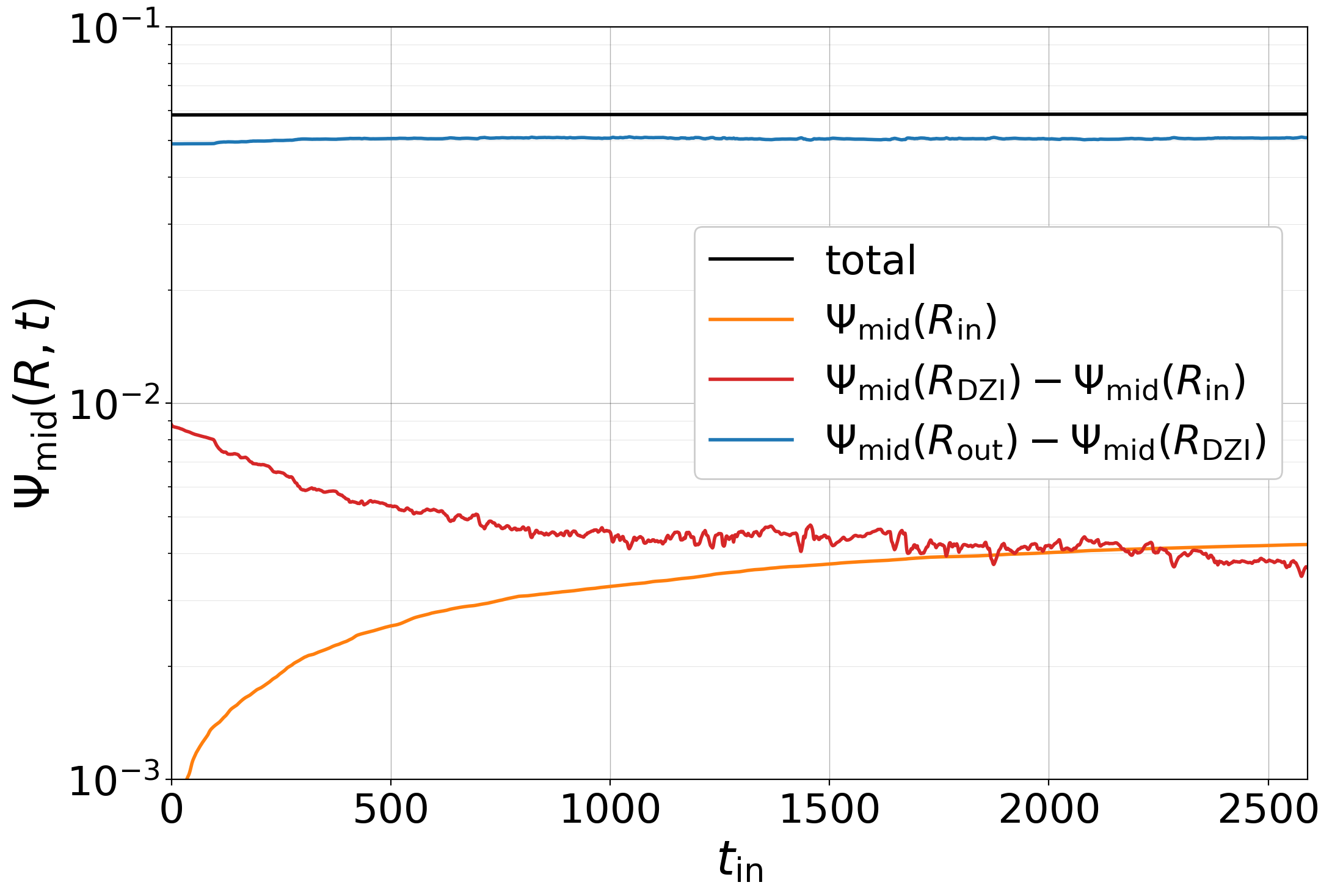}
    \caption{Evolution of the total magnetic flux threading the midplane $\Psi_{\text{mid}}(R,t)$ (black), for \texttt{NF-BAZ}. The flux is split into three components: the flux on the inner shell (orange), which increases as flux is advected inward; the flux threading the active zone (red), which decreases as no flux is resupplied from the dead zone; and, the flux threading the dead zone (blue).}
    \label{fig:BAZ_psi_mid_split_shell}
\end{figure}

\begin{figure}
    \centering
    \includegraphics[width=\columnwidth]{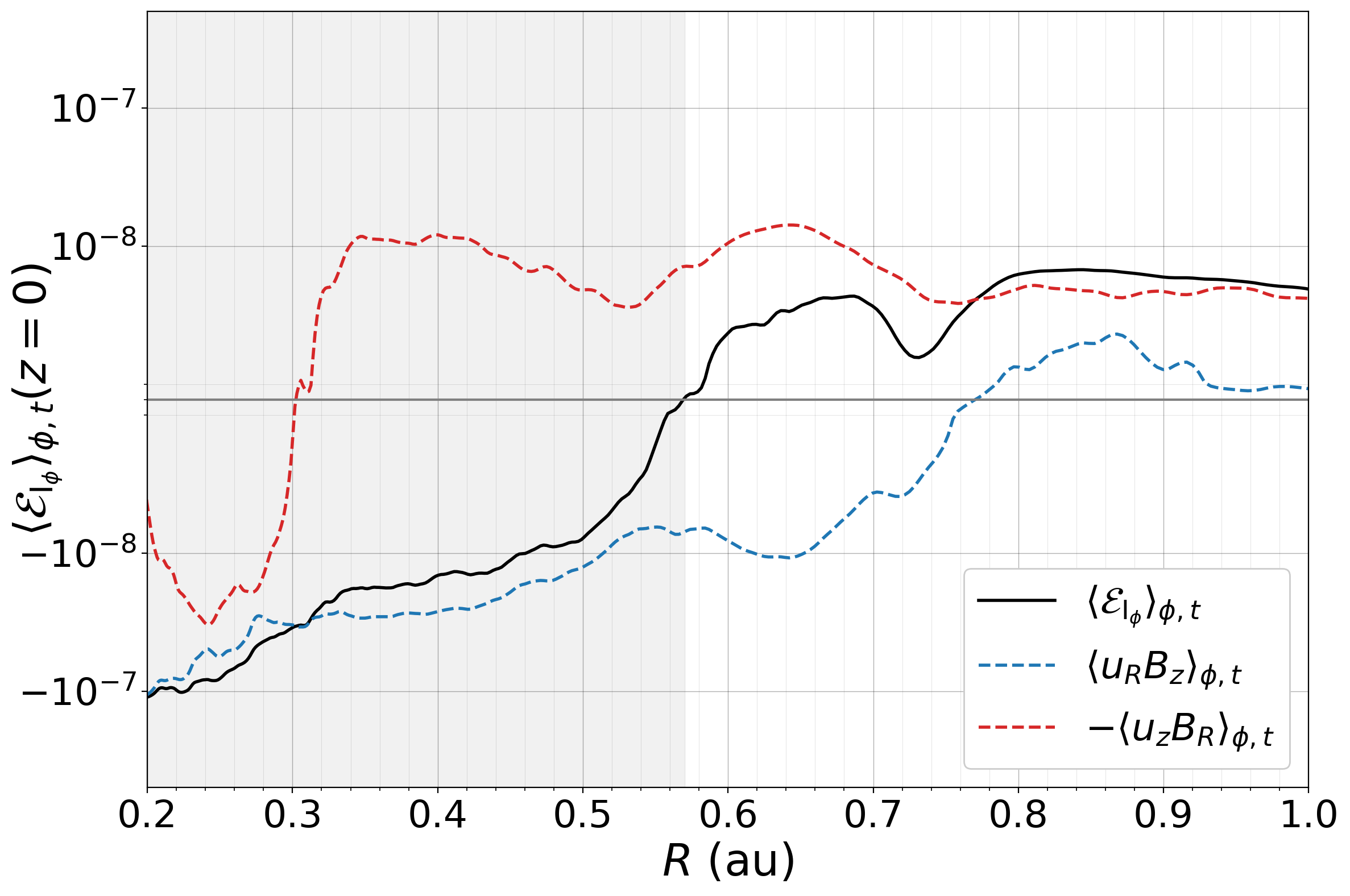}
    \caption{Midplane decomposition of the ideal toroidal electromotive field, $\langle \mathcal{E}_{\text{I}_\phi}\rangle _{\phi,t}$ (black) in the active zone for \texttt{NF-BAZ} averaged over the interval $t_{\text{in}}\in[1500,2500]$. The grey-shaded region denotes sustained inward transport, which is dominated by advective radial transport of vertical flux, $\langle u_R B_z \rangle_{\phi,t}$ (dashed blue), highlighting the split direction of transport in the active zone.}
\label{fig:mean_field_paper_magnetic_flux_speed_compare}
\end{figure}

Magnetic flux is gradually depleted from most of the active zone in all dead--active zone models, consistent with the prediction of \citet{lesur_hydro-_2023} and the simulations of \citetalias{iwasaki_dynamics_2024}, because magnetic flux is transported inward in the inner part of the active zone and is not resupplied from the dead zone (see Fig.~\ref{fig:baz_magnetic_flux_transport_plot}). In the inner part of the active zone, the magnetic field piles-up at the inner domain edge (see orange line in Fig.~\ref{fig:BAZ_psi_mid_split_shell}), tentatively suggesting that in a physical system the flux should accumulate in some region of small radius, possibly at the disc-magnetosphere boundary.
\footnote{In fact, recent unpublished simulations by G. Lesur of star--disc interactions -- with an MRI-active disc threaded by a VNF that is distinct from the stellar dipole -- reveal a magnetic flux pile-up at the magnetospherical truncation radius.} \\
\indent The sustained inward transport in this ideal-MHD region is governed by the components of the ideal electromotive field, $\langle \mathcal{E}_{\text{I}_\phi} \rangle_{\phi,t}$, outlined in equation~\eqref{eqn:decompose_ephi}. Fig.~\ref{fig:mean_field_paper_magnetic_flux_speed_compare} presents this midplane decomposition over the interval $t_{\text{in}} \!\in\![1500, 2500]$, revealing that radial advection of vertical field, $\langle u_R B_z \rangle_{\phi,t}$ (dashed blue), dominates the transport in the inner part of the active zone. A further decomposition (not shown) shows that the turbulent component, $\langle u_R B_z \rangle_{\phi,t} - \langle u_R \rangle_{\phi,t} \langle B_z \rangle_{\phi,t}$, is the main driver of inward transport.  

\subsubsection{Impact of flux depletion in the inner part of the active zone}
\label{section:impact_flux_draining}

\begin{figure}
    \centering
    \includegraphics[width=\columnwidth]{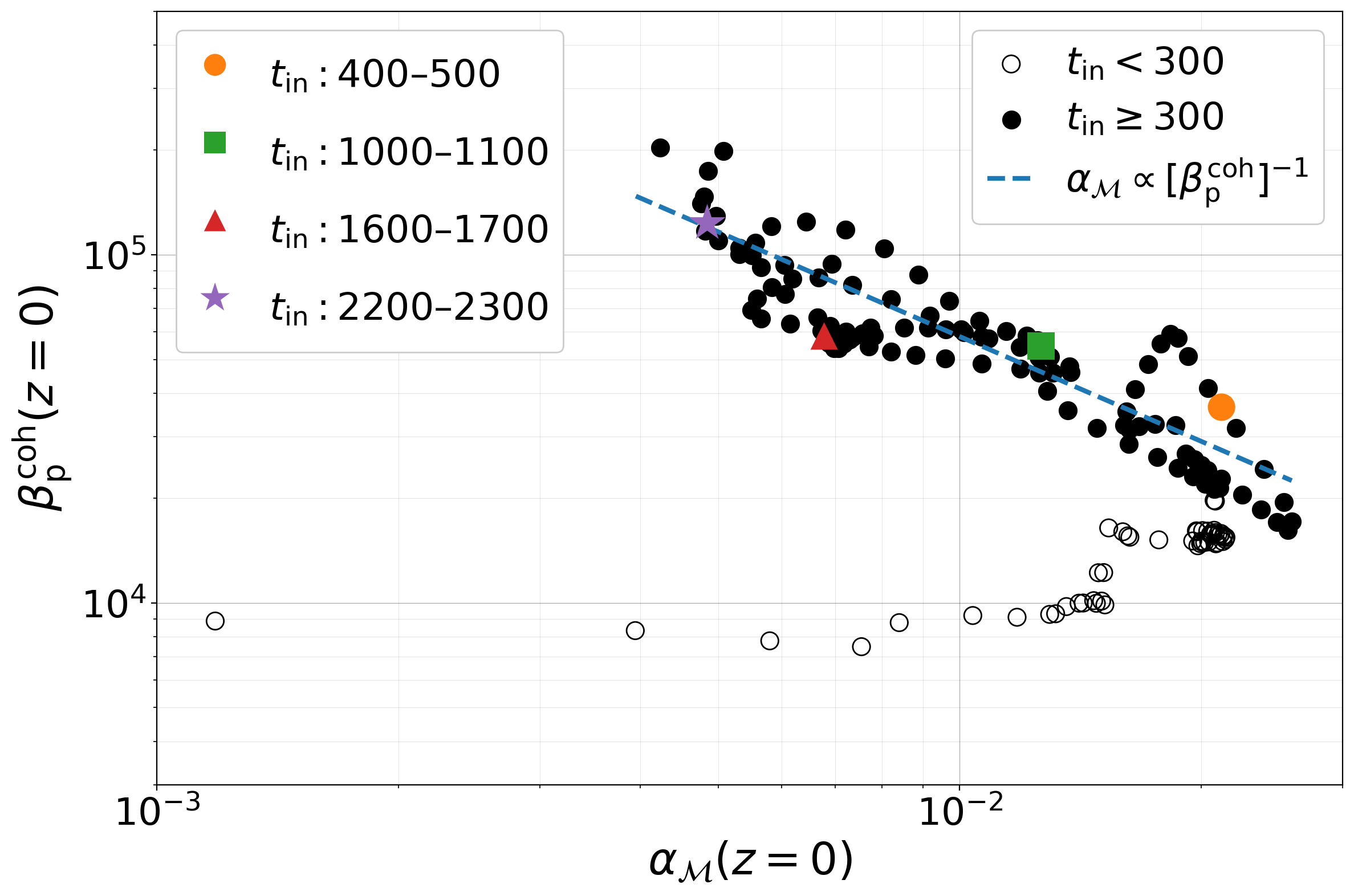}
    \caption{Correlation between the coherent poloidal plasma-$\beta$, $\beta_\text{p}^{\text{coh}}$, and the Maxwell $\alpha$ component, $\alpha_{\mathcal{M}}$, at the midplane for \texttt{NF-BAZ}, averaged over $R \!\in\![0.3, 0.5]\,\text{au}$. The data are smoothed with a Gaussian filter with mean $t_{\text{in}}=100$ and standard deviation $t_{\text{in}}=50$. This yields a (tentative) least-squares regression fit of $\alpha_{\mathcal{M}}(z=0)\propto [\beta_\text{p}^{\text{coh}}(z=0)]^{-1}$.} 
    \label{fig:alpha_betap_BAZ_evolution}
\end{figure}

\begin{figure}
    \centering
    \includegraphics[width=\columnwidth]{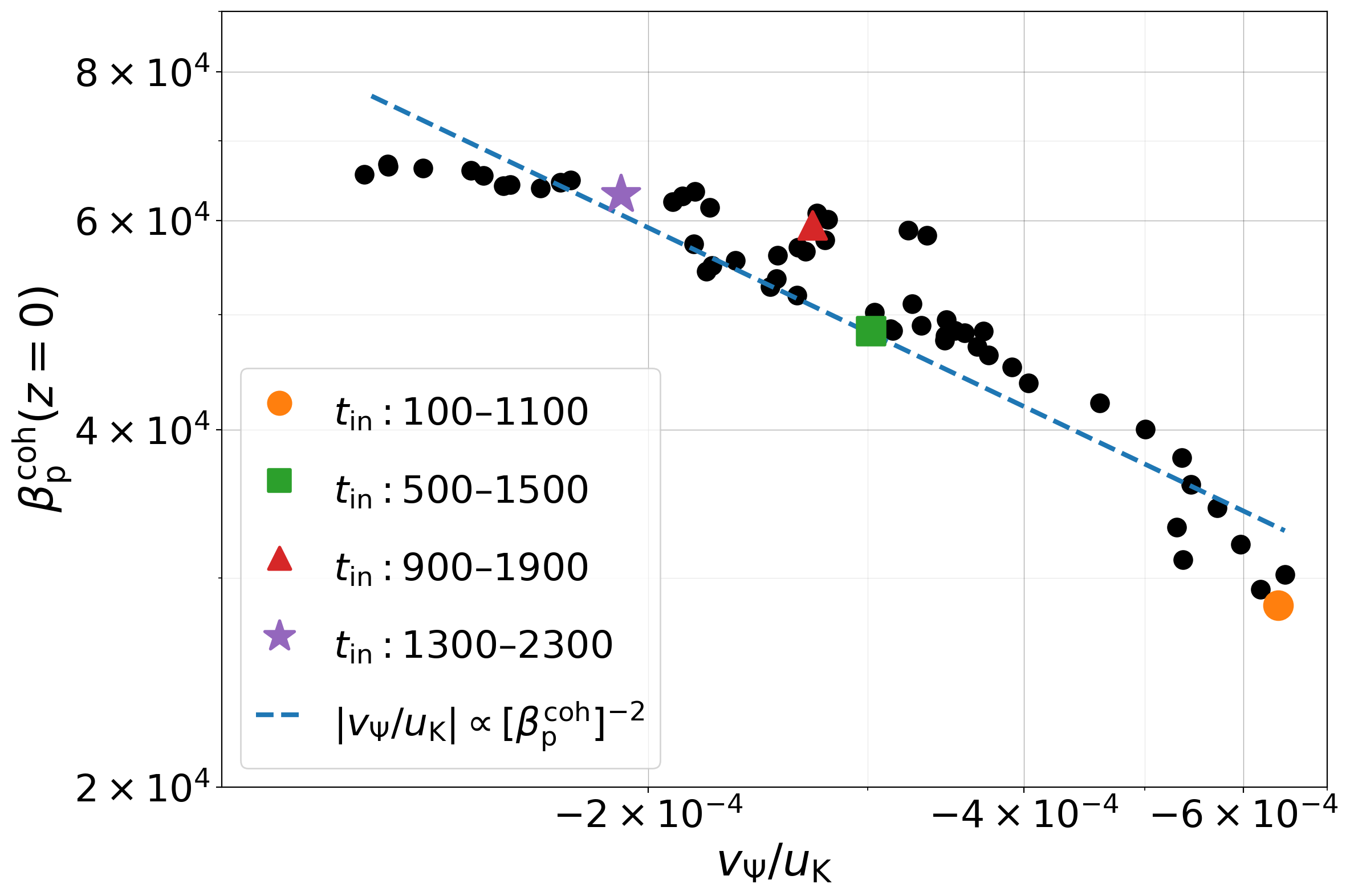}
    \caption{Correlation between the coherent poloidal plasma-$\beta$, $\beta_\text{p}^{\text{coh}}$, and the normalised flux transport velocity, $v_\Psi/u_\text{K}$,  at the midplane for \texttt{NF-BAZ}, averaged over $R \!\in\![0.3, 0.5]\,\text{au}$ and intervals of $1000t_\text{in}$. This yields a (tentative) least-squares regression fit of $|v_\Psi/u_\text{K}|(z=0)\propto [\beta_\text{p}^{\text{coh}}(z=0)]^{-2}$.} 
\label{fig:BAZ_v_psi_betap_active_zone}
\end{figure}

The gradual depletion of magnetic flux from the inner active zone as it piles up at the inner boundary, introduces a feedback loop, since the relative rates at which mass and magnetic flux are evacuated inward both depend on $\beta^\text{coh}_\text{p}$ (e.g. \citetalias{jacquemin-ide_magnetic_2021}). As shown in Fig.~\ref{fig:alpha_betap_BAZ_evolution}, this quantity gradually increases from $10^4$ to $\sim\!2\times10^5$ in the inner part of the active zone, indicating a progressive weakening in the dynamical influence of the large-scale poloidal field -- consistent with the steadily declining accretion rate in this region shown in Fig.~\ref{fig:BAZ_accretion_rate_time} -- since magnetic flux is advected more rapidly than the mass. \\
\indent Additionally, Fig.~\ref{fig:alpha_betap_BAZ_evolution} shows a scatter plot of the correlation between $\alpha_{\mathcal{M}}(z=0)$ and $\beta^\text{coh}_\text{p}(z=0)$. To smooth the data, a Gaussian-weighted average is applied with mean $t_{\text{in}}=100$, and standard deviation $t_{\text{in}}=50$. Prior to MRI saturation, $\beta_\text{p}^{\text{coh}}(z=0)$ remains constant at $10^4$ (unfilled circles), highlighting an important subtlety: although magnetic flux is transported inward during this epoch, the relative rate compared to the mass flux is key. After MRI saturation, the inverse relationship is clear (filled circles, $t_{\text{in}}\geq 300$), with a least-squares regression fit yielding $\alpha_{\mathcal{M}}(z=0) = 6\times10^{2}\, [\beta_\text{p}^{\text{coh}}(z=0)]^{-1}$. Whilst this is only at the midplane, it is steeper than the vertically integrated measurements of $\alpha_{\mathcal{M}} \propto (\beta_\text{p}^{\text{coh}})^{-0.5}$ (e.g. \citealt{salvesen_accretion_2016}; \citetalias{jacquemin-ide_magnetic_2021}). \\
\indent Meanwhile, flux evacuation throughout most of the active zone also self-regulates its own inward transport. Fig.~\ref{fig:BAZ_v_psi_betap_active_zone} shows the correlation between $\beta_\text{p}^{\text{coh}}$ and $v_\Psi/u_\text{K}$ at the midplane for \texttt{NF-BAZ}, averaged over $R\!\in\![0.3, 0.5]\,\text{au}$ and time intervals of $1000t_\text{in}$. The inward transport velocity clearly decreases as flux is evacuated -- $\beta_\text{p}^{\text{coh}}$ increases -- yielding a tentative least-squares regression fit of $|v_\Psi/u_\text{K}|(z=0)=7\times10^5\, [\beta_\text{p}^{\text{coh}}(z=0)]^{-2}$, which is steeper than the $-1$ power-law reported by \citetalias{jacquemin-ide_magnetic_2021}. The implications of these power-law scalings for turbulent, large-scale magnetic flux transport are examined using a simple analytic model in Appendix~\ref{section:turbulent_transport_magnetic_flux}. Further evidence of this scaling appears in the left half of Fig.~\ref{fig:baz_magnetic_flux_transport_plot}, where, at fixed cylindrical radius, the angle of the white contour lines relative to the vertical -- which corresponds to the flux transport speed -- decreases over time. This contrasts with the pure MRI-active model introduced in Appendix~\ref{appendix:turbulent_versus_laminar_discs}, which sustains a constant flux transport velocity at a given cylindrical radius. This difference is discussed further in Appendix~\ref{section:impact_daz_flux_transport} and is consistent with timescale expectations: the maximum radius of magnetic flux resupply to the active disc now corresponds to the outer disc radius, orders of magnitude larger than in models with a dead--active zone interface. \\
\indent Additionally, placing these results in the context of previous work, the active model of \citetalias{jacquemin-ide_magnetic_2021} attains $v_\Psi / u_\text{K} \!\sim\! -10^{-3}$, whereas both the model presented here and that of \citetalias{iwasaki_dynamics_2024} have slower, evolved transport velocities of roughly $-(3$--$6)\times 10^{-4}$ and $-5 \times 10^{-5}$, despite all starting from an initial midplane $\beta$ of $10^4$. The order-of-magnitude lower value in \citetalias{iwasaki_dynamics_2024} is consistent with their aspect ratio of  $\varepsilon\!\sim\!0.06$ at $R_{\text{DZI}}$. \\
\indent The self-regulated flux evacuation also impacts the timescale on which magnetic flux is drained from most of the active zone. With no replenishment from the dead zone, and assuming the early-time flux transport velocity is constant, the depletion timescale would be roughly $(R-R_0)/|v_\Psi|\!\sim\!3 \times 10^3 t_\text{in}$ at $R=0.5$ au, which corresponds to $\sim\!10^2$ yr. Fortunately, this rapid, unrealistic depletion timescale is moderated by several effects in the active zone: $v_\Psi$ declines and is self-regulated as flux is evacuated (e.g. Fig.~\ref{fig:BAZ_v_psi_betap_active_zone}, Appendix D); $v_\Psi$ decreases by roughly an order of magnitude when the disc aspect ratio is reduced to a more realistic inner-disc value of $\varepsilon=0.05$ (\citetalias{jacquemin-ide_magnetic_2021}); and interactions at the star--disc boundary may further slow the loss. 

\subsection{Outer part of the active zone}
\label{section:mft_outer_az}

In contrast, magnetic flux gradually concentrates at the outer edge of the active zone, as shown in Fig.~\ref{fig:baz_magnetic_flux_transport_plot}, helping to carve out a strong, axisymmetric density minimum (see Section~\ref{section:hd_morphology_evolution}). This sustained outward transport in the outer part of the active zone is independent of the local current-sheet evolution, persisting throughout \texttt{NF-BAZ}. \\
\indent The outward-directed transport is driven by the radial variation in $\langle u_R B_z \rangle_{\phi,t}$, as shown by the dashed blue line in the right panel of Fig.~\ref{fig:mean_field_paper_magnetic_flux_speed_compare}, which causes $\langle \mathcal{E}_{\text{I}_\phi} \rangle_{\phi,t}$ -- and by extension $v_\Psi$ -- to change sign. Physically, it results from sustained midplane decretion around the interface (see Section~\ref{section:results_accretion}) and from the self-feedback of a strong surface-density minimum deepened by the flux concentration, which likely restricts flux leakage into the dead zone in this epoch, in a manner akin to a zonal flow or planet-carved gap \citep[e.g.][]{riols_spontaneous_2019, wafflard-fernandez_planet-disk-wind_2023}. \\
\indent Together with the lack of flux transport from the active zone into the dead zone during this epoch,\footnote{
Although the contour lines in Fig.~\ref{fig:baz_magnetic_flux_transport_plot} indicate negligible radial flux transport just inside the dead--active zone interface for \texttt{NF-BAZ}, $v_\Psi$ does not vanish there in Fig.~\ref{fig:mean_field_paper_magnetic_flux_speed_compare}. This discrepancy suggests that the steady-state, axisymmetric framework that depends on $\mathcal{E}_{\text{I}_\phi}$ breaks down in this extremely volatile region.} this leads to a fourfold accumulation of poloidal magnetic field. At the same time, the density decreases (see Fig.~\ref{fig:SAZ_BAZ_compare_HD_density}), lowering the local $\beta_\text{p}^\text{coh}$ to $\sim\!10^{2}$, two orders of magnitude below its initial value (see Section~\ref{section:multi_zone_outflow_inheritance} for supporting evidence). This places the outermost part of the active zone at $R\!\gtrsim\!0.8\,\text{au}$ into the strong-field regime $(\beta_\text{p}^\text{coh} \lesssim 10^2)$. Consistent with \citetalias{jacquemin-ide_magnetic_2021}, this state is characterised by a midplane current sheet without MRI-driven turbulent eddies, as shown in the right panel of Fig.~\ref{fig:baz_connection_magnetic}; further evidence with a shorter time average is provided in Section~\ref{section:magnetic_flux_burps_intro}. Consequently, the active zone is divided into a weak-field interior and a strong-field exterior, providing a second mismatch of magnetic-field states in the inner disc. This promotes further variability and is also partly reflected in the outflow structure (see Section~\ref{section:results_outflows}). \\
%\indent Finally, the flux concentration may alter the local thermodynamics. The reduction in surface density and the suppression of MRI-driven turbulent eddies both lower the viscous-heating-controlled temperature in this region. Such cooling could promote the inward migration of the dead--active zone interface, as explored in models of Solar System formation by \citet{ueda_early_2021}, thereby disrupting the flux-concentration mechanism  and introducing additional inner-disc variability. 

\subsection{Dead zone}
\label{section:mft_dead_zone}

In the dead zone, magnetic flux is transported outward and does not cross into the active zone, effectively isolating the active zone from the flux reservoir in the rest of the disc and altering its long-term dynamics (see Section~\ref{section:impact_flux_draining}). \\
\indent The outward-directed magnetic flux transport velocity is constant at $v_\Psi/u_\text{K}\!\sim\! 1\times 10^{-4}$ across $R\!\in\![2, 8]\,\mathrm{au}$, in close agreement with \citet{lesur_systematic_2021}, who report $v_\Psi/u_\text{K}\!\sim\!1 \times 10^{-4}$ for an initial midplane $\beta_\text{p}=10^4$, $\Lambda_{\text{A}_0} = 1$, and a similar functional form for $\text{R}_\text{m}$.\footnote{\citet{lesur_systematic_2021} normalise $v_\Psi$ by $H\Omega_\text{K}$ instead of $u_\text{K}$, so a factor of $\varepsilon$ is required to make a direct comparison.} However, as outlined in equation~\eqref{eqn:decompose_ephi}, $v_\Psi$ is highly sensitive to small changes in the non-ideal MHD diffusivities, which themselves can vary by multiple orders of magnitude depending on the adopted grain and metal chemistry assumptions \citep[e.g.][]{lesur_magnetohydrodynamics_2021}. Consequently, other global VNF simulations -- not necessarily modelling the inner disc -- report a wide range of values for $v_\Psi / u_\text{K}$ despite all assuming an initial midplane $\beta_\text{p} = 10^4$. For example, \citet{gressel_global_2020} and \citetalias{iwasaki_dynamics_2024} report faster speeds, $v_\Psi/u_\text{K} \!\sim\! 10^{-2}$ and $\sim\!10^{-3}$ respectively, consistent with $\Lambda_{\text{A}_0}$ values up to six orders of magnitude lower than those in our model; in contrast, \citet{bai_hall_2017} report a smaller value of $v_\Psi/u_\text{K}\!\sim\!3\times 10^{-5}$ consistent with $\Lambda_{\text{A}_0}$ half of ours and no Ohmic diffusion. Meanwhile, the Hall effect, which is dynamically relevant in the inner dead zone, can reverse the direction of magnetic flux transport \citep[e.g.][]{bai_hall_2017} and is a caveat that remains to be tested for this global problem. \\
\indent Nevertheless, the outward flux transport is set by a positive non-ideal toroidal electromotive field, $\langle \mathcal{E}_{\text{NI}_\phi} \rangle_{\phi,t}(z=0)$, which exceeds its ideal counterpart, $\langle \mathcal{E}_{\text{I}_\phi} \rangle_{\phi,t}(z=0)$, by just $1$ per cent in agreement with \citet{lesur_systematic_2021}. Physically, $\langle \partial_z {B_R}\rangle_{\phi,t}$ dominates, so diffusion of the hourglass-shaped poloidal field (see right panel of Fig.~\ref{fig:baz_wide_scale_early_time}) outcompetes the inward transport, primarily driven by the radial advection of vertical field, $\langle u_RB_z \rangle_{\phi,t}$. \\
\indent Additionally, zonal fields only develop weakly in the dead zone compared to other studies \citep[e.g.][]{riols_spontaneous_2019}, due to strong Ohmic resistivity \citep{suriano_formation_2018} and the relatively immature dead zone, which completes only 10 local orbits at $R = 4\,\text{au}$.

\subsection{Long-term magnetic flux variability in \texttt{NF-SAZ}}
\label{section:mft_burps}
Finally, we examine flux transport in \texttt{NF-SAZ}, which behaves similarly to \texttt{NF-BAZ}, except in the innermost part of the active zone, and exhibits additional interface-localised flux variability. 
Recall from Table~\ref{table:list_net_flux_global_simulations} that \texttt{NF-SAZ} evolves for roughly three times longer at the dead--active zone interface than \texttt{NF-BAZ}.

\subsubsection{Comparison of flux transport in \texttt{NF-SAZ} and \texttt{NF-BAZ}}
\begin{figure}
    \centering
    \includegraphics[width=\columnwidth]{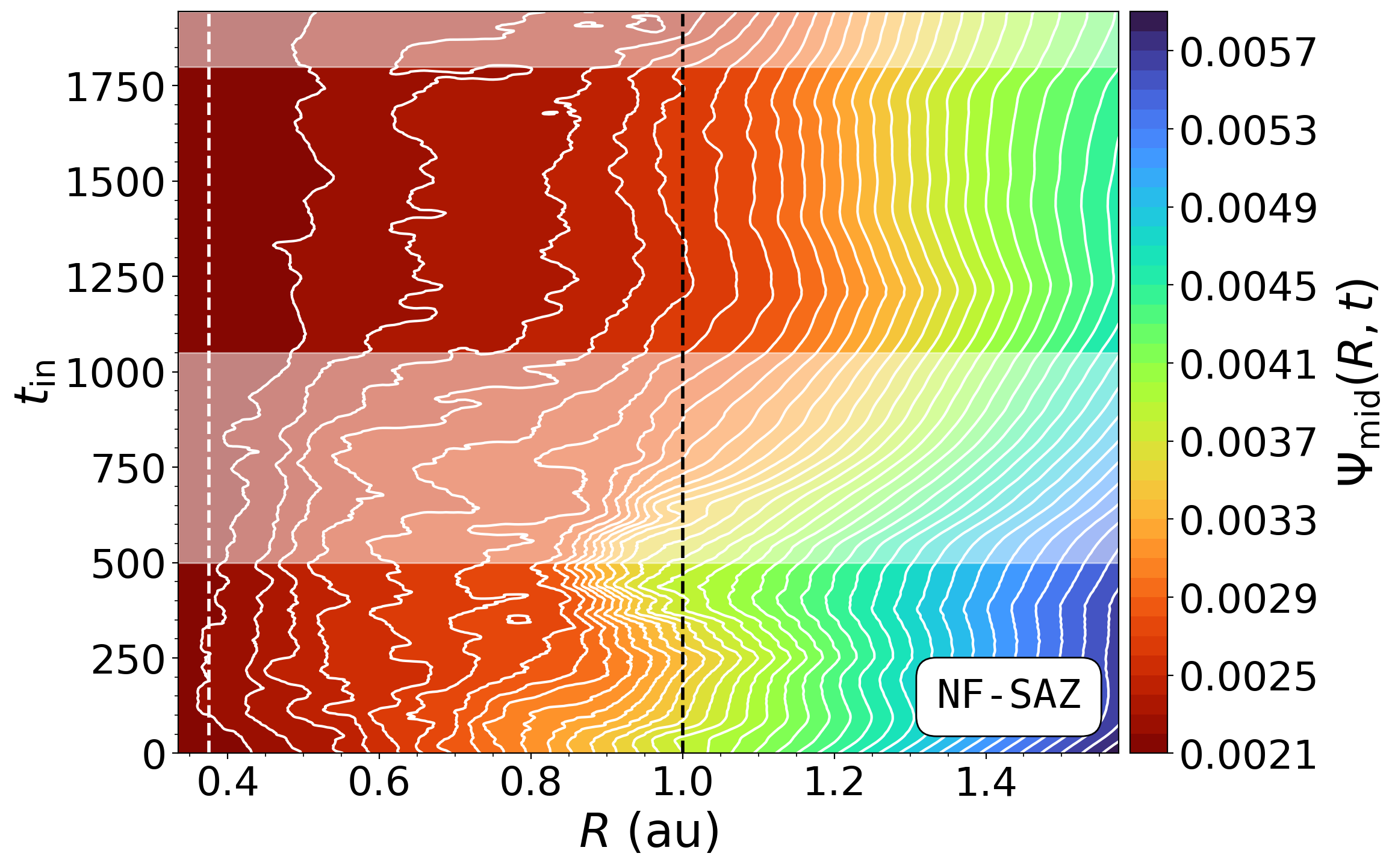}
    \caption{Evolution of the magnetic flux threading the midplane, $\Psi_{\text{mid}}(R,t)$, as defined in equation \eqref{eq:define_magnetic_flux_mid}, for \texttt{NF-SAZ}. Strong variability at the dead--active zone interface (dashed black line) is marked by two periods (white-shaded regions) where magnetic field lines (white contours) pass from the active to the dead zone. The corresponding plot for \texttt{NF-BAZ}, which only evolves to $t_{\text{in}}=658$ at the interface in these units, is shown in Fig.~\ref{fig:baz_magnetic_flux_transport_plot}. The dashed white line denotes the edge of the inner radial buffer.}
	\label{fig:SAZ_magnetic_flux_transport_variability}
\end{figure}
\begin{figure}
    \centering
    \includegraphics[width=\columnwidth]{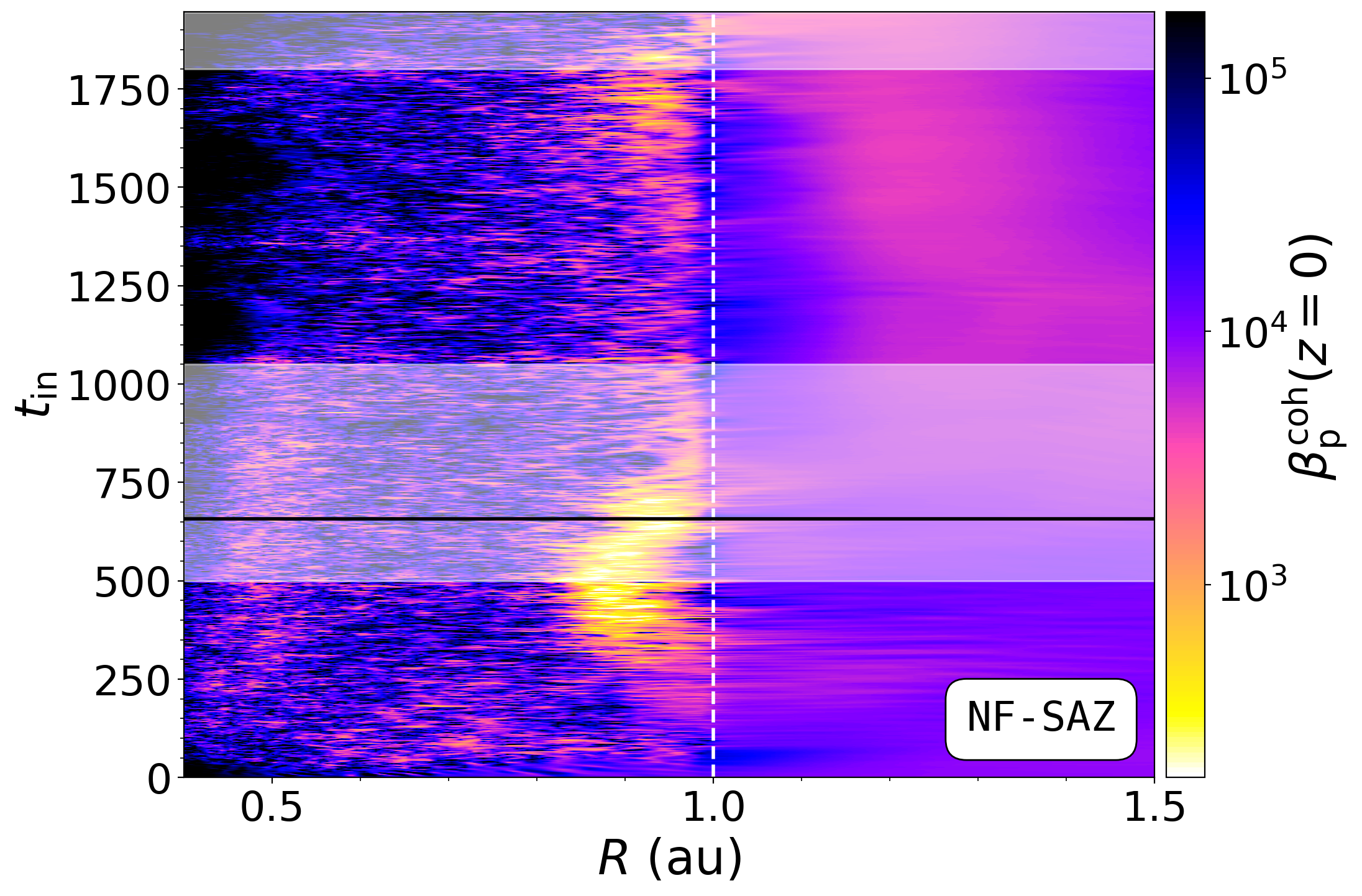} 
    \caption{Space--time ($R,t$) diagram of the coherent poloidal component of the plasma-$\beta$ parameter, $\beta_\text{p}^{\text{coh}}$, defined in equation~\eqref{eq:beta_poloidal_coherent}, for \texttt{NF-SAZ}. Magnetic flux accumulates in the outermost part of the active zone and dissipates rapidly in two distinct epochs (white-shaded regions), as shown in Fig.~\ref{fig:SAZ_magnetic_flux_transport_variability}. The end run-time of \texttt{NF-BAZ} at the interface is indicated by the horizontal black line.}
\label{fig:SAZ_magnetic_flux_periodic_evidence}
\end{figure}
Fig.~\ref{fig:SAZ_magnetic_flux_transport_variability} shows the evolution of magnetic flux threading the midplane, $\Psi_{\text{mid}}(R,t)$, to be compared with the corresponding plot for \texttt{NF-BAZ} in Fig.~\ref{fig:baz_magnetic_flux_transport_plot}. Again, no magnetic flux passes from the dead zone (beyond $R=1.1\,\text{au}$) into the active zone. Nevertheless, there are two notable differences. \\
\indent First, inward flux transport ceases beyond $t_{\text{in}}=200$ in the inner part of the active zone. This likely stems from the restricted radial extent of the active region (see Fig.~\ref{fig:BAZ_SAZ_grid_comparison}), indicating an implicit dependence on the inner BC that is confirmed by the sustained inward flux transport in \texttt{NF-SAZ-BC}, which applies a different BC (see Appendix~\ref{appendix:boundary_test}). Second, \texttt{NF-SAZ} has two episodes of strong flux transport -- beyond the initial relaxation -- from the outer part of the active zone into the dead zone, leading to the rapid expulsion of concentrated flux that we term a \textit{flux burp}. The first commences at $t_\text{in}=500$, which is close to the end-runtime of \texttt{NF-BAZ} at the interface, whilst the second (weaker) burp starts at approximately $t_\text{in}=1800$; both are highlighted by the white-shaded regions in Fig.~\ref{fig:SAZ_magnetic_flux_transport_variability}. 
Neither event originates explicitly from the inner boundary, since the flux threading the inner shell monotonically increases, $\partial_t[\Psi_{\text{mid}} (R_0)]>0$ (not shown), in contrast to fig.~24 of \citetalias{iwasaki_dynamics_2024}.\\
\indent Finally, after each burp there is no subsequent trident-like current-sheet configuration at the interface (see \texttt{SAZ\_bphi.mp4}), confirming that the re-accumulation of flux is independent of the process examined in Section~\ref{section:currents_connection_mechanism}. \\
\begin{figure*}
    \centering
    \includegraphics[width=\textwidth]{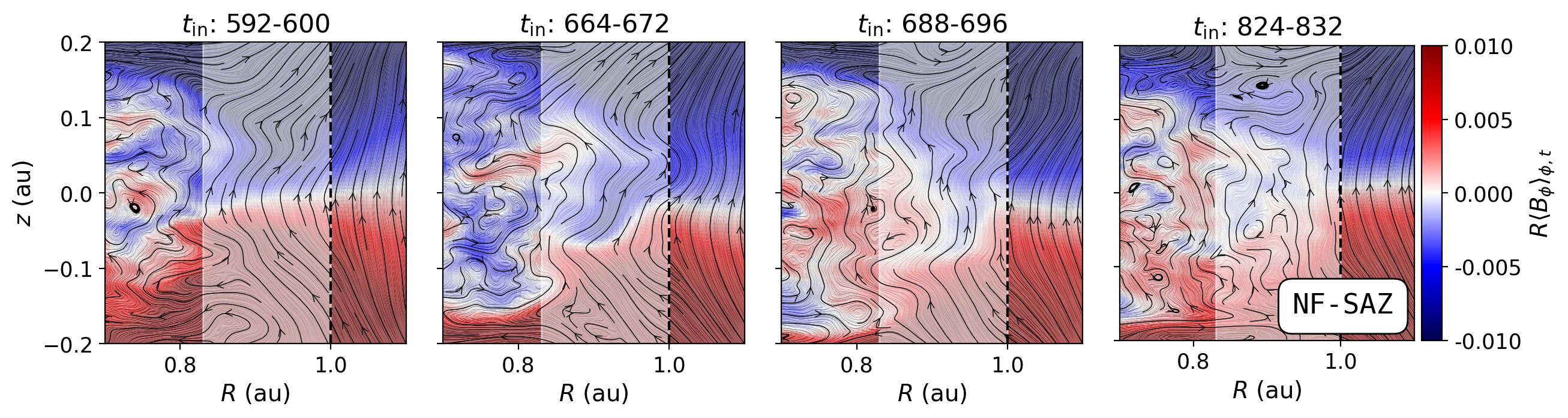}
    \caption{Meridional ($R,z$) plots of the poloidal magnetic field $\langle \mathbf{B_p} \rangle _{\phi,t}$ (black lines and LIC), overlaid on 
    $R \langle B_\phi \rangle _{\phi,t}$, during the first flux burp in \texttt{NF-SAZ}. These are azimuthally, and time-averaged over one orbit at the interface. The white-shaded regions highlight: the breaking of midplane symmetry in the strong-field state (left panels) and the re-emergence of MRI-driven turbulent eddies as the region transitions to the weak-field state (right panels).}
	\label{fig:SAZ_bphi_around_first_burp_epoch}
\end{figure*}
\subsubsection{Magnetic flux burps}
\label{section:magnetic_flux_burps_intro}

 The evidence for a second flux burp in Fig.~\ref{fig:SAZ_magnetic_flux_transport_variability} is relatively weak due to the scaling of the contours, as flux is depleted from the active zone. Therefore, to confirm the existence of two burps in \texttt{NF-SAZ}, Fig.~\ref{fig:SAZ_magnetic_flux_periodic_evidence} shows the space--time ($R,t$) diagram of $\beta_\text{p}^{\text{coh}}(z=0)$, which increases from approximately $2 \times 10^{2}$ to $5 \times 10^{3}$ during the first; and $8 \times 10^{2}$ to $3 \times 10^{4}$ during the second. These burps reduce the local midplane Maxwell stress from $\alpha_{\mathcal{M}} \!\sim \!10^{-1}$ -- consistent with strong-field MRI simulations \citep{salvesen_accretion_2016, mishra_strongly_2020} -- by over an order of magnitude, driving strong local stress variations around the interface that last 40--50$t_\text{DZI}$ with a recurrence timescale of $\sim\!150$$t_\text{DZI}$. Therefore, alongside the evolution of current sheets across the interface, flux burps might plausibly be connected to the ubiquitous accretion variability observed in the inner disc \citep[e.g.][]{cleaver_magnetically-activated_2023, fischer_accretion_2023}.\\
\indent Focusing on the first burp, Fig.~\ref{fig:SAZ_bphi_around_first_burp_epoch} shows four meridional ($R,z$) plots of the poloidal magnetic field $\langle \mathbf{B_p} \rangle _{\phi,t}$, overlaid on its normalised toroidal component $R \langle B_\phi \rangle _{\phi,t}$ and averaged over $1t_{\text{DZI}}$. The far-left panel provides additional evidence for a strong-field state $(\beta_\text{p}^{\text{coh}}\!\lesssim \!10^2)$ in the outermost part of the active zone during the flux-concentrated epoch -- comparable to fig. 18 in \citetalias{jacquemin-ide_magnetic_2021} -- where enhanced magnetic compression produces an hourglass-shaped poloidal field, unlike the more complex morphology driven by the surface-layer accretion in the weak-field state (e.g. left panel of Fig.~\ref{fig:compare_magnetic_LIC_laminar_turbulent}). \\
\indent Over the course of the first burp, two consecutive features are visible in Fig.~\ref{fig:SAZ_bphi_around_first_burp_epoch}. First, the breaking of midplane symmetry, whereby the midplane-located current sheet, characteristic of the strong-field state, moves towards the lower disc surface (left panels). Second, the re-emergence of MRI-driven turbulent eddies in the outer part of the active zone, accompanied by a turbulent poloidal field configuration (right panels), as the local region reverts to a weak-field state. Whilst both features are also evident during the second flux burp, the direction of causality between them cannot easily be determined.\\
\indent We tentatively interpret flux burps as a generic feature of the inner disc because a double mismatch in magnetic-field states over a narrow radial extent naturally engenders pronounced variability. Indefinite flux accumulation at the interface we regard as unlikely, and indeed the inner-disc model of \citetalias{iwasaki_dynamics_2024} exhibits a `flux burp' at $t_{\text{DZI}}\approx90$ (see their fig. 19), albeit not originating from a flux-concentrated configuration. With the inner-boundary caveat in mind, a plausible explanation for why flux burps only occur in \texttt{NF-SAZ}, and not in \texttt{NF-BAZ} or \texttt{NF-SAZ-BC}, is the shorter evolution time at the interface in these models, which may prevent a sufficiently strong mismatch in magnetic-field states from developing. 

\subsubsection{Mechanism of magnetic flux burps}

% Comment on speculate + different to I24
We hypothesise that flux burps arise from difficulty in maintaining the current across the weak- and strong-field MRI-active regions -- comparable, but \emph{distinct}, to the interface-localised current-sheet evolution examined in Section~\ref{section:nf_flux_current_sheet}, especially as the strong-field state continually strengthens. \\
% Comment on mechanism directly
\indent In the strong-field state, the midplane symmetry breaks (see Fig.~\ref{fig:SAZ_bphi_around_first_burp_epoch}), altering the local balance of magnetic flux transport. This asymmetry likely occurs because it becomes increasingly difficult to maintain a constant poloidal current flux between the weak- and strong-field MRI-active regimes as the latter strengthens, breaking the current-sheet configuration and producing a downward-displaced sheet. In the pre-burp strong-field state, the sheet provides an inward-directed contribution to the mass flow, thereby supporting magnetic flux transport in a similar manner. When the downward-displaced configuration sets in, this inward contribution diminishes, triggering rapid outward transport of large-scale poloidal flux into the dead zone: the flux burp. The outermost part of the active zone then reverts to a weak-field state, after which flux accumulation recommences, gradually returning it to the strong-field state. This cycle naturally produces intermittent flux burps in the outermost part of the active zone.

%%%%%%%%%%%%%%%%%%%%%%%%%%%%%%%%%%%%%%%%%%%%%%%%%%%%%%%%%%%%%%%%%%%%%%%%%%%%%%%%%%%%%%%%%%%%%%%%%%%%

\section{Outflows}
\label{section:results_outflows}

The active-zone outflow evolves as magnetic flux is gradually depleted and field lines are dragged inward in its surface layers. Consequently, its properties are unsteady, so to broadly characterise them we adopt the late-time average used previously for \texttt{NF-BAZ}, $t_{\text{in}}\!\in\![2200,2400]$. The analysis is also restricted to the upper half-plane due to the time-averaged outflow symmetry about the midplane. 

\subsection{Multi-zone outflow structure}
\label{section:multi_zone_outflow_inheritance}
\begin{figure}
    \centering
    \includegraphics[width=\columnwidth]{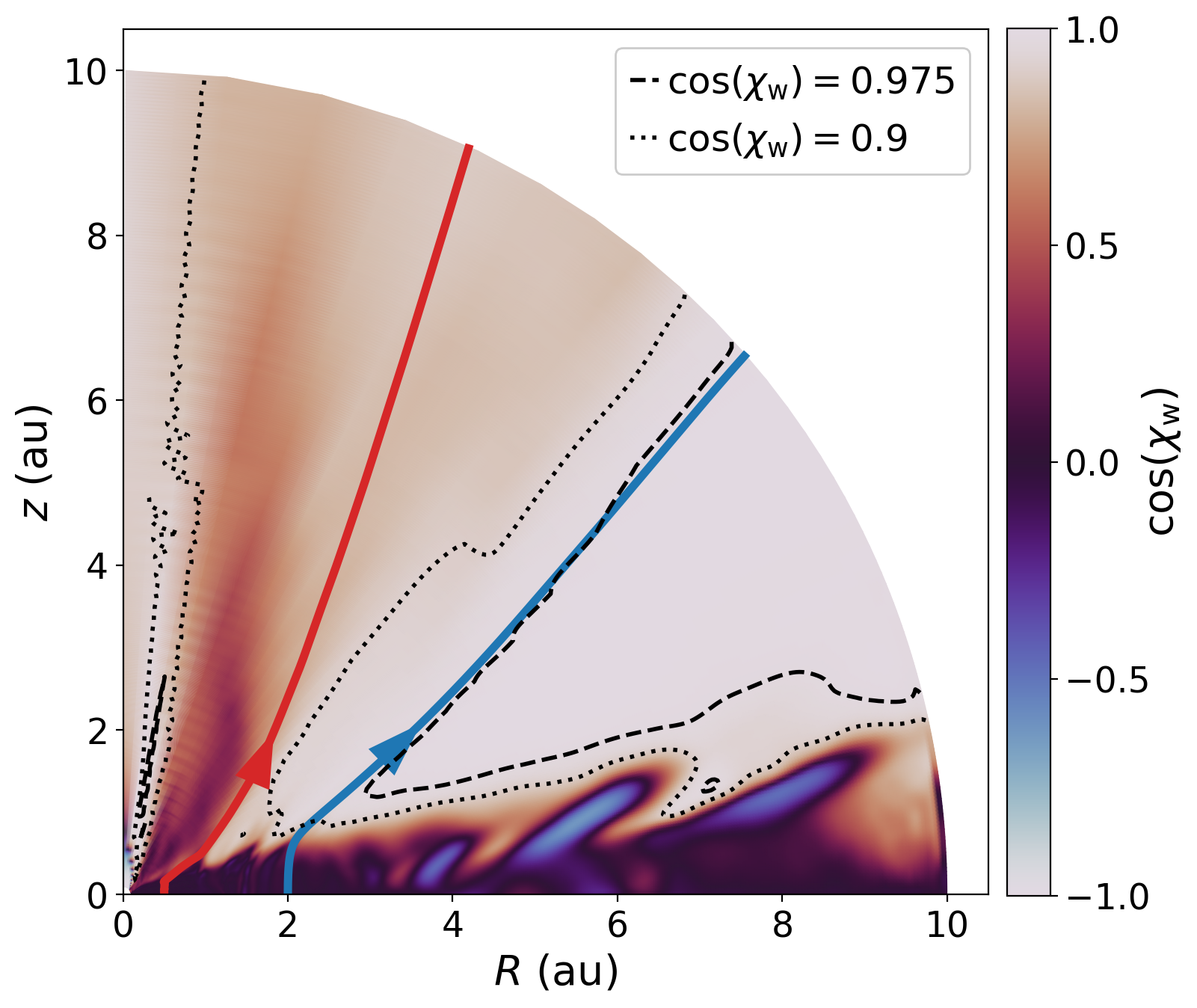} \\
    \includegraphics[width=\columnwidth]{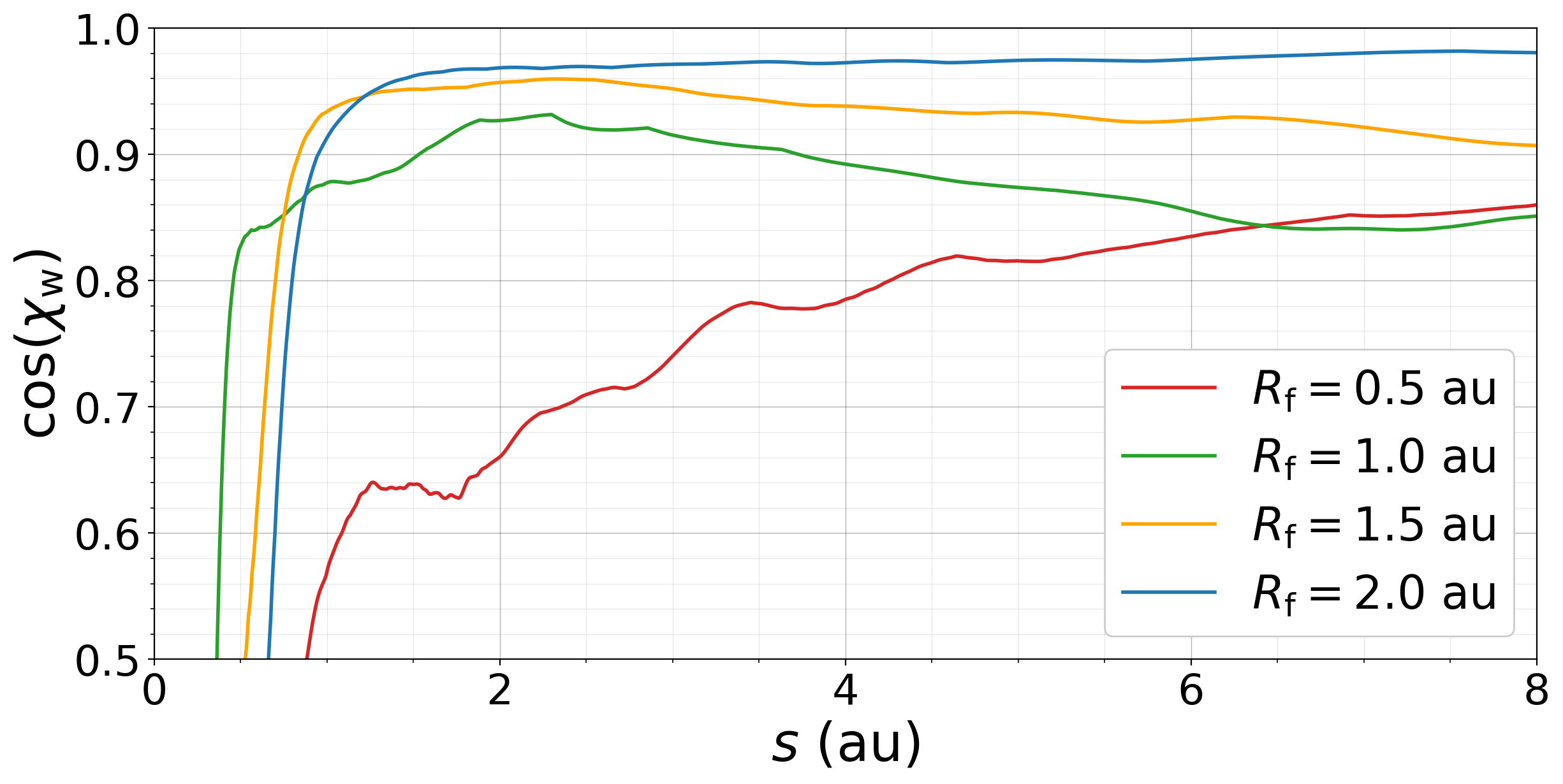}
    \caption{Top: Meridional ($R,z)$ plot of the upper half-plane, averaged in azimuth and over the interval $t_\text{in}\!\in\![2200,2400]$ for \texttt{NF-BAZ}, showing the angle between the poloidal velocity and magnetic fields, $\cos(\chi_\text{w})$ as defined in equation~\eqref{eqn:cos_chi_definition}, and magnetic field lines with midplane foot points in the active zone at $R_{\text{f}} = 0.5\,\text{au}$ (red line) and in the dead zone at $R_{\text{f}} = 2.0\,\text{au}$ (blue line). The dotted and dashed black lines delineate the regions where $\cos(\chi_\text{w})\!\geq\!0.9$ and $\cos(\chi_\text{w})\!\geq\!0.975$ respectively. Bottom: Evolution of $\cos(\chi_\text{w})$ along four field lines, where $s$ is the distance along the field line.}
\label{fig:BAZ_magnetic_field_lines_cos_chi}
\end{figure}

\begin{figure}
    \centering
    \includegraphics[width=\columnwidth]{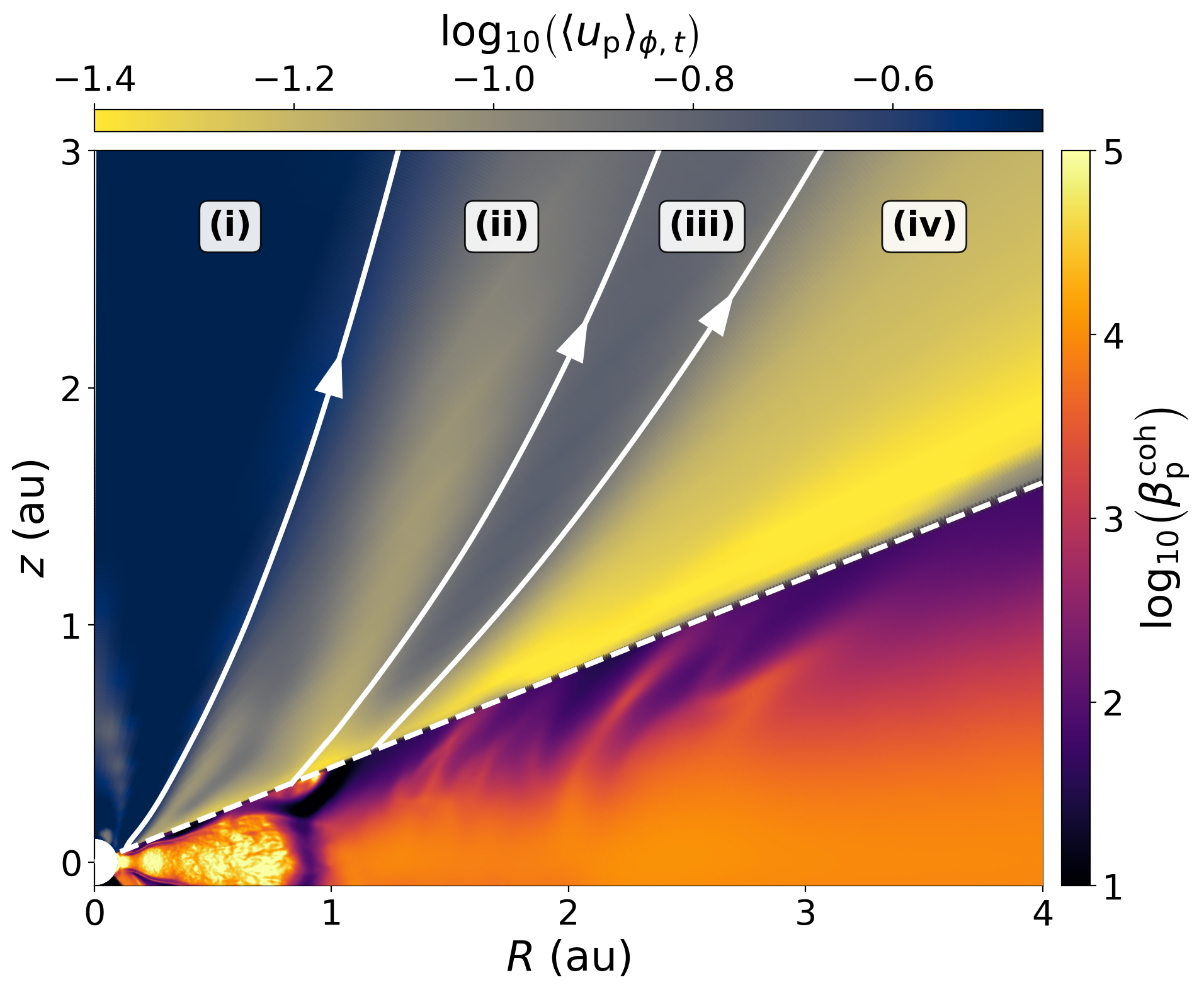}
    \caption{Meridional $(R,z)$ plot, averaged in azimuth and over the interval $t_\text{in}\!\in\![2220,2400]$ for \texttt{NF-BAZ}, showing that the local magnetic-field state within the disc, represented by $\log_\text{10}(\beta_\text{p}^\text{coh})$, affects the poloidal velocity, $\log_\text{10}(\langle u_\text{p}\rangle_{\phi,t})$, within the corona (shown above the dashed white line at $z=4H$). This gives rise to distinct outflow zones that are delineated by three poloidal fluid streamlines, $\langle\mathbf{u_p}\rangle_{\phi,t}$ (white lines). Moving outward in cylindrical radius these states are: (i) $\beta_\text{p}^\text{coh}<1$ (not shown) due to flux advection onto the inner core, (ii) $\beta_\text{p}^\text{coh}\gtrsim10^5$ in the flux depleted part of the active zone, (iii) $\beta_\text{p}^\text{coh}\!\sim\!10^2$ near the interface, and (iv) $\beta_\text{p}^\text{coh}\!\sim\!10^4$ in the laminar dead zone. The poloidal speeds can be converted into physical units of km s$^{-1}$ by multiplying by 94.2.}
    \label{fig:u_total_betap_coh_22002400}
\end{figure}

The outflow operates across all disc radii, and exhibits a time-variable morphology that reflects the radially varying conditions -- turbulence level, magnetic-field strength and density -- at its launch point. \\
\indent The turbulent inheritance of the outflow, previously shown in the right panel of Fig.~\ref{fig:baz_wide_scale_early_time}, can be quantified by the angle between the averaged poloidal velocity and magnetic fields, $\cos(\chi_\text{w})$, defined in equation~\eqref{eqn:cos_chi_definition}. The top panel of Fig.~\ref{fig:BAZ_magnetic_field_lines_cos_chi} shows this, demonstrating that the active-zone-launched outflow remains turbulent and unsteady out to the outer boundary ($r=10\,\text{au}$), with $\cos(\chi_\text{w}) \neq \pm 1$. The adjacent outflow launched in the dead-zone also deviates from its normal laminar form, as shown by the misalignment between the magnetic field and the fluid streamlines, which would be aligned in a purely laminar outflow. This effect is illustrated in the bottom panel of Fig.~\ref{fig:BAZ_magnetic_field_lines_cos_chi}, which shows that for a field line with a foot point at $R_{\text{f}} = 1.5\,\text{au}$ (orange), $\cos({\chi_\text{w}})$ steadily decreases as the line extends through the corona. In contrast, a field line anchored farther from the turbulent region, at $R_{\text{f}} = 2.0\,\text{au}$ (blue), maintains $\cos({\chi_\text{w}})\!\approx\!0.975$, with the deviation from unity likely due to numerical diffusion. This result suggests that the boundary between the laminar and turbulent wind may be unstable, or that it permits turbulent fluctuations to impinge upon the laminar region. \\
\indent The magnetic field strength at the launch point further partitions the outflow structure into zones with different poloidal speeds. Fig.~\ref{fig:u_total_betap_coh_22002400} shows the averaged poloidal velocity, $\langle u_\text{p} \rangle_{\phi,t}$, in the corona, and the coherent poloidal component of $\beta$ in the disc. This exhibits a four-zone outflow-velocity structure that connects with distinct magnetic-field states in the disc, and which is retained as the outflow streams through the corona. Moving outward in cylindrical radius from the rotation axis, these states split the active-zone outflow into three -- (i) a high-velocity (numerical) jet from the magnetic-flux advected onto the inner core $(\beta_\text{p}^\text{coh}\!<1\!)$; (ii) a low-velocity turbulent outflow from the flux depleted part of the active zone $(\beta_\text{p}^\text{coh}\!\gtrsim\!10^5)$; and (iii) a medium-velocity turbulent outflow from the strong-field, concentrated state near the interface $(\beta_\text{p}^\text{coh}\!\sim\!10^2)$. Outflow (iv) corresponds to the much slower, laminar wind launched from the dead zone. The shear between these adjacent wind zones likely seeds Kelvin--Helmholtz instabilities, providing a secondary source of turbulence that can intrude into the otherwise laminar region, consistent with the observed impingement.\footnote{The poloidal velocity configuration shown in the left panel of Fig.~\ref{fig:baz_wide_scale_early_time} is time-averaged and shown at an early time, and therefore does not reveal the turbulent velocity field above the active zone.} Therefore, the local magnetic-field state provides a second channel through which the outflow inherits its structure from its foot point. An analogous mechanism is also likely to operate in other radially varying magnetic flux concentrations, such as zonal fields or  planet-carved gaps \citep[e.g.][]{wafflard-fernandez_planet-disk-wind_2023, hsu_rossby_2024}. \\
\indent Finally, the density maximum just outside the interface leaves a signature that persists along fluid streamlines originating from that region, all the way to the outer boundary. In theory, it also should be feasible to detect the vortex-associated density asymmetry \citep{petrov_density_2023}; however, this remains unattainable in our model due to the restricted azimuthal domain and the short lifetime of the vortex, which decays due to numerical diffusion \citepalias{roberts_global_2025}.

\subsection{Conserved quantities}
\label{section:vnf_outflows_cons}

\begin{figure}
    \centering
    \includegraphics[width=\columnwidth]{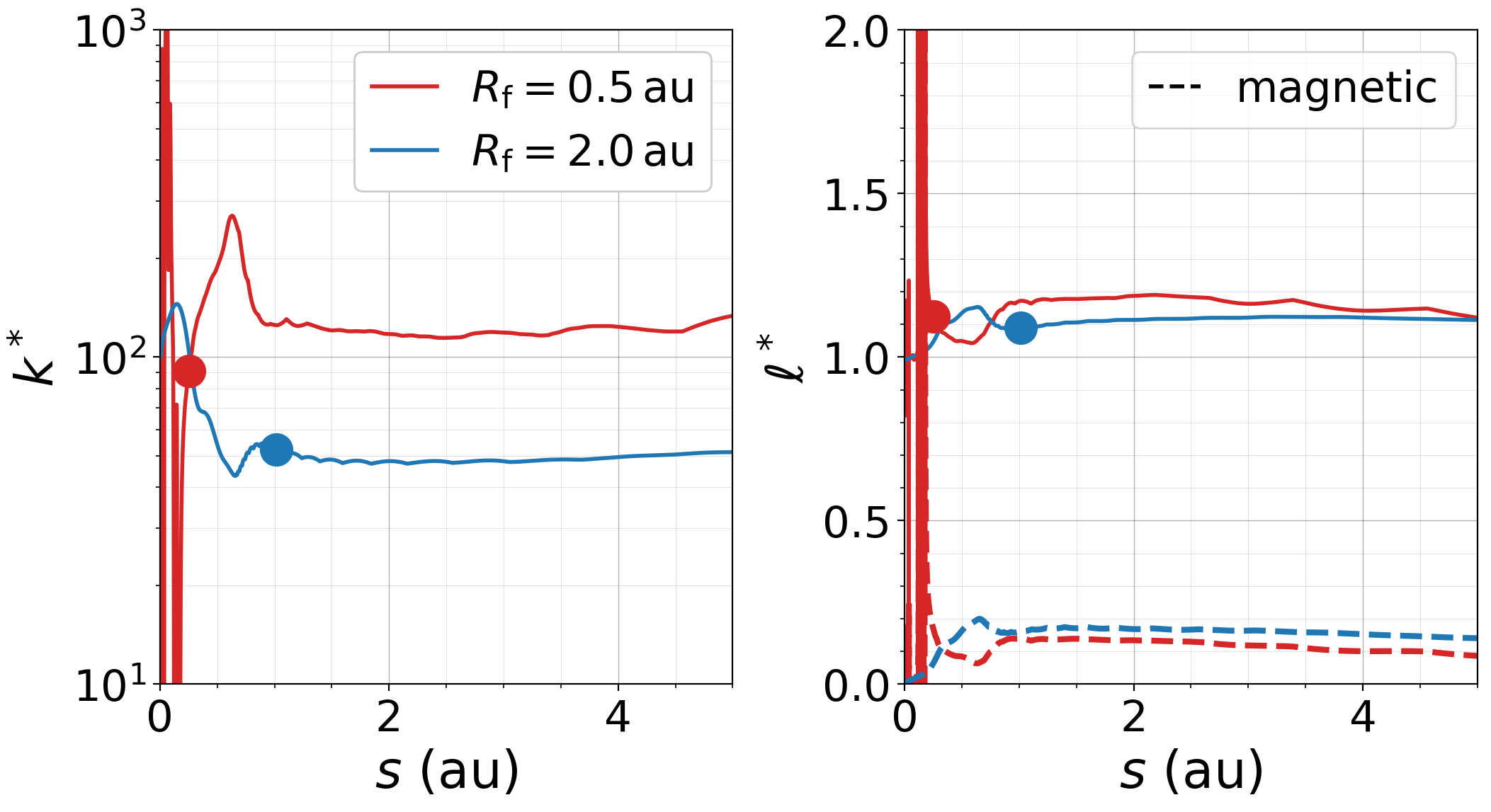}
    \caption{Normalised mass loading $k^*$ (left) and angular momentum $\ell^*$ (right) parameters for field lines anchored at $R_{\text{f}} = 0.5\,\text{au}$ (red) and $2.0\,\text{au}$ (blue), shown in Fig.~\ref{fig:BAZ_magnetic_field_lines_cos_chi}, for \texttt{NF-BAZ}. Dashed lines in the right panel denote the magnetic component of $\ell^*$, outlined in equation~\eqref{eqn:ell_normalised}, dots denote the disc surface at $|z|=4H$, and $s$ is the distance along the field line.} 
\label{fig:BAZ_outflow_mass_loading_magnetic_arm_compare}
\end{figure}

\begin{figure}
    \centering
    \includegraphics[width=\columnwidth]{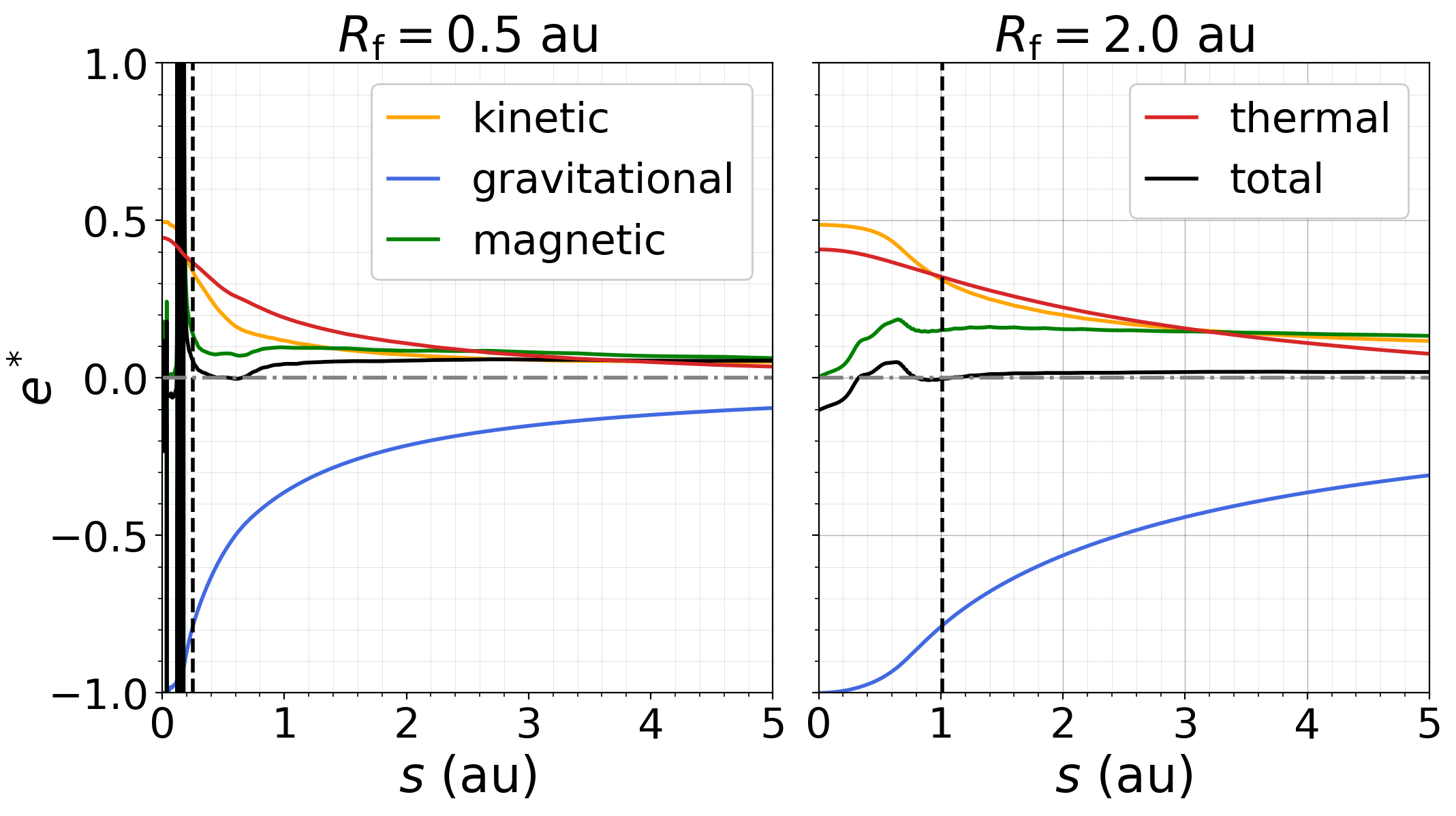}
    \caption{Decomposition of the normalised energy $e^*$, defined in equation~\eqref{eqn:bernoulli_function_normalised}, for the field lines shown in the top panel of Fig.~\ref{fig:BAZ_magnetic_field_lines_cos_chi} for \texttt{NF-BAZ}. The thermal component (red) is non-negligible due to the hot corona, and the asymptotic values $e^*>0$ show that both outflows reach $r\to \infty$. The dashed black line marks the disc surface, and $s$ is the distance along the field line.}
\label{fig:BAZ_outflow_energy_comparison}
\end{figure}

\indent Although the poloidal magnetic field and fluid streamlines are not aligned when issuing from the active zone, the outflow can nevertheless be broadly characterised by the axisymmetric, steady-state, conserved quantities defined in Section \ref{section:diagnostics_outflows_current_sheets}. \\
\indent The normalised mass loading parameter $k^*$, angular momentum extraction parameter $\ell^*$, and total energy content $e^*$, are computed along the field lines shown in the top panel of Fig.~\ref{fig:BAZ_magnetic_field_lines_cos_chi}. The profiles for the active-zone field line ($R_\text{f} = 0.5$ au) and dead-zone field line ($R_\text{f} = 2.0$ au) are shown in Figs.~\ref{fig:BAZ_outflow_mass_loading_magnetic_arm_compare} and \ref{fig:BAZ_outflow_energy_comparison}. These parameters are conserved along both field lines once sufficiently above the disc surface, with asymptotic values (for this epoch) of ($k^*$, $\ell^*$, $e^*$) = $(115, 1.16, 0.08)$ for the active-zone field line and $(47, 1.12, 0.03)$ for the dead-zone field line. Crucially, $e^* > 0$ across all disc radii (not shown), establishing that the magnetothermal outflows always escape to $r \to \infty$ -- even in the inner dead zone -- in contrast to the model of \citetalias{iwasaki_dynamics_2024}. \\
\indent In both regions, $\ell^*$ is significantly below the classical magnetocentrifugal-wind threshold ($\ell^* > 1.5$; \citealt{blandford_hydromagnetic_1982}), since the high coronal temperature \citep[e.g.][]{aresu_x-ray_2011}, which is set to $T_\text{c}=16T_\text{d}$ in our model, provides a strong thermal component to the outflow (see red lines in Fig.~\ref{fig:BAZ_outflow_energy_comparison}, compared with fig. 8 in \citealt{lesur_systematic_2021}). This is indicative of a weak magnetothermal wind, consistent with \citet{bai_global_2017} who report $\ell^* = 1.15$. Comparable unheated-corona MRI-active and dead models with an initial midplane $\beta$ of $10^4$ give ($k^*$, $\ell^*$) = $(0.2, 5)$ (\citetalias{jacquemin-ide_magnetic_2021}) and $(12, 1.75)$ \citep{lesur_systematic_2021} respectively.\footnote{\citetalias{jacquemin-ide_magnetic_2021} normalise by field-line values at the slow magnetoacoustic point instead of the midplane, which has a negligible impact on the values.} This confirms that a heated corona limits both outflows' angular momentum extraction whilst increasing the mass loading, thereby affecting large-scale HD structure formation at the interface by altering the local torque mismatch (see Section~\ref{section:results_accretion}). \\
\indent Although this strong thermal component increases the wind opening angle -- consistent with the low angular momentum extraction efficiency shown in the right panel of Fig.~\ref{fig:BAZ_outflow_mass_loading_magnetic_arm_compare} -- the outflows subsequently collimate towards the axis via a magnetic hoop stress. That said, the underlying mechanism of the purely conical outflows observed from the inner disc ($0.7$--$3.4\,\text{au}$) of DG Tau B, which extend to $z = 1200\,\text{au}$ \citep{de_valon_alma_2020} above the disc surface, remains unclear, since these outflows occur on scales much larger than those modelled here.
%%%%%%%%%%%%%%%%%%%%%%%%%%%%%%%%%%%%%%%%%%%%%%%%%%%%%%%%%%%%%%%%%%%%%%%%%%%%%%%%%%%%%%%%%%%%%%%%%%%%

\section{Comparison with Iwasaki et al.\ (2024)}
\label{section:discussion}

% 1. Introduction. Only comparable point in parameter space + research at the same time 
Amidst the vast global parameter space, \citetalias{iwasaki_dynamics_2024} is the only work directly comparable with our simulations, which include a VNF and both active and dead zones. Since the investigations were carried out independently but concurrently, their results offer a valuable opportunity for testing the robustness of both studies’ findings. \\
% 2 Results similarities (all four main ones)
\indent There is agreement with four key conclusions from our study. First, despite a different inner-disc model, the dead--active zone interface is still a one-way barrier to inward transport of large-scale magnetic flux from the dead zone into the active zone (in the absence of the Hall effect), leading to gradual flux depletion throughout most of the active zone. Second, there is sustained midplane decretion in the outermost part of the active zone (see figs. 23 and 25 in \citetalias{iwasaki_dynamics_2024}; note that their negative $\dot{M}$ indicates accretion). Third, their simulation forms a pressure maximum at the interface. Fourth, the active region with a heated corona is not `puffed up' (see fig. 7 in \citetalias{iwasaki_dynamics_2024}), in contrast to unheated-corona, MRI-active disc models (e.g. \citealt{zhu_global_2018}, \citetalias{jacquemin-ide_magnetic_2021}). \\
% 3 Model differences
\indent Despite the similarities in the results and setup, two key differences in their model drastically alter the evolution: an initial hourglass-shaped poloidal field, and an ambipolar Elsässer number profile with a complex radial form (see fig. 2 in \citetalias{iwasaki_dynamics_2024}) -- equal to unity at $R_\text{DZI}$, more than five orders of magnitude lower than ours in the inner dead zone ($\sim\!2R_\text{DZI}$) before rising back to unity by $\sim\!10$ au. Whilst this discrepancy is partly due to their heavier disc and the high sensitivity of non-ideal diffusivity calculations to assumptions about grain and metal chemistry \citep[e.g.][]{bai_wind-driven_2013, lesur_magnetohydrodynamics_2021}, we find no evidence of such intricate $\Lambda_\text{A}$ profiles in other calculations (e.g. middle panels of fig.~8 in \citealt{thi_radiation_2019} and bottom panel of fig.~10 in \citealt{lesur_magnetohydrodynamics_2021}). Moreover, their assumption that dust grains are fully sublimated ($T\!\gtrsim\!1400$ K), is unrealistic given the interface temperature ($T\!\sim\!800$--1000 K), leading to an underestimated interface sharpness \citep[e.g.][]{desch_high-temperature_2015, williams_ionization_2024}. Therefore, even though our non-ideal MHD model is simpler, we believe that the subsequent evolution of the inner disc is more realistic. \\
% 4 Results difference
\indent Their modelling choices generate a magnetic-flux-devoid `transition zone’ just outside the interface within five local orbits. This is magnetic-wind free, with suppressed accretion, and dynamically isolates the active region from the dead region, thus controlling the evolution of their entire simulation. It lies in the region where $\Lambda_\text{A}$ varies rapidly with radius, and is a result of sharply bent poloidal fields that have abruptly reconnected early in the simulation and then formed a large-scale poloidal-field loop (see figs.~15 and 19 in \citetalias{iwasaki_dynamics_2024}). This striking feature is absent from the inner-disc models in this paper and in \citetalias{roberts_global_2025}. \\
\indent To our knowledge, the only other examples of rapid, closed poloidal-field loop formation within the global dead-zone literature are the following: (i) \citet{yang_global_2021}, who employ $\Lambda_\text{A}\!\sim\!1$, initialise a sharply bent poloidal field, and truncate the disc at the outer radius within the model; and (ii) \citet{martel_magnetised_2022-1}, who also employ $\Lambda_\text{A}\!\sim\!1$ but initialise a purely vertical field to model a transition disc. We therefore suggest that the ‘transition zone’ is not a generic feature of discs, but rather a consequence of very specific initial conditions that include strong ambipolar diffusion. \\

%%%%%%%%%%%%%%%%%%%%%%%%%%%%%%%%%%%%%%%%%%%%%%%%%%%%%%%%%%%%%%%%%%%%%%%%%%%%%%%%%%%%%%%%%%%%%%%%%%%%
%%%%%%%%%%%%%%%%%%%%%%%%%%%%%%%%%%%%%%%%%%%%%%%%%%%%%%%%%%%%%%%%%%%%%%%%%%%%%%%%%%%%%%%%%%%%%%%%%%%%

\section{Conclusion}
\label{section:conclusion}

We performed five three-dimensional global VNF non-ideal MHD simulations, three of which included a dead--active zone interface, to investigate the inner regions of protoplanetary discs within the weak-magnetic-wind paradigm. \\ 
\begin{figure*}
    \centering
    \includegraphics[width=0.8\textwidth]{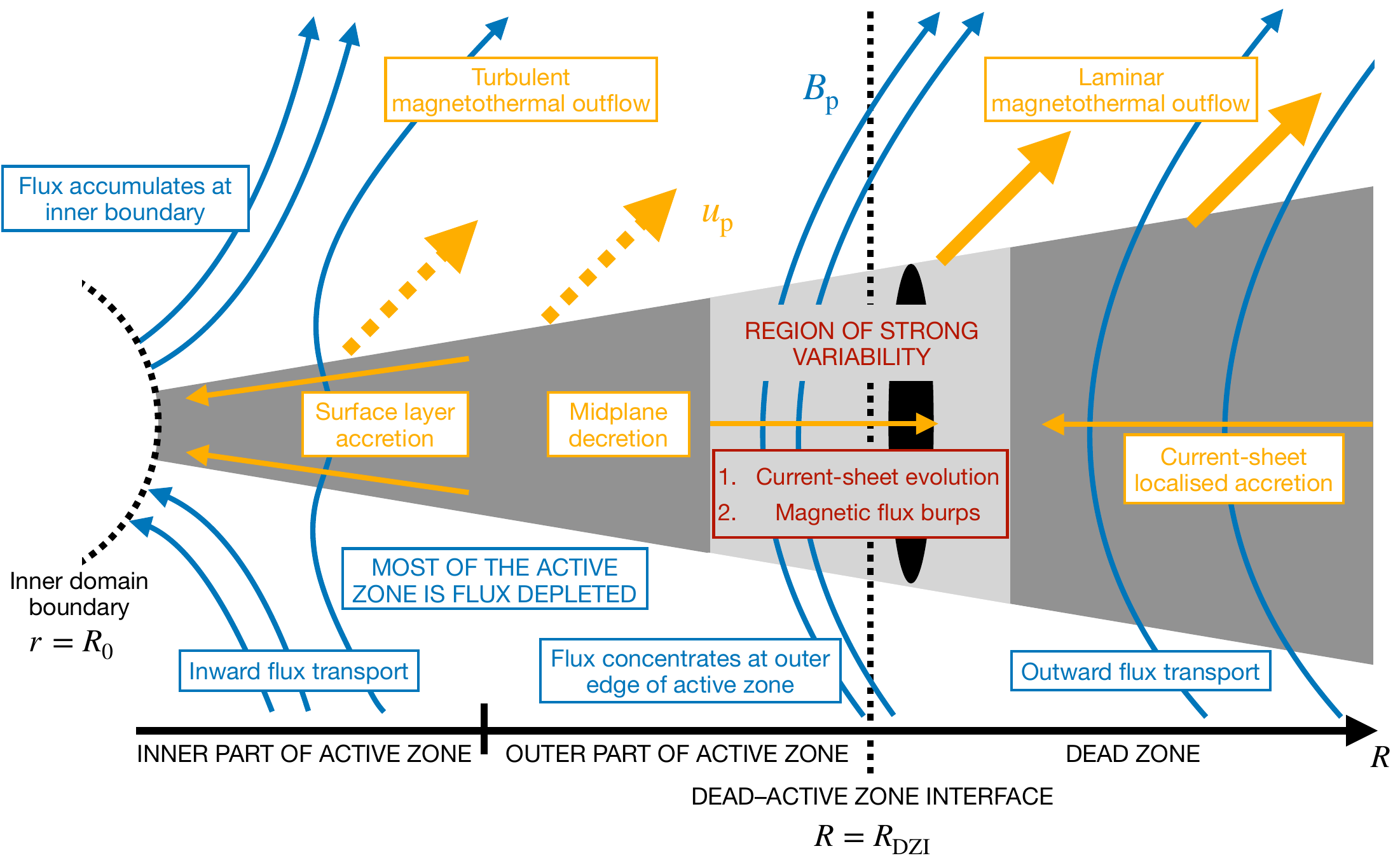}
    \caption{Schematic of the results, illustrating the mass flux (yellow) and the poloidal magnetic field (blue). The main findings are: (a) the dead--active zone interface (dashed black line) is a one-way barrier to inward transport of magnetic flux from the dead zone; (b) the emergence of two sources of interface-localised variability (red), likely associated with the challenge of maintaining a constant, vertically integrated current across distinct magnetic-field states; (c) the formation of a local pressure maximum (black oval) at the interface; and (d) the formation of a multi-zone outflow structure operating across all disc radii. This schematic should be compared with fig. 6 of \citetalias{iwasaki_dynamics_2024} and the outflows are only shown in the upper half for illustrative purposes.} 
    \label{fig:BAZ_summary_conclusion_schematic}
\end{figure*} 
\indent Our analysis focused on accretion structure, variability, magnetic flux transport, and outflows. The two principal findings are: (a) the dead--active zone interface is a one-way barrier to inward transport of magnetic flux from the dead zone, leading to the gradual depletion of magnetic flux throughout most of the active zone; and (b) the emergence of two novel sources of interface-localised variability, associated with the challenge of maintaining a constant, vertically integrated current across distinct magnetic-field states -- weak- and strong-field MRI, and MRI-dead and MRI-active -- over limited radial extents. The summary of our results is presented below.

\begin{enumerate}[align=left, labelindent=0pt, itemsep=5pt]
    \item The accretion structure exhibits 
    significant variability near the interface, as the vertically integrated current must remain matched across the evolving and distinct magnetic-field states. This constitutes a novel conceptual framework that can be applied to a broad range of MHD disc models. More specifically, we find the matching underlies (a) a progression through different magnetic epochs, as the poloidal current circuits seeks a dynamically favourable configuration, and (b)   intermittent magnetic flux expulsion events from the active zone into the dead zone (in one long-time numerical model). Overall, the accretion structure is characterised by sustained surface-layer accretion in the inner part of the active zone, dynamic but persistent midplane decretion across the dead--active zone interface, and current-sheet-localised accretion within the dead zone.
    \item Large-scale HD structure formation persists at the dead--active zone interface in the VNF regime in our models. The pressure maximum remains coherent through to the surface layers despite the local variability, suggesting that the interface can act as an effective site for dust accumulation throughout the disc's vertical extent. However, it remains necessary to evaluate the ubiquity of these HD structures elsewhere in the global VNF parameter space.
    \item The dead--active zone interface is a one-way barrier to inward transport of large-scale poloidal magnetic flux from the dead zone, leading to continual flux depletion throughout most of the active zone, and a concomitant reduction in the  accretion rate and flux transport velocity. Overall, magnetic flux (a) is transported radially inward within the inner part of the active zone, (b) is transported outward and accumulates at the dead zone boundary, within the outer part of the active zone, a process punctuated by episodic expulsion into the dead zone in one model, and (c) is transported outward within the dead zone. However, the sensitivity to the inclusion of the Hall effect, which is dynamically relevant in the dead zone, and the magnetosphere remains to be tested.
    \item In contrast to \citetalias{iwasaki_dynamics_2024}, there is no magnetic-flux-devoid, magnetic-wind-free `transition zone'. Instead, magnetothermal outflows operate across all disc radii, with a multi-zone outflow structure that partly reflects the radially varying launch conditions -- magnetic-field strength, turbulence level and density. The inner turbulent wind impinges upon the outer, more laminar wind. 
    \item A heated corona prevents the `puffing up' of weak VNF MRI-active disc regions, restricting the altitude of surface-layer accretion and inhibiting the vertical lofting of dust within the active region.
\end{enumerate}
To aid the interpretation of these results, Fig.~\ref{fig:BAZ_summary_conclusion_schematic} presents a schematic of the inner disc summarising our key conclusions. It is formatted similarly to fig. 6 in \citetalias{iwasaki_dynamics_2024} to facilitate direct visual comparison. \\
\indent Whilst we have proposed that both sources of strong variability at the dead--active zone interface are a consequence of matching the vertically integrated current across distinct and evolving magnetic-field states, this framework merits further investigation in more controlled settings. The impact of a heated corona on the VNF MRI-active disc structure also warrants further study. Beyond these open questions, our results suggest several directions for future work: incorporating the Hall effect, given its known dynamical influence in the dead zone \citep[e.g.][]{bai_hall_2017}; adopting a temperature-dependent ionisation model, and thus a non-static interface \citep[e.g.][]{latter_dynamics_2012,faure_thermodynamics_2014}; including a magnetosphere with an independent stellar dipole; and implementing dust--gas feedback. \\
\indent Whilst fully realistic long-term global simulations spanning the Class 0 to Class II stages remain unfeasible, the targeted simulations presented so far in this series have unveiled critical MHD-mediated processes in the inner regions of protoplanetary discs. These mechanisms are fundamental to the disc's secular evolution, the ubiquitously observed disc variability, and planetesimal formation within the putative habitable zone. 

%%%%%%%%%%%%%%%%%%%%%%%%%%%%%%%%%%%%%%%%%%%%%%%%%%%%%%%%%%%%%%%%%%%%%%%%%%%%%%%%%%%%%%%%%%%%%%%%%%%%
%%%%%%%%%%%%%%%%%%%%%%%%%%%%%%%%%%%%%%%%%%%%%%%%%%%%%%%%%%%%%%%%%%%%%%%%%%%%%%%%%%%%%%%%%%%%%%%%%%%%

\section*{Acknowledgements}

The authors would like to thank the referee, Zhaohuan Zhu, for useful comments that helped improve the manuscript. \\ 
\indent This research was supported by a PhD studentship from the Science and Technology Facilities Council (STFC), grant number 2603337. This work was performed using resources provided by the Cambridge Service for Data Driven Discovery (CSD3), operated by the University of Cambridge Research Computing Service (www.csd3.cam.ac.uk), and funded by Dell EMC, Intel, and Tier-2 funding from the Engineering and Physical Sciences Research Council (capital grant EP/T022159/1), as well as DiRAC funding from the Science and Technology Facilities Council (www.dirac.ac.uk). This work also used the DiRAC Extreme Scaling service (Tursa) at the University of Edinburgh, managed by the Edinburgh Parallel Computing Centre on behalf of the STFC DiRAC HPC Facility. The DiRAC service at Edinburgh was funded by BEIS, UKRI, and STFC capital and operations grants. DiRAC is part of the UKRI Digital Research Infrastructure. \\
\indent The simulations were run with version \href{https://github.com/idefix-code/idefix/releases/tag/v2.0.03}{v2.0.03} of \textsc{idefix}. The data were processed with Python via various libraries, in particular \href{https://github.com/numpy/numpy}{\textsc{numpy}}, \href{https://github.com/matplotlib/matplotlib}{\textsc{matplotlib}}, \href{https://github.com/scipy/scipy}{\textsc{scipy}}, and \href{https://github.com/volodia99/nonos}{\textsc{nonos}}.

%%%%%%%%%%%%%%%%%%%%%%%%%%%%%%%%%%%%%%%%%%%%%%%%%%
\section*{Data Availability}
The data underlying this article will be shared on reasonable request to the corresponding author.

% \FloatBarrier



\bibliographystyle{mnras}
\bibliography{bib} % if your bibtex file is called example.bib

\FloatBarrier
%\begin{thebibliography}{99}
%\end{thebibliography}

%%%%%%%%%%%%%%%%%%%%%%%%%%%%%%%%%%%%%%%%%%%%%%%%%%

%%%%%%%%%%%%%%%%% APPENDICES %%%%%%%%%%%%%%%%%%%%%

\appendix

\section{MRI-active and MRI-dead discs}
\label{appendix:turbulent_versus_laminar_discs}

% Introduction. Why ? and How ?
The investigation in this paper joins together a fully MRI-active and a fully MRI-dead disc in the weak VNF regime. Therefore, it is prudent to understand each case separately within the framework of our numerical model (see Section~\ref{section:methods_and_model}), establish a clear foundation for the mismatches -- magnetic field morphology, current-sheet configuration and flux transport direction -- between these regimes (see Section~\ref{section:introduction}), and verify that \textsc{idefix} reproduces results obtained with other codes (e.g. \citealt{zhu_global_2018}; \citealt{gressel_global_2020}; \citetalias{jacquemin-ide_magnetic_2021}; \citealt{lesur_systematic_2021}). \\
% Magnetic flux transport comparison. Introduce figure

\subsection{Magnetic morphology and flux evolution}
\begin{figure}
    \centering
	\includegraphics[width=0.98\columnwidth]{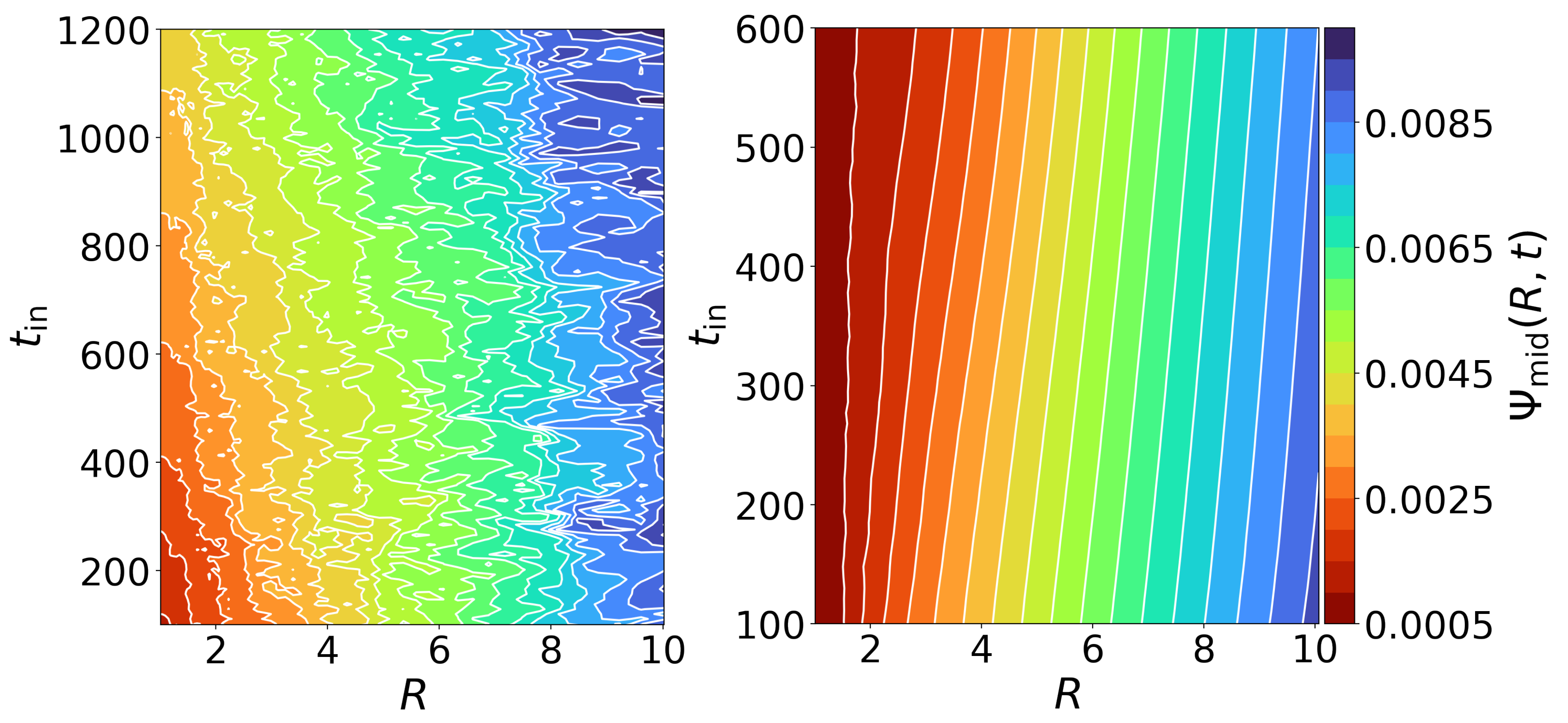}
    \caption{Evolution of the magnetic flux threading the disc midplane $\Psi_{\text{mid}}(R,t)$, as defined in equation \eqref{eq:define_magnetic_flux_mid}. The contour lines represent magnetic field lines and they are transported inward in \texttt{NF-AZ} (left), and outward in \texttt{NF-DZ} (right). The cylindrical radius $R$ is in code units ($R_0$).}
    \label{fig:compare_magnetic_flux_transport_laminar_turbulent}
\end{figure}

\begin{figure}
    \centering
	\includegraphics[height = 0.36\textwidth]{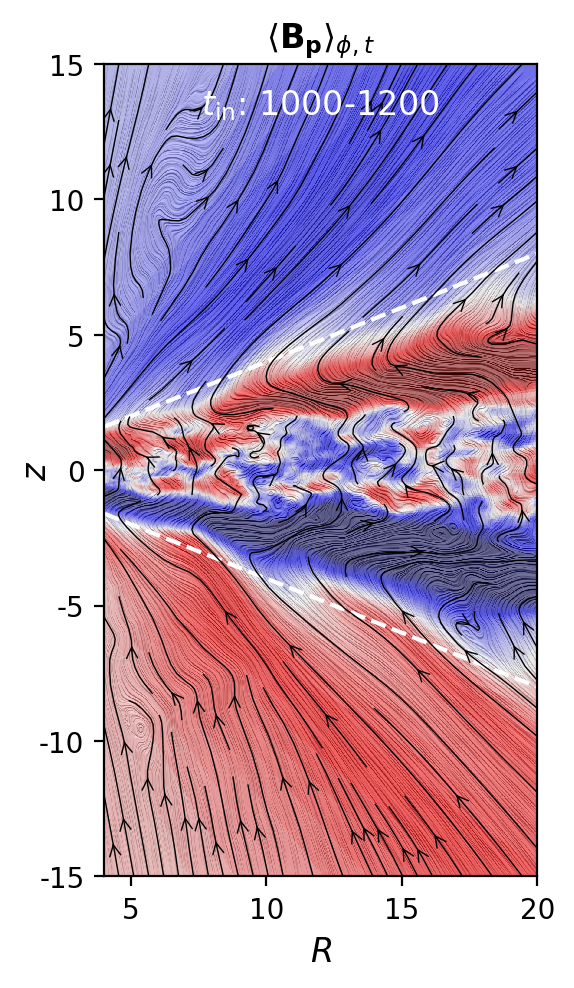}
	\includegraphics[height = 0.36\textwidth]{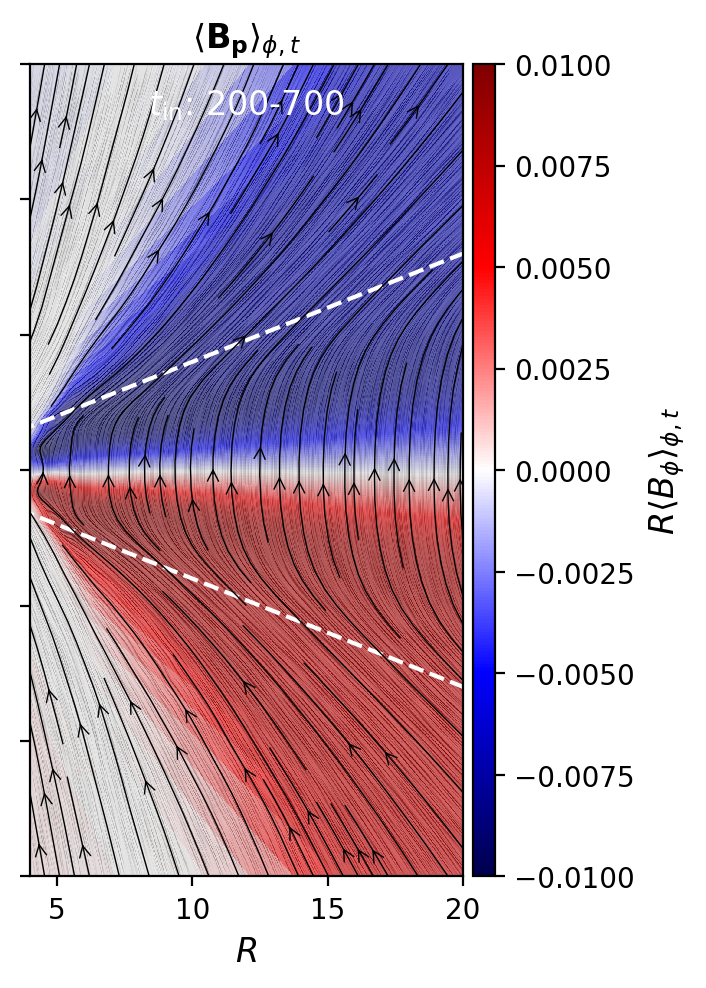}
    \caption{Magnetic field structure in the meridional ($R,z$) plane for \texttt{NF-AZ} (left) and \texttt{NF-DZ} (right). In the MRI-active disc (left) the toroidal field (background colour) forms a coherent surface-layer structure, which is vertically constrained by the disc--corona temperature transition (dashed white line). The poloidal magnetic field lines (black lines and LIC) are pinched inward in the surface-layers in the active disc. All quantities have been averaged in azimuth and time, and $R$ is in code units ($R_0)$.}
    \label{fig:compare_magnetic_LIC_laminar_turbulent}
\end{figure}

Fig.~\ref{fig:compare_magnetic_flux_transport_laminar_turbulent} confirms that global magnetic flux is transported in opposite directions: inward in the fully ideal model (\texttt{NF-AZ}) and outward in the fully non-ideal model (\texttt{NF-DZ}), in the absence of the Hall effect. The figure shows space--time ($R,t$) diagrams of the magnetic flux threading the midplane, $\Psi_{\text{mid}}$, defined in equation~\eqref{eq:define_magnetic_flux_mid}. \\
% Magnetic field structure comparison. Emphasise current-sheet and mismatch. Introduce figure
\indent Meanwhile, Fig.~\ref{fig:compare_magnetic_LIC_laminar_turbulent} compares the magnetic field morphology of the two models, showing the meridional structure of the normalised toroidal field, $R\langle {B_\phi} \rangle_{\phi,t}$, and the poloidal magnetic field configuration, $\langle\mathbf{B_p}\rangle_{\phi,t}$, averaged in azimuth and time. The MRI-active model (left panel) exhibits a complex morphology with a pronounced vertical structure; an MRI-turbulent disc interior ($|z| \lesssim 2H$), with coherent field structures in magnetostatic equilibrium in the surface layers ($2H \lesssim |z| \lesssim 4H$) \citepalias{jacquemin-ide_magnetic_2021}. The poloidal field is pinched inward in these surface layers and there are two surface-layer current sheets where the accretion is localised. In contrast, the MRI-dead disc (right panel) is laminar, with an hourglass-shaped poloidal field structure and a single, diffuse midplane current sheet where accretion is localised. The vertical location of this sheet depends on the non-ideal MHD diffusivities and need not coincide with the disc midplane (e.g. fig. 1 in \citealt{riols_ring_2020}; fig. 12 in \citealt{lesur_systematic_2021}). Thus, the two unconnected regimes differ clearly in both the number of current sheets and the vertical disc structure.

\subsection{Impact of heated corona on MRI-active disc structure}

% What is different 
The corona is hot due to a combination of strong stellar irradiation \citep[e.g.][]{aresu_x-ray_2011} and inefficient gas cooling, the latter arising from reduced thermal accommodation on dust particles in this dust-sparse region \citep[see fig. 10 in][]{thi_radiation_2019}. However, this is often unaccounted for in global VNF MRI-active models (e.g. \citealt{zhu_global_2018}; \citetalias{jacquemin-ide_magnetic_2021}). \\
\indent In these comparable unheated models -- with $\varepsilon=0.1$ and an initial midplane $\beta$ of $10^4$ -- shown in the right panel of fig. 19 in \citet{zhu_global_2018} and fig. 2 in \citetalias{jacquemin-ide_magnetic_2021}, the system is `puffed up', with large-scale coherent field structures located at elevated heights of roughly $|z|=8$--$10H$. In contrast, in our heated-corona model (left panel of Fig.~\ref{fig:compare_magnetic_LIC_laminar_turbulent}), the coherent magnetic field structure is confined to the disc--corona temperature transition at $|z|=4H$. This confinement restricts the altitude of accretion and the extent of vertical dust lofting, which would otherwise shadow the inner disc or alter the local temperature structure. \\
\indent Despite the sensitivity of current-sheet morphology and MRI-driven turbulence to resolution, the differences are unlikely to be resolution dependent, because \citet{zhu_global_2018} use a coronal resolution that is roughly twice as coarse as ours and \citetalias{jacquemin-ide_magnetic_2021}, and the meridional grid of \citetalias{jacquemin-ide_magnetic_2021} is geometrically stretched from $|z|\gtrsim5.5H$, significantly below the altitude of their coherent structures. Therefore, the confinement is likely a physical consequence of the imposed temperature transition. We hypothesise that the modified vertical gas-pressure gradient, alters the magnetostatic equilibrium and prevents the vertical expansion of the MRI-active disc. \\
\indent To our knowledge, this is the first global VNF MRI-active protoplanetary disc simulation to include a temperature transition at the corona, highlighting the need for future dedicated studies.

\section{Inner boundary condition test} 
\label{appendix:boundary_test}

\begin{figure}
    \centering
    \includegraphics[height=0.4635 \columnwidth]{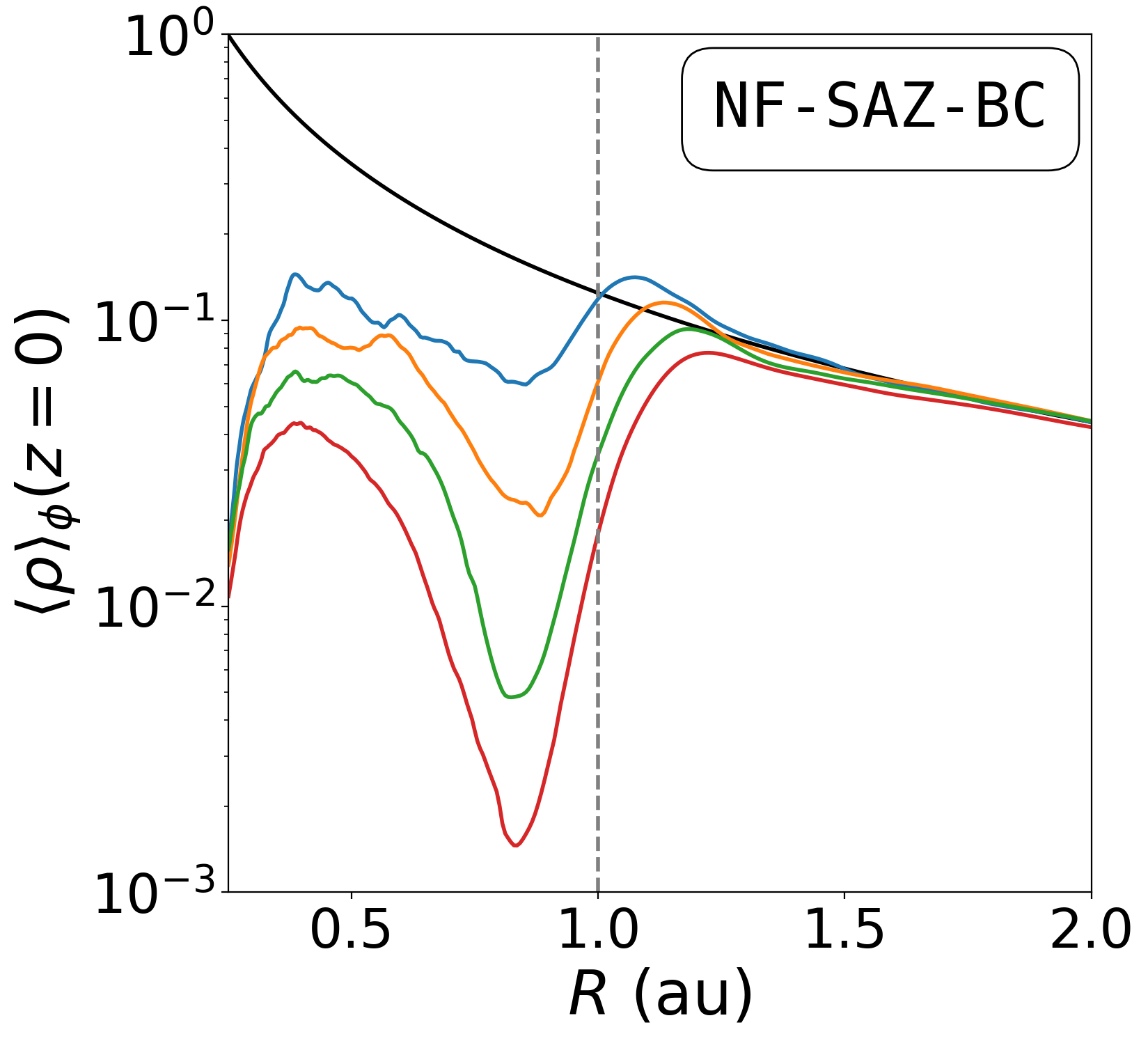}
    \includegraphics[height=0.454745 \columnwidth]{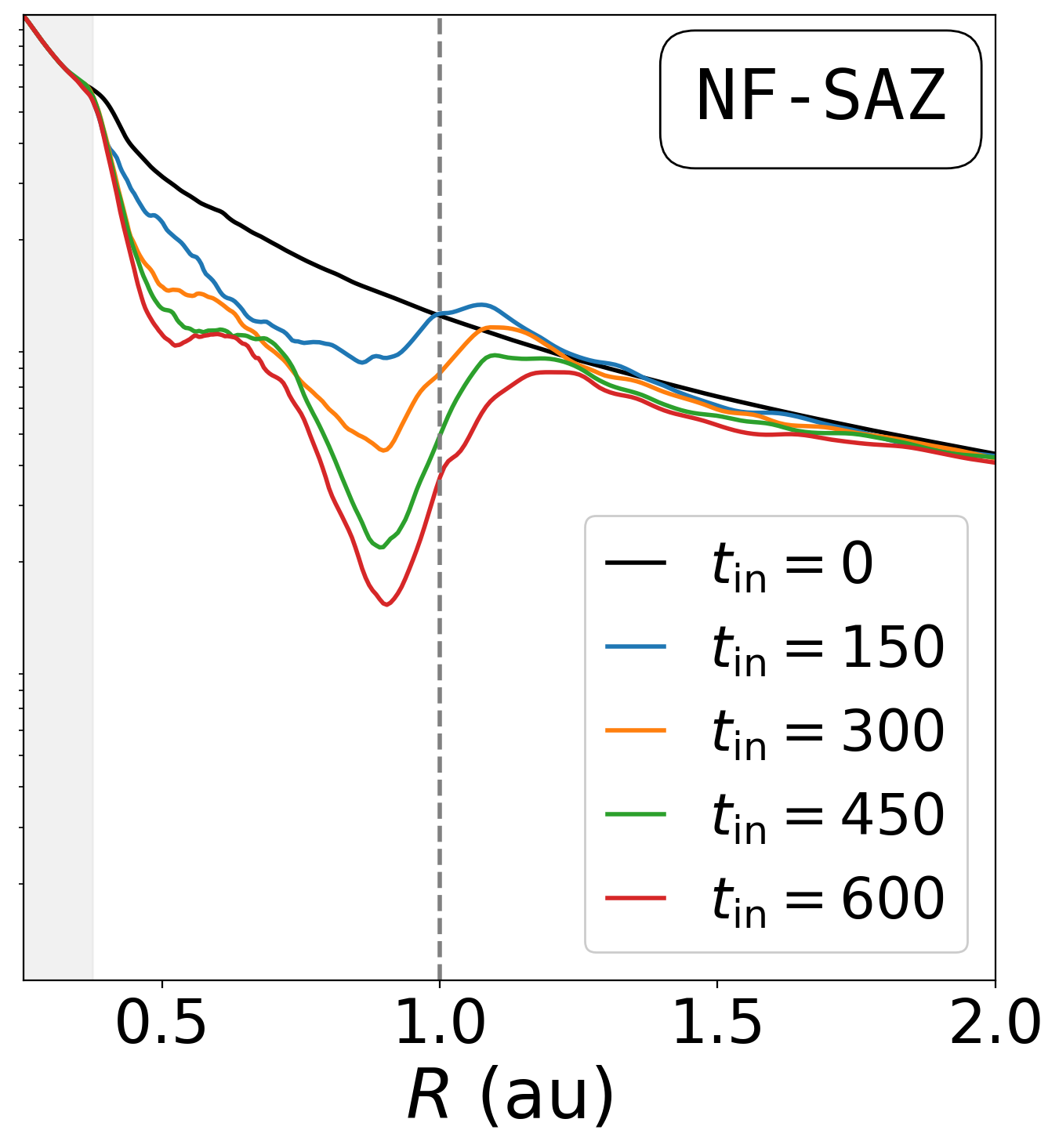}
    \caption{Radial profiles of the azimuthally averaged density, $\langle \rho \rangle _\phi$, at the midplane. \texttt{NF-SAZ-BC} (left) allows unrestricted evolution of the density, whilst \texttt{NF-SAZ} (right) restores the density within the inner radial buffer (grey-shaded region).}
    \label{fig:SAZBC_SAZ_compare_HD_density}
\end{figure}

% Introduce
The inner domain boundary lies near the dead--active zone interface, so we use the model \texttt{NF-SAZ-BC} to test the robustness of our results to an alternative, physically motivated inner radial BC. \\
% Describe BC
\indent In contrast to the BC described in Section~\ref{section:boundary_conditions}, we now use the other extreme limit: rather than restricting the loading of mass onto the protostar, the magnetospheric accretion rate is unimpeded, allowing for the draining of mass from the extreme inner disc. Specifically, we alter the BC to mimic \citetalias{jacquemin-ide_magnetic_2021} by neglecting the entire inner buffer region, which is evanescent, density restoring and resistive, and setting the ghost cells to copy $B_\theta$ and $B_\phi$ from the innermost active cell. Removing restrictions on the density evolution permits mass to drain freely from the extreme inner disc, and the impact of this change is evident in the radial profiles of the azimuthally averaged midplane density, $\langle \rho \rangle_\phi$, for \texttt{NF-SAZ-BC} and \texttt{NF-SAZ} shown in Fig.~\ref{fig:SAZBC_SAZ_compare_HD_density}.
\FloatBarrier
% Results. Three and brevity
\indent The three key results from this BC test (not shown for brevity) are: (i) the interface continues to act as a one-way barrier to inward transport of magnetic flux from the dead zone to the active zone. Consequently, magnetic flux is depleted from most of the active zone, and it still accumulates in the outer part of the active zone, reaching $\beta_\text{p}^\text{coh}\!\sim\!10^1$; (ii) in contrast with \texttt{NF-SAZ} but consistent with \texttt{NF-BAZ}, no flux burp occurs (up to $t_\text{in}^\text{end}=800$) and magnetic flux is transported inward in the inner part of the active zone for the entire simulation, accumulating at the inner domain boundary, albeit at a faster initial rate of $v_\Psi/u_\text{K}\!\sim\!-8\times10^{-2}$; and (iii) large-scale HD structures still form at the interface, followed by the complex interface-localised current-sheet evolution. \\
% Conclusion
\indent Therefore, whilst this test supports all the conclusions of \texttt{NF-BAZ}, it highlights the limitations of drawing exact quantitative measurements from global simulations and the sensitivity of some processes to BCs, even when physically motivated.

\section{Physical unit conversions}
\label{appendix:units}

The results are presented in code units to assist comparison between the different models. The only physical scaling provided is that the fixed dead--active zone interface corresponds to $R=1\,\text{au}$. \\
\indent Table~\ref{table:code_physical_unit_conversion} presents the conversion from code to physical units for all the quantities shown in this paper, using the case of a disc around a 1M$_{\odot}$ star (typical for a T Tauri star), with a surface density of $\Sigma(R) = 300\,(R/1\,\text{au})^{-1/2}$ g$\,$cm$^{-2}$ \citep[e.g.][]{miotello_setting_2023}. The scaling with $R$ corresponds to the initial conditions: a self-similar density scaling and a constant aspect ratio of $\varepsilon = 0.1$.\\
\indent The physical units for temperature assume an ideal gas in a sufficiently dense environment \citep[e.g.][]{flock_3d_2017}, with $T = \mu m_\text{u} c_\text{s}^2 / k_\text{B}$, where $\mu = 2.34$ is the mean molecular weight, $m_\text{u}$ the atomic mass unit and $k_\text{B}$ the Boltzmann constant. The factor of $\varepsilon^2$ absent in $T_0$ in Table~\ref{table:code_physical_unit_conversion} is accounted for in the definition of the disc temperature $T_\text{d}$ in Section~\ref{section:temp_prescription}. The choice of $\varepsilon = 0.1$, which is a numerical compromise to adequately resolve the MRI, renders the disc temperature artificially high (e.g. $2500$ K at $R_\text{DZI}$).
\begin{table}
\centering
\captionof{table}{Conversion between code and physical units for a disc around a 1M$_{\odot}$ star, with a (gas) surface density of $\Sigma(R) = 300\,(R/1\,\text{au})^{-1/2}$ g$\,$cm$^{-2}$ and a constant aspect ratio of $\varepsilon=0.1$. This is evaluated at two representative inner radii, determined by setting $R_{\text{DZI}}=1\,\text{au}$: $ R_0 = 0.1\,\mathrm{au}$ for \texttt{NF-BAZ} and $ R_0 = 0.25\,\mathrm{au}$ for \texttt{NF-SAZ}. The factor of $\sqrt{4\pi}$ from the Alfvén speed is absorbed into the magnetic field code units in \textsc{idefix}. The disc temperature at $R_0$ is $T_\text{d}=T_0\varepsilon^2$ (see Section~\ref{section:temp_prescription}).}

\renewcommand{\arraystretch}{1.1}
\begin{tabular}{ c c c} 
\hline
Code unit & Physical unit  & Physical unit \\ 
\hline
$R_0$ & $0.1\,\text{au}$ & $0.25\,\text{au}$  \\
$t_{\text{in}}=2\pi\Omega_{\text{K}_0}^{-1}$ & 0.0316 yr & 0.125 yr \\
$u_0 = \Omega_{\text{K}_0} R_0$ & 94.2 km$\,$s$^{-1}$ & 59.6 km$\,$s$^{-1}$   \\
$\Sigma_0$ & $949\,$g$\,$cm$^{-2}$ & $600\,$g$\,$cm$^{-2}$ \\
$\rho_0 = \Sigma_0/(\sqrt{2\pi}\varepsilon R_0)$ & $2.53\times10^{-9}\,$g$\,$cm$^{-3}$ & $6.40\times10^{-10}\,$g$\,$cm$^{-3}$ \\
$\dot{M_0} = 2\pi R_0 \Sigma_0 u_0$ & $1.33\times 10^{-3}$ M$_\odot\,$yr$^{-1}$ & $1.33\times 10^{-3}$ M$_\odot\,$yr$^{-1}$\\
$B_0 = u_0\sqrt{\rho_0}$ & 475 G & 151 G  \\
$T_0 = \mu m_\text{u} u_0^2/k_\text{B}$ & $2.50\times 10^6$ K & $1.00\times 10^6$ K \\
\hline 
\end{tabular}
\label{table:code_physical_unit_conversion}
\end{table}

\section{Turbulent transport of magnetic flux}
\label{section:turbulent_transport_magnetic_flux}

Here we present a simple analytic model for turbulent magnetic flux transport, extending the analysis from Section~\ref{section:mft_inner_az}. It is motivated by the power-law scalings of $|v_\Psi|$ with $\beta_z$ reported in Section~\ref{section:impact_flux_draining} and \citetalias{jacquemin-ide_magnetic_2021}.

\subsection{Model setup}
In this one-dimensional model, the poloidal magnetic flux function $\Psi$ is defined via $B_z= (1/R)\d_R \Psi$ and obeys \citep[e.g.][]{guilet_transport_2012}
\begin{equation} \label{flux1}
\d_t\Psi + v_\Psi \d_R \Psi = 0.
\end{equation}
We assume the advection speed is $v_\Psi= - A u_K \beta_z^{-a},$ where $A$ and $a$ are dimensionless constants. Furthermore, we suppose that the pressure $P$ is steady in time and has a power law profile: $P= P_0 (R/R_0)^{-b}$, for constant $b$. Typically, $c_s^2\propto R^{-1}$ and $\rho\propto R^{-3/2}$, which yields $b= 5/2$. The simulations of \citetalias{jacquemin-ide_magnetic_2021} suggest $a=1$ and $A=10$, whilst our simulations of the active zone yield $a=2$ and $A=7\times 10^5$ (see Fig.~\ref{fig:BAZ_v_psi_betap_active_zone}). The initial condition is a constant $\beta_z$ with respect to radius. Finally, to ease notation, in the following we scale space by $R_0$, time by $1/\Omega_0$, and $B_z$ by $\sqrt{8\pi P_0}$.

%Thus at $t=0$, $B_z\propto r^{-b/2}$, 
%and $\Psi\propto r^{-b/2+2}$. 

%$v_K= r_0\Omega_0 \left(r/r_0\right)^{-1/2}$, with $r_0$ being the inner radius of our domain and $\Omega_0=\Omega(r_0)$, and finally $\beta= 2P/B_z^2$. Furthermore,  with $b$ another dimensionless constant and $P_0$ the pressure at the inner boundary. Obviously, this is not what's happening in the simulations, as the density profile will vary in time. 

%Note that these assumptions mean that
%$$ \dfrac{v_\Psi}{r_0\Omega_0} = -2^{-a} A \left(\dfrac{r}{r_0}\right)^{^{ab-1/2}} \left( \dfrac{B_z^2}{P_0}\right)^a. $$

\subsection{Transformations and solution via characteristics}

Equation \eqref{flux1} is difficult to solve because it is nonlinear in the spatial derivatives. We hence operate on \eqref{flux1} with $(1/R)\d_R$ and get back
\begin{equation} \label{flux2}
\d_t B_z + \dfrac{1}{R}\d_R (R v_\Psi B_z) = 0.
\end{equation}
Next, we introduce the variables:
\begin{equation}
    \Theta= R^\frac{2ab+1}{2(2a+1)} B_z, \qquad \xi = \dfrac{2 R^\frac{8a-2ab+3}{2(2a+1)}}{A(8a-2ab+3)}.
\end{equation}
These reduce equation \eqref{flux2} to the more manageable
\begin{equation} \label{flux3}
\d_t \Theta - \Theta^{2a} \d_\xi \Theta = 0,
\end{equation}
which, though nonlinear, can be tackled using the method of characteristics.

We select curves $\xi=\xi(t)$ so that $d\xi/dt = - \Theta^{2a}$. On these curves $\Theta$ is a constant. If $\xi(0)=\xi_i$ and $\Theta = \Theta_i$ on the $i$'th characteristic, then the latter are given by straight lines
$\xi= -\Theta_i^{2a} t + \xi_i.$ Using the initial condition, we transform back to cylindrical radius to find:
\begin{equation} \label{char}
R(t)= R_i \left( 1- \dfrac{t}{t_i}  \right)^{\frac{2(2a+1)}{8a-2ab+3}},\quad t_i= \dfrac{2}{8a-2ab+3}\cdot \dfrac{\beta_0^a R_i^{3/2}}{A} ,
\end{equation}
where $R_i=R(0)$ and $\beta_0$ is the initial uniform plasma beta. Note that $t_i\propto R_i/v_\Psi(R_i)$, and supplies the characteristic timescale for these curves. However, $B_z$ is not a constant on the characteristics (only $\Theta$ is). In fact $B_z^2\propto R(t)^{-\frac{2ab+1}{(2a+1)}}$, which \emph{increases} more slowly along the curve compared with the steady $P\propto R(t)^{-b}$ for $b>1$. Thus $t_i$ only provides a lower bound on the typical timescale of magnetic flux transport. 

%The characteristic timescale for these curves is $t_i$. Suppose $\beta_0=10^4$. For Jona's sims and for $r_i\lesssim 10$, we get $t_i\sim 10^4 \Omega_0^{-1}$. For Matt's sim, we get $t_i\sim 10^5 \Omega^{-1}_0$. However, $t_i$ is only a lower bound for the typical time of magnetic field transport. That is because 

%To obtain $B_z(r,t)$ we must first solve \eqref{char} for $r_i$:
%\begin{equation} \label{ugly}
%(r r_i^{-1})^{\dfrac{8a-2ab+3}{2(2a+1)}} = 1 - %\dfrac{A(8a-2ab+3)t}{2\beta_0^2} r_i^{-3/2},
%\end{equation} 
%which generally is a transcendental equation, though if $\dfrac{8a-2ab+3}{2(2a+1)}$ is a rational number it can be reduced to a polynomial equation. Once we obtain $r_i$, we obtain $B_z$ from $\Theta=\Theta_i$. Using the initial conditions, this can be manipulated into 
%$$ B_z = \sqrt{2/\beta_0}{}r^{-\dfrac{2ab+1}%{2(2a+1)}} r_i^\dfrac{1-b}{2(2a+1)}.$$
%Alternatively, one can solve this equation for $r_i$ and substitute into \eqref{ugly} and then obtain an equation for $B_z$ directly. 

 \subsection{Solving for $B_z$ and $\Psi$}

To obtain $B_z$, we first must use the initial condition to find $R_i$ in terms of $B_z$, i.e. solve $\Theta=\Theta_i$. Then we insert $R_i$ into \eqref{char}. Doing so supplies a nonlinear algebraic equation for $B_z$, which in general must be solved numerically. Then $\Psi$ can be achieved by integrating.

\begin{figure}
    \centering
	\includegraphics[width=\columnwidth]{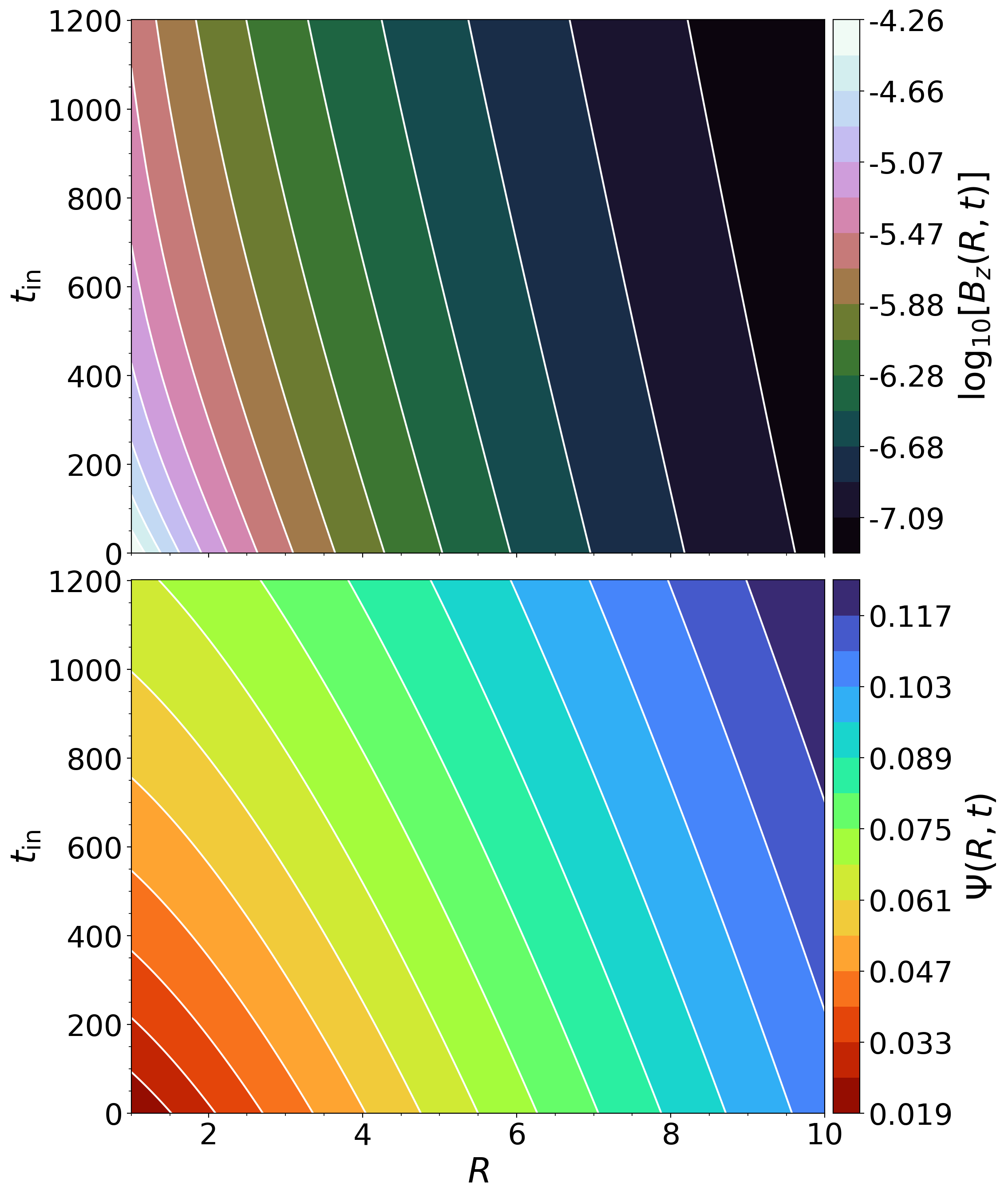}
    \caption{Top: Contours of $\log_{10}(B_z)$ in a space--time diagram, according to the early-phase analytical model, equation \eqref{Bz_analytic}. Bottom: Evolution of the associated midplane flux function $\Psi$, see equation~\eqref{Psi}. In both cases $\beta_0=10^4$, $A=10$ and $a=1$. As in the text a factor of $\sqrt{4\pi}$ has been absorbed into $B$ (and $\Psi$), and a factor of $\varepsilon=0.1$ is required to enable comparison with the global models shown in Figs.~\ref{fig:baz_magnetic_flux_transport_plot} and \ref{fig:compare_magnetic_flux_transport_laminar_turbulent}.}
    \label{fig:analytic_flux_transport}
\end{figure}

When $a=1$ and $b=5/2$ \citepalias[as in][]{jacquemin-ide_magnetic_2021}, the problem reduces to
\begin{equation}
X^3 -  X - \dfrac{3A t}{\beta_0 R^{3/2}} =0,
\end{equation}
where $X= \beta_0 R^{-5/2} B_z^{-2}$. One finds two regimes of evolution, an early-time `fast' and a later-time `slow' inward drift of $B_z$. 
 During the `early evolution' phase, i.e. for $t \ll R^{3/2}\beta_0/A$, we have
\begin{equation} \label{Bz_analytic}
B_z \approx \beta_0^{-1/2} R^{-5/4} \left(1+ \tfrac{3}{2} R^{-3/2} t A \beta_0^{-1}\right)^{-1/2}.
\end{equation}
During the `later evolution' phase, $t\gg R^{3/2}\beta_0/A$, we have instead 
\begin{equation}
B_z \approx \beta_0^{-1/2} R^{-1} \left(\dfrac{3 A t}{\beta_0}\right)^{-1/6}.
\end{equation}
 For larger $\beta_0$, the later phase begins after quite some time, especially at larger radii, and thus the assumption of a static background pressure is likely violated. Thus we only focus on the early phase, which for $\beta_0\!\sim\!10^4$ and $R\!\lesssim\!10$ extends for several thousand inner orbits. In the top panel of Fig. \ref{fig:analytic_flux_transport} we plot the contours of $B_z$ in a space--time diagram for the early-phase evolution. 
 
 %Moreover, in simulations, the absence of any flux resupply from the outer boundary will impact on the solution; but in the above analysis we have assumed the disk is infinitely long. So comparison with simulations will fail on this count too.
 
 Next, by integrating $RB_z$, one can easily obtain $\Psi$ in the early phase: \begin{equation} \label{Psi}
 \Psi = \dfrac{4 R^{3/4}}{3\sqrt{\beta_0}} \sqrt{1+ \tfrac{3}{2} A t \beta_0^{-1} R^{-3/2}}.
 \end{equation}
A representative space--time diagram is plotted in the bottom panel of Fig.~\ref{fig:analytic_flux_transport}. The agreement with the left panel of Fig.~\ref{fig:compare_magnetic_flux_transport_laminar_turbulent} and fig.~5 from \citetalias{jacquemin-ide_magnetic_2021} is very good.
Note that the flux contours, for fixed $\Psi$, are then
  \begin{equation}
  R(t) = R_i \left(1- \dfrac{3}{2}\dfrac{|v_\Psi(R_i) | }{u_\text{K}(R_i)} R_i^{-3/2} t \right)^{2/3},
  \end{equation}
 where $R_i= R(0)$ again. Although we adopt a different model for $v_\Psi$, this approximate formula agrees with equation (26) in \citetalias{jacquemin-ide_magnetic_2021}, who assume simply that $v_\Psi \propto u_\text{K}$ and therefore neglect their measured dependence of $v_\Psi$ on $\beta_z$ (based on only two data points). In this  early-time evolutionary phase, it would appear that the $\beta_z$ dependence is not important and that the advection of magnetic flux is controlled by the local physics at the contour foot point.
 
 %(Recall, though, that at later times, this formula will become inaccurate.) The J-Ide formula assumes, essentially, that local physics (and hence the local $v_\Psi$) is controlling the advection of flux at any moment, which makes sense to me. This doesn't seem quite the case here, as it is the $v_\Psi$ at the footpoint that carries through... maybe I've done something wrong or haven't thought through this properly yet.... or perhaps in this early-evolution regime, the local $v_\Psi$ is well approximated by its value at the footpoint, with corrections coming in at next order (which I haven't computed). It is true that the contours don't move very far from their footpoint in this regime... I dunno.

 %Anyway, in Figure 1, I've plotted some of these results. The flux contours do seem to agree loosely with Matt's sims (see left panel of Fig A1) and also J-Ide (his Fig 5 from the 2021 paper).  I haven't yet coded up the exact (numerical) solution for $\Psi$. Will do that if I can.
 
 %I confess, I don't understand the later time asymptotic formula for $\Psi$, but maybe it doesn't matter. The assumption of fixed $P$ is completely crap at that point.

Finally, if we set $a=2$ and $b=5/2$ as suggested by \texttt{NF-BAZ} (see Section~\ref{section:impact_flux_draining}), which incorporates both active and dead zones, we obtain the following quintic equation
 \begin{equation}
 X^5 - X^2 - \dfrac{9A t}{2\beta_0^2R^{3/2}} =0.
 \end{equation}
This also yields an early- and a late-phase evolution. In fact, the early-stage asymptotic expressions are identical to the $a=1$ case, only that $\beta_0$ is replaced by $\beta_0^2$. Thus, the early-time evolution is the same, but is conducted on a timescale that reflects the difference in the initial magnetic flux advection speed. Our $v_\Psi$ power-law scaling yields an initial $|v_\Psi|$ at $\beta_0=10^4$ that is seven times higher than \citetalias{jacquemin-ide_magnetic_2021}, rendering an exact comparison between these two models futile. 

\subsection{Impact of dead--active zone interface}
\label{section:impact_daz_flux_transport}
Ultimately, including a dead--active zone interface slows the inward-directed magnetic flux transport in the inner part of the active zone in our models. Representative contours at $R_i=3, 4$ and 5$R_0$ in \texttt{NF-BAZ} evolve roughly twice as slowly as the equivalent contours in the MRI-active model \texttt{NF-AZ} (compare Fig.~\ref{fig:baz_magnetic_flux_transport_plot} to the left panel of Fig.~\ref{fig:compare_magnetic_flux_transport_laminar_turbulent}), despite the models using the same grid and BCs. \\
\indent In conjunction with the (possibly) different $v_\Psi$ power-law scaling with $\beta_z$, this is likely due in part to the maximum cylindrical radius of magnetic-flux resupply -- which acts as the outer boundary in these analytic calculations -- being much smaller in models with an interface: in a fully MRI-active disc, it corresponds to the outermost edge of the disc ($\sim\!100$--500 au) rather than $\sim\!0.5$ au. Eventually, even fully MRI-active discs might experience the influence of limited magnetic flux resupply. Therefore, despite the complexities of global magnetic flux transport, the evolution in the inner part of the active zone ($R\lesssim R_\text{DZI}/2$) can perhaps be interpreted, to leading-order, as analogous to an extremely small, fully MRI-active disc embedded interior to the rest of the model.

%%%%%%%%%%%%%%%%%%%%%%%%%%%%%%%%%%%%%%%%%%%%%%%%%%

% Don't change these lines
\bsp	% typesetting comment
\label{lastpage}
\end{document}